\documentclass[%
aip,
reprint,
superscriptaddress,
nobibnotes,
amsmath,amssymb, 
]{revtex4-1}

\usepackage[utf8]{inputenc}

\usepackage{layouts}
\usepackage{soul}

\newlength\fwidth
\usepackage{graphicx}
\usepackage{bm}
\usepackage[separate-uncertainty]{siunitx}
\newcommand{\highlightcorrect}[1]{{#1}}
\newcommand{\correcttext}[2]{{#2}}

\begin{document}
	
	\title{Stability of ionisation-injection-based laser-plasma accelerators}
	
	\author{Simon Bohlen}
	\email{simon.bohlen@desy.de}
	\affiliation{%
		Deutsches Elektronen-Synchrotron DESY, Notkestr. 85, 22607 Hamburg, Germany
	}%
	\affiliation{%
		Universit\"at Hamburg, Luruper Chaussee 149, 22761 Hamburg, Germany
	}
	
	\author{Jonathan C. Wood}%
	\email{jonathan.wood@desy.de}
	\affiliation{%
		Deutsches Elektronen-Synchrotron DESY, Notkestr. 85, 22607 Hamburg, Germany
	}%
	
	\author{Theresa Br\"ummer}%
	\affiliation{%
		Deutsches Elektronen-Synchrotron DESY, Notkestr. 85, 22607 Hamburg, Germany
	}%
	
	\author{Florian Gr\"uner}%
	\affiliation{%
		Universit\"at Hamburg, Luruper Chaussee 149, 22761 Hamburg, Germany
	}
	\author{Carl. A. Lindstr\o m}%
	\affiliation{%
		Deutsches Elektronen-Synchrotron DESY, Notkestr. 85, 22607 Hamburg, Germany
	}%
	
	\author{Martin Meisel}%
	\affiliation{%
		Deutsches Elektronen-Synchrotron DESY, Notkestr. 85, 22607 Hamburg, Germany
	}%
	\affiliation{%
		Universit\"at Hamburg, Luruper Chaussee 149, 22761 Hamburg, Germany
	}

	\author{Theresa Staufer}%
	\affiliation{%
		Universit\"at Hamburg, Luruper Chaussee 149, 22761 Hamburg, Germany
	}
	
	\author{Richard D'Arcy}%
	\affiliation{%
		Deutsches Elektronen-Synchrotron DESY, Notkestr. 85, 22607 Hamburg, Germany
	}%
	
	\author{Kristjan P\~oder}%
	\affiliation{%
		Deutsches Elektronen-Synchrotron DESY, Notkestr. 85, 22607 Hamburg, Germany
	}%

	\author{Jens Osterhoff}%
	\affiliation{%
		Deutsches Elektronen-Synchrotron DESY, Notkestr. 85, 22607 Hamburg, Germany
	}%
	
	
	\begin{abstract}
		Laser-plasma acceleration (LPA) is a compact technique to accelerate electron bunches to highly relativistic energies, making it a promising candidate to power radiation sources for industrial or medical applications. 
		We report on the generation of electron beams from an \SI{80}{\mega\electronvolt}-level LPA setup based on ionisation injection (II) over a duration of 8 hours at a repetition rate of 2.5 Hz, resulting in 72000 consecutive shots with charge injection and acceleration.
		Over the full operation time the moving averages of the total beam charge of \SI{14.5}{\pico\coulomb} and the charge between 70--\SI{80}{\mega\electronvolt} did not drift on a detectable level.
		The largest source of shot-to-shot jitter was in the beam charge (26\% standard deviation), which was most strongly correlated with fluctuations in the plasma density (3.6\% standard deviation). 
		Particle-in-cell simulations demonstrate that this was chiefly caused by stronger laser self-focusing in higher density plasmas, which significantly increased the ionised charge along with the emittance of the beam.
		The nonlinearity of this process imposes tight constraints on the reproducibility of the laser-plasma conditions required for a low jitter II-LPA output if self-focusing plays a role in the laser evolution.
	\end{abstract}
	
	\maketitle
	
	\section{Introduction}
	\label{sec:intro}
	Laser-plasma acceleration (LPA) enables the acceleration of electron bunches with gradients often exceeding $\SI{100}{\giga\volt\per\metre}$ by exploiting the large electric fields of a plasma wave \cite{Tajima1979}.
	As these fields are around three orders of magnitude stronger than those provided by conventional radio-frequency technology, LPAs have long been proposed as future compact electron accelerators.
	So far, experiments of applications of these sources have concentrated on secondary radiation such as the production of undulator radiation \cite{Fuchs2009}, hard x-rays from Thomson or Compton scattering \cite{Compton1923,Catravas_2001,Schwoerer2006,TaPhuoc2012} or bremsstrahlung \cite{Bethe1954,Edwards2002,Glinec2005,DOPP2016515}, and high resolution tomographic imaging \cite{Cole2015,Wenz2015,Cole6335} using betatron radiation produced by oscillations of the electron beam within the LPA cavity \cite{Rousse2004,Kneip2010}.
	Recently, free electron lasing of LPA produced electron beams was demonstrated for the first time \cite{Wang2021}.\\
	The development of these applications has gone hand-in-hand with progress in realising high quality electron beams via a range of techniques.
	It has been shown that electron beams with few per-cent energy spread or less can be produced \cite{Mangles2004a,Geddes2004a,Faure2004a,Wang2016}.
	Maximum reported energy gains have continued to increase to as high as \SI{8}{\giga\electronvolt} \cite{Wang2013,Kim2013,Gonsalves2019}, while nanocoulomb level beams with $\sim \SI{10}{\pico\coulomb\per\mega\electronvolt}$ at \SI{0.25}{\giga\electronvolt} have been demonstrated \cite{Couperus2017}. 
	Even microcoulomb class beams have been derived from highly energetic laser drivers \cite{Shaw2021}.
	Sub 1\,mm-mrad normalised emittance beams have been measured \cite{Weingartner2012}, while the naturally few tens of micrometers size of the plasma wakefield, in typical conditions for most experiments, ensures an electron beam duration of a few tens of femtoseconds \cite{Mangles2006} to a few femtoseconds \cite{Lundh2011}.
	Combining the above features leads to the generation of beams with an estimated 6D brightness close to those produced at modern large-scale linacs \cite{Wang2016}.
	To date, however, there has been relatively little work on high repetition rate operation over long timescales or on reproducibility, which are equally important factors when considering a `workhorse' accelerator to drive secondary sources, as  highlighted in the recently published roadmap for the field \cite{Albert2020}.\\
	Almost all studies of LPAs have focused on results ranging from single shots to of order 100 shots with only a handful of exceptions: a 100,000 shot run at \SI{1}{\Hz} with a stabilised laser path \cite{Maier2020}, electron beam optimisation using active feedback at \SI{5}{\Hz} \cite{Dann2019}, and experiments with milli-Joule, kHz class lasers producing moderately relativistic electron beams \cite{He2015,Rovige2020,Gustas2018}.
	\begin{figure*}[h!tb]
		\centering        
		\includegraphics[width=0.90\textwidth]{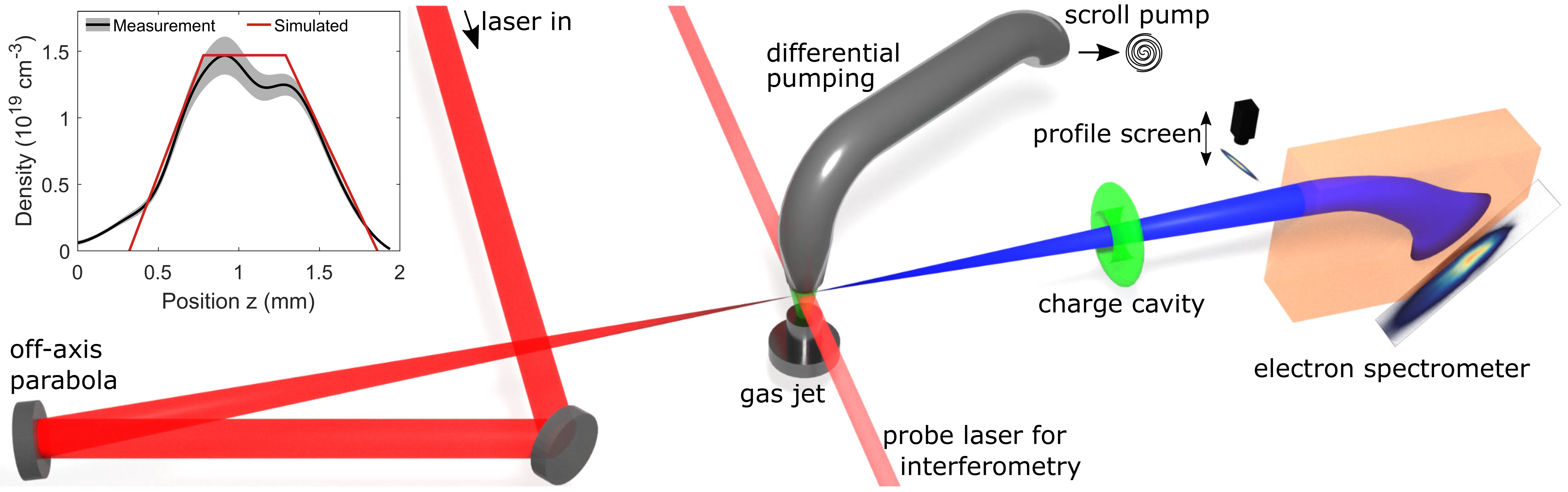}
		\caption[Experimental Setup]{Schematic of the experimental setup. Inset top left: measured plasma electron density (black), its rms deviation (grey shaded area) and the density profile used in the simulations (red line).}
		\label{fig:setup}
	\end{figure*}\\
	In this work we report on the generation and acceleration of highly relativistic electron bunches peaking at an average energy of \SI{78}{\mega\electronvolt} from an LPA using ionisation injection (II) running at \SI{2.5}{\hertz} for \SI{8}{\hour}.
	Charge was injected on every one of 72,000 consecutive shots and no decrease in beam charge was detected over the run, while the spectral peak dropped gradually by \SI{1.3}{\mega\electronvolt} per 10,000 shots.
	This was achieved with no stabilisation or active feedback, and using a simple gas jet target, in the setup described in section \ref{sec:setup}.
	The properties of the electron beams are presented and discussed with reference to particle-in-cell simulations in sections \ref{sec:electronresults} and \ref{sec:analysis} respectively.
	The charge per shot was the parameter with the largest variance and was strongly correlated with plasma density. 
	The trend was reproduced in PIC simulations which showed that density changes alter the self-focusing of the laser pulse \cite{Sprangle1987} and thus the volume from which tightly bound electrons are ionised and injected in to the LPA.
	In this way small plasma density fluctuations can significantly alter the charge and phase space of the injected beam.
	While ionisation injection reliably produces electron beams, we show that the plasma density and laser energy should be very tightly controlled for it to be considered to be a low jitter injection platform.\\
	An immediate consequence of high repetition rate, reproducible beams would be an improvement in the average dose rate of relativistic electrons or energetic photons from secondary sources by multiple orders of magnitude.
	For the electron energies reported here, this system could (relative to other all-optical set-ups) serve as the foundation for a high flux source of few keV x-rays from betatron radiation \cite{Rousse2004}, multi-MeV gamma rays from bremsstrahlung or sub-100\,keV x-rays from Thomson scattering.
	An additional application for such beams could be very high energy electrons for the treatment of deep-seated tumours \cite{Kokurewicz2019,Palma2015,Papiez2002,DesRosiers2008,Subiel2014}, where several nanocoulombs of charge should be delivered in a per-cent level energy bandwidth in an acceptably short time for the patient.
	The naturally ultra-high peak dose rate from LPAs may prove to be beneficial for treatment via reduced damage to healthy tissue, as shown in FLASH radiography studies \cite{Favaudon245ra93,Durante2018}.
	In this case, we will use these electron beams in an all-optical Thomson scattering setup\cite{Staufer2019} to produce hard x-rays for use in x-ray fluorescence imaging \cite{Cheong2010,Manohar2016,Gruner2018,Schmutzler2021,Sanchez-Cano2021,Kahl2021}, an application in which the imaging time is shortened proportionally with repetition rate.
	Implementations of such high-repetition rate LPA sources at GeV levels would, for example, lead to a reduction of betatron x-ray tomography scan times from a day to minutes, with photon energies of multiple tens of keV. 
	Another important benefit would be a large increase in the signal to noise ratio of data from electron-photon scattering experiments in the nonlinear regime of QED, where more precise statistical studies are required to better discriminate between the available models \cite{Cole2018,Poder2018}.
	
	\section{Experimental setup}
	\label{sec:setup}
	The experimental setup is depicted in Fig.~\ref{fig:setup}.
	The electrons were accelerated by focusing a high-intensity laser into a pulsed, supersonic jet of gas.
	The gas jet was opened for \SI{7}{ms} and the laser was fired \SI{5}{ms} after the opening to ensure a continuous flow regime.
	The gas was evacuated by a differentially pumped cone with an opening of \SI{8}{\mm} placed \SI{4}{\mm} above the nozzle. 
	This enabled the ambient pressure in the interaction chamber to remain at a constant level of \SI{2.0e-4}{\milli\bar} and the vacuum compressor at \SI{2.1e-5}{\milli\bar} when pulsing the gas jet at a repetition rate of \SI{2.5}{\Hz}.\\
	To diagnose the electron beam charge a cavity charge monitor (CCM) was used. 
	A CCM is a cavity that non-invasively measures the charge of the electron bunch passing through it from the amplitude of the induced TM01 electromagnetic mode \cite{Lipka2011, Lipka2013}. 
	It was placed \SI{1.1}{\m} behind the gas jet. 
	At the end of the beamline the electron bunch was deflected by a dipole magnet with a length of \SI{0.5}{\m} and a field of \SI{0.14}{\tesla}. 
	The dispersed electrons were detected using a DRZ-type phosphor screen \cite{Kurz2018, Schwinkendorf2019} which was imaged with a camera to measure the energy spectrum of the beam. 
	For further characterisation a DRZ profile screen could be driven into the electron beam \SI{1.25}{\m} from the gas jet to measure the transverse profile and pointing of the electron bunch.
	A vacuum beam pipe between the LPA chamber and the CCM stopped electrons travelling at angles $|\theta| > \SI{13.3}{\milli\radian}$ from propagating downstream to any of the diagnostics.\\
	The LPA was driven by a high power titanium sapphire laser system delivering \SI{170}{\milli\J} pulses centered on \SI{805}{\nano\metre} at \SI{2.5}{\Hz}, with an rms energy variation of 1.3\%.
	To maintain the laser energy at a constant level over the full \SI{8}{\hour} run only a single manual increase in the pump laser flashlamp voltage was required around shot number 10,000.
	This was the only adjustment made during the run.
	The full width at half maximum (FWHM) pulse duration was \SI{26.9 \pm 0.2}{\fs}.
	This and all errors here represent the standard deviation unless stated otherwise. 
	The laser was focused with an $f/12$ off-axis parabola resulting in a 1/$e^2$ of intensity radius of \SI{8.0 \pm 0.1}{\um}.
	After accounting for the energy not contained in the central focal spot (35.3\% of the energy was contained within the FWHM) the peak normalised vector potential of the pulse was $a_0 = \SI{1.41(3)}{}$.
	The laser was not operated at its maximum repetition rate of \SI{10}{\Hz} to keep the energy deposition on the compressor gratings and beamline optics at a moderate level, as excessive grating heating has been shown to degrade the spatio-temporal properties of the compressed pulse \cite{Fourmaux2009, Leroux2018,Dresselhaus2020}. \\
	The gas was helium with a 0.5\% nitrogen dopant by weight (equivalent to approx.~7 N$_2$ molecules per 10,000 He atoms) and the backing pressure was \SI{3.9 \pm 0.1}{\bar}. A dedicated transverse probe beam was used to produce interferograms of the interaction.
	After converting the phase shift from the interferograms to a line integrated plasma density, the 3D plasma density was found assuming cylindrical symmetry via the Abel inversion.
	The peak density at the interaction height of \SI{1.2}{\milli\metre} above the nozzle was \SI{1.47(1)e19}{\per\cubic\centi\metre}, where the error is the standard error, and its longitudinal profile is shown in the inset of Fig.~\ref{fig:setup}.
	The shot-to-shot variation of the plasma density was estimated from the standard deviation of the peak phase shift of the probe laser beam.
	This is less susceptible to noise which can be amplified during the Abel inversion, especially close to the laser axis.
	The resulting rms uncertainty of 3.6\% was close to that of the backing pressure at 2.6\%.\\
	The gas species was chosen so that electrons would be injected into the wakefield via ionisation injection \cite{Chen2006,Oz2007,Pak2010,McGuffey2010,Clayton2010}.
	In this scheme the helium electrons and the first five nitrogen electrons are ionised by the front of the laser pulse and form the plasma wakefield. 
	The core electrons of nitrogen are only ionised close to the peak of the laser pulse and can be trapped inside the wakefield potential.
	In this experiment self-focusing of the laser pulse was required to trigger ionisation and injection.
	
	\section{Stable long-term electron acceleration}
	\label{sec:electronresults}
\begin{figure}[h!tb]
		\centering        
		\includegraphics[width=\columnwidth]{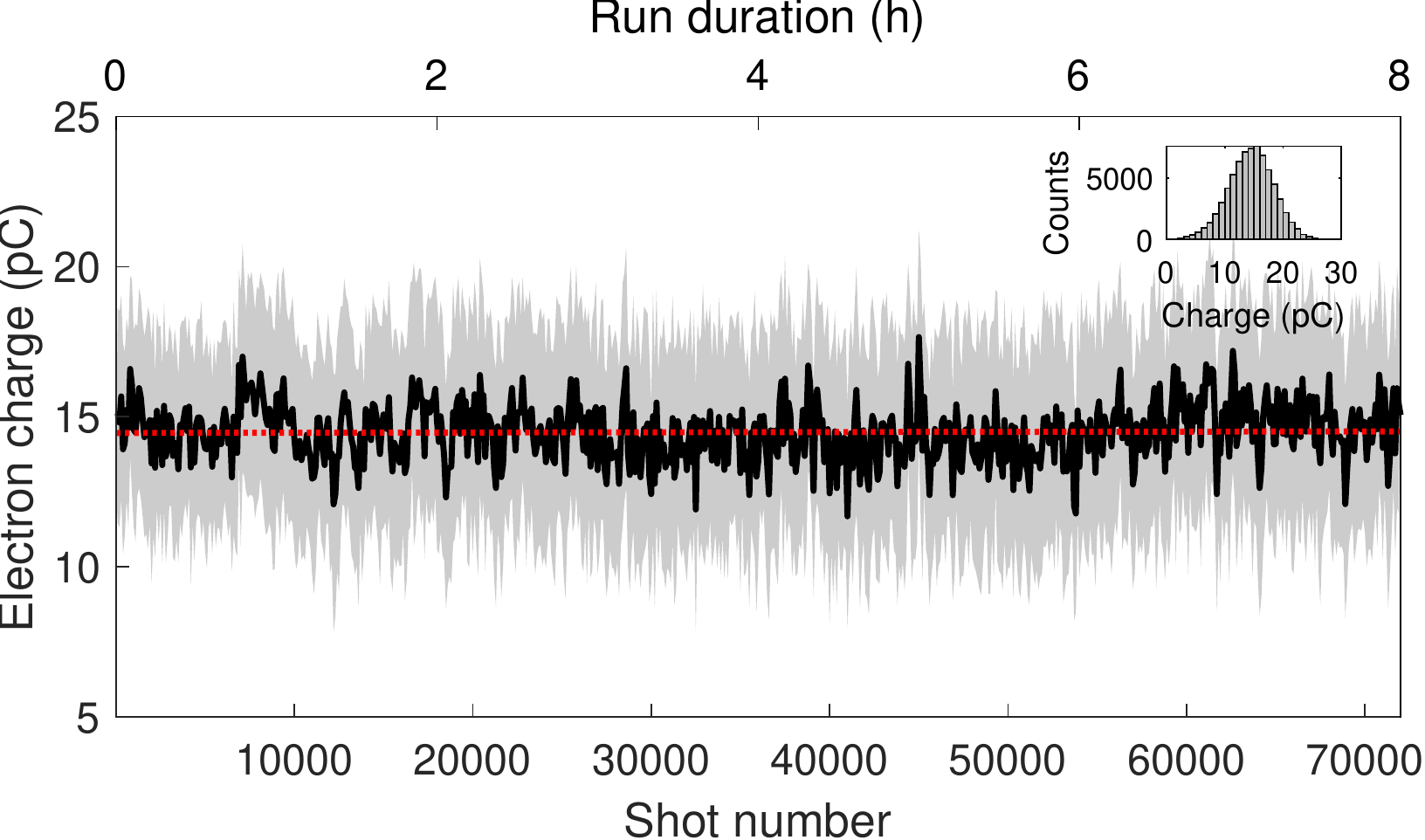}
		\caption[ChargeOverTime]{Measured charge as a function of shot number and run time. The black line shows the 100 shot moving average of the charge with its standard deviation as grey band. A linear fit of all charge measurements is shown in red. Inset: a histogram of the injected charge.}
		\label{fig:chargeplot}
	\end{figure}

	Before the main run the LPA was optimised for high charge and reliable injection by varying the gas jet backing pressure, the valve opening time, the laser energy and the laser focal position relative to the gas jet.  
	After the optimisation, the LPA was run at a repetition rate of \SI{2.5}{\Hz} for \SI{8}{\hour}, totalling 72,000 shots.
	Electron beams were successfully injected and accelerated on every shot. 
	The average charge of the electron beams measured by the CCM was constant over the entire dataset as shown in Fig. \ref{fig:chargeplot}. 
	The mean charge was \SI{14.5 \pm 3.8}{\pico\coulomb}, corresponding to a flux of approx.~\SI{2.2}{\nano\coulomb} per minute, adding up to more than \SI{1000}{\nano\coulomb} for the 8 hour run. A linear fit of the charge as a function of shot number shows a negligible increase of the mean charge by \SI{0.05}{\pico\coulomb} after 72,000 shots. 
	The minimum measured charge of \SI{1.5}{\pico\coulomb} is well above the noise level of \SI{0.5}{\pico\coulomb} of the CCM, demonstrating injection on every shot.
	Note that the CCM measures the charge of all of the electrons passing through it, while the electron spectrometer only detects electrons with a minimum energy of \SI{37}{\mega\electronvolt}.
	The number of counts registered on the electron spectrometer screen increased by 2.1\% over the whole run.\\
	\begin{figure}[h!t]
		\begin{center}
			\includegraphics[width=\columnwidth]{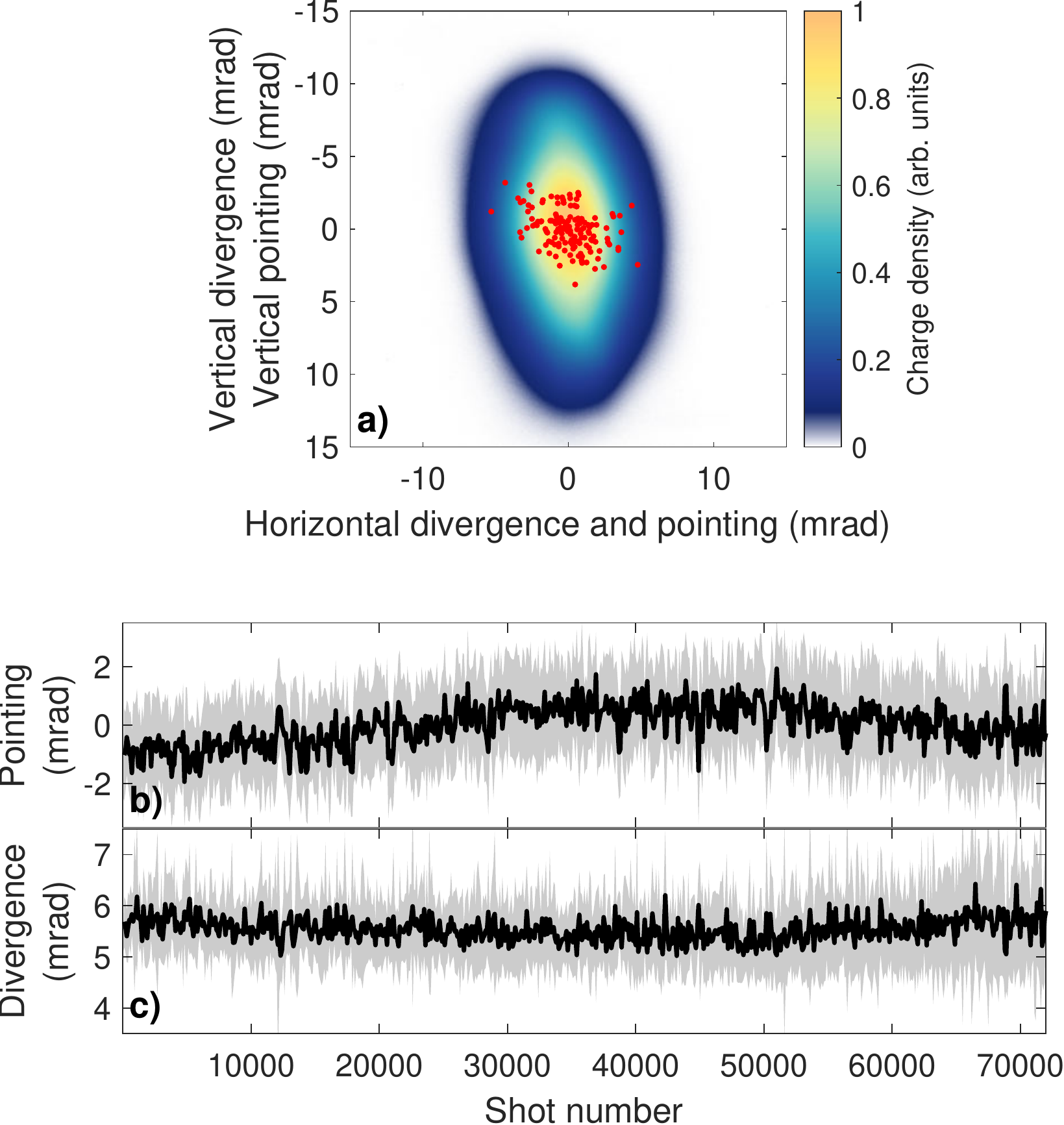}	
			\caption{Pointing and divergence of the electron beams. (a) Average 2D-divergence of 200 shots on the profile screen and the corresponding pointing of the single shots (red dots). (b) and (c) 100 shot moving average of the pointing and divergence (black lines) in the non-dispersive axis, with their 100 shot standard deviations shown as grey bands.}
			\label{fig:ProfilePointing}
		\end{center}
	\end{figure} 
	\begin{figure*}[h!t]
		\centering
		\includegraphics[width=0.95\textwidth]{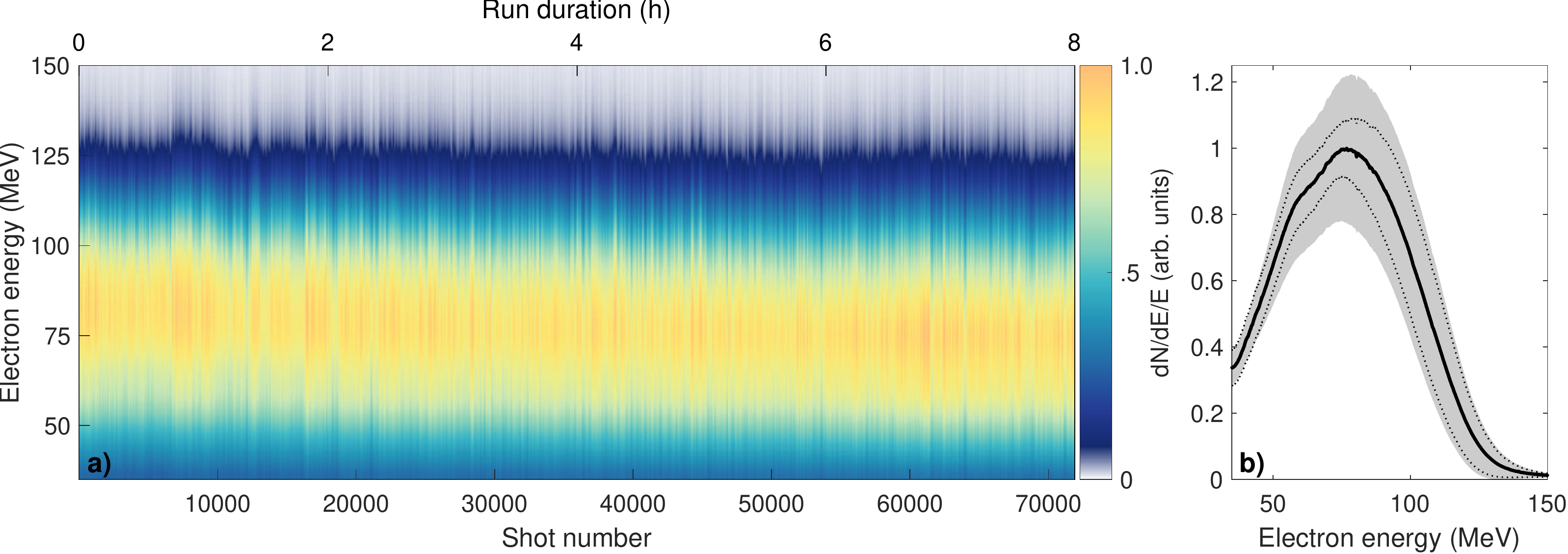}
		\caption[Waterfall and Peaks]{\correcttext{Electron energy spectra $(\mathrm{d}N/\mathrm{d}E/E)$ as function of shot number and time. The average spectrum is shown in Fig. 6a}{Electron energy spectra $(\mathrm{d}N/\mathrm{d}E/E)$. (a) As function of shot number and time (waterfall). (b) Average spectrum with its standard deviation (grey area) and the standard deviation of normalised spectra indicating fluctuations independent of charge fluctuations (dotted line).}}
		\label{fig:wtfallpeaks}
	\end{figure*} 
	The 2D electron beam pointing and FWHM divergence were measured using the profile screen in an initial 200 shot run, shown in Fig.~\ref{fig:ProfilePointing}a.
	The pointing and divergence were monitored in the horizontal axis via the electron spectrometer throughout the main run, and they are plotted in Fig.~\ref{fig:ProfilePointing}b and c. 
	The FWHM divergence of the beams was \SI{6.4 \pm 0.4}{mrad} in the horizontal axis and \SI{12.3 \pm 0.5}{mrad} in the vertical axis, which was also the dispersion axis of the electron spectrometer. 
	The divergence was larger in the polarisation axis of the laser, typical of ionisation injection experiments \cite{McGuffey2010}. 
	The pointing stability was found to be almost symmetric with a standard deviation of \SI{1.7}{mrad} and \SI{1.3}{mrad} in horizontal and vertical axis respectively. 
	The standard deviation of the horizontal pointing over the entire run was \SI{1.9}{mrad} with a mean divergence in the same axis of \SI{5.5 \pm 0.9}{mrad}, within the error of the initial measurement. \\
	The measured electron energy spectra ($\mathrm{d}N/\mathrm{d}E/E$, where $N$ is the number of electrons and $E$ their energy) are plotted as a waterfall in Fig.~\ref{fig:wtfallpeaks}\,(a).
	The average spectrum is shown in Fig.~\ref{fig:wtfallpeaks}\,(b) along with its standard deviation and the standard deviation of the normalised spectra, which is independent of the charge fluctuation.  
	The figure shows that the spectra were broadband ranging from the low energy cut-off of the spectrometer up to approximately \SI{130}{\mega\electronvolt}.
	The mean peak energy, defined as the energy at which $\mathrm{d}N/\mathrm{d}E/E$ is maximised, was \SI{78 \pm 7}{\mega\electronvolt}.
	A slow drift of the peak electron energy over time towards lower values can be observed in Fig.~\ref{fig:wtfallpeaks}. 
	A linear fit to this shows that the peak energy decreases by \SI{1.3}{\mega\electronvolt} per 10,000 shots, falling from \SI{82.3}{\mega\electronvolt} to \SI{72.8}{\mega\electronvolt} over 72,000 shots, while the jitter remained approximately constant at 8--9\% over the run. \\
	While the peak energy, as defined, allows for easy comparison with other works or simulations, it is not necessarily the most important figure of merit for some applications.
	For most quasi-monochromatic secondary sources or radio-biological studies one may employ an energy selective beam transport system to take the electron beam from the accelerator to the insertion device or target.
	Thus, one wishes to produce a constant charge within some energy bandwidth, integrated over a number of shots.
	In this experiment $\mathcal{F} = 15.1 \pm 1.4$~\% of the total charge per shot lay between 70 and \SI{80}{\mega\electronvolt}, denoted $q$. 
	This was constant over the full run \correcttext{; a linear fit to $\mathcal{F}$ shows a reduction of order $10^{-7}$~\%}{and no resolvable evolution was observed} over 72,000 shots.
	Therefore, the deviation in $\sum_N q_N = \mathcal{F} Q$ would be dominated by the fluctuation of the total charge $Q$, which decreases as $1/\sqrt{N}$ with the number of integrated shots $N$.
	For a single shot $q = \SI{2.2 \pm 0.6}{\pico\coulomb}$.
	Increasing to a larger number of shots with a practical example, this system would deliver \SI{1}{\nano\coulomb} in the range 70-\SI{80}{\mega\electronvolt} in 3 minutes with a dose error of 1.2\%.
	Due to the long term stability of $Q$ and $\mathcal{F}$, this would not change perceptibly over \SI{8}{\hour}.
	
	\section{Analysis of the interaction}
	\label{sec:analysis}
\begin{figure}
		\centering        
		\includegraphics[width=\columnwidth]{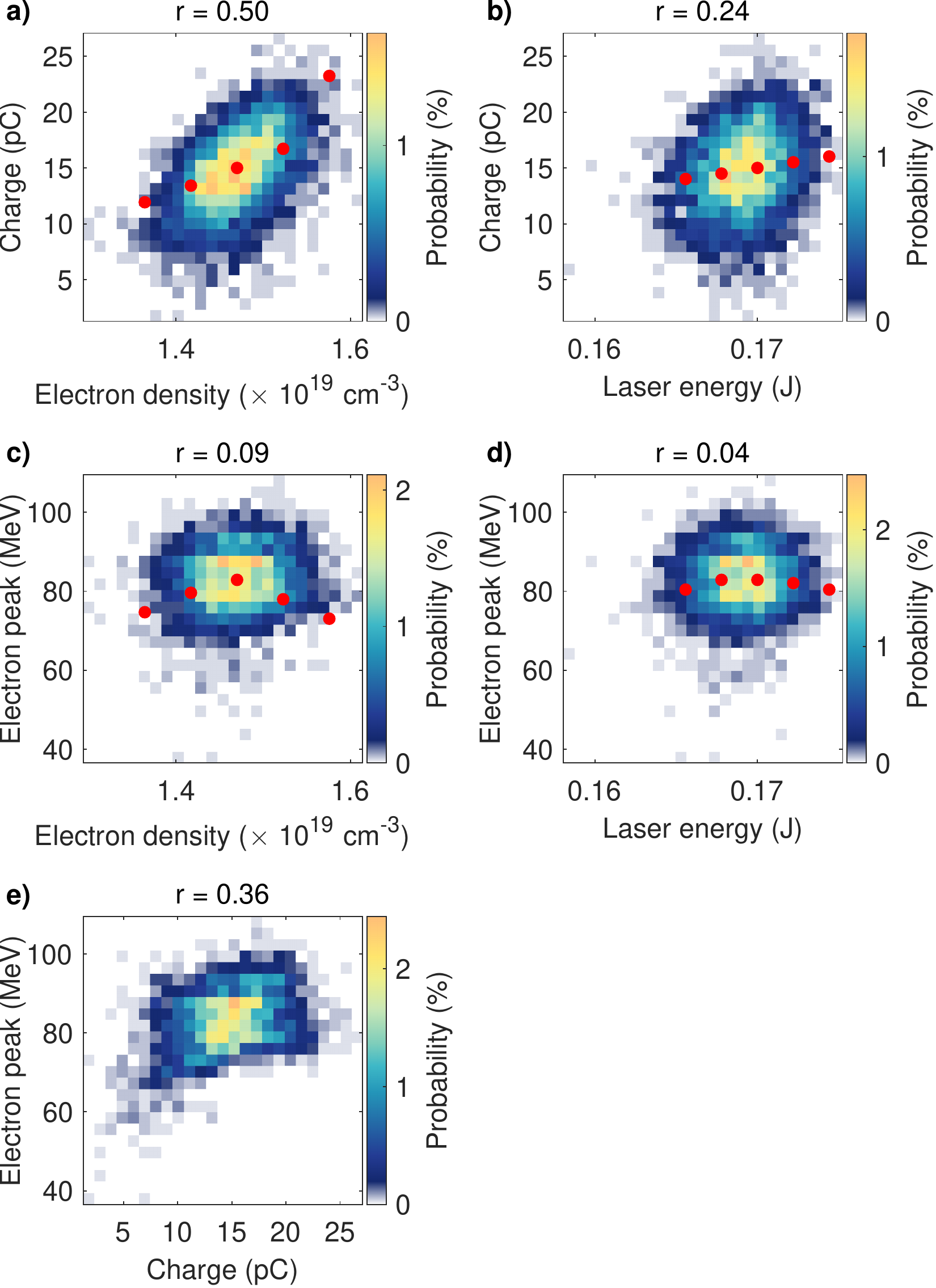}
		\caption[Correlations]{Correlations of the data using a subset of 3600 shots (5\% of total shots). 
			The sub-plot titles show the correlation coefficients (Pearson's $r$). 
			Red markers represent PIC simulation results.}
		\label{fig:correlations}
	\end{figure}

\begin{figure*}[h!t]
		\centering
		\includegraphics[width=0.85\textwidth]{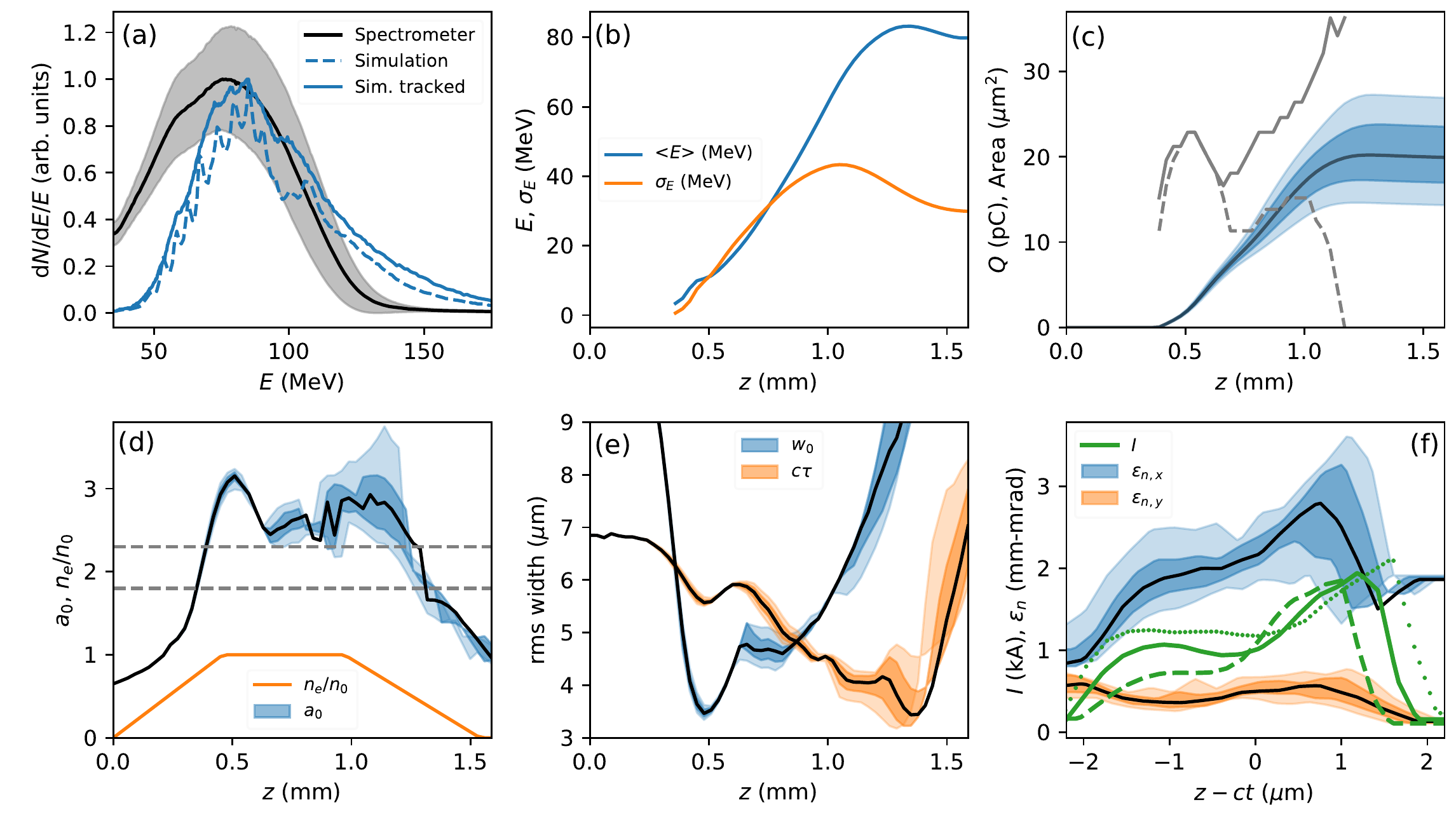}
		\caption[Simulation results]{PIC simulation results. (a) Normalised experimental electron spectrum (black) with its standard deviation (grey), the raw simulated spectrum (blue dashes), and the simulated spectrum propagated to the spectrometer screen (solid blue line).
			(b) Simulated mean energy (solid blue) and rms energy spread (orange) vs.~$z$.
			(c) Simulated charge from II vs.~$z$ at $n_e \pm (1,2)\sigma_{n_e}$ (bounded by the blue and pale blue shaded areas).
			The behaviour at $n_e$ is shown in black, as it is in all following plots.
			The peak area of the laser pulse envelope with sufficient intensity to produce N$^{6+}$ is shown for the lowest (dashed line) and highest (solid line) density simulations in grey, in the region where charge is trapped.
			(d) Simulated $a_0$ evolution at densities $n_e \pm (1,2)\sigma_{n_e}$.  Normalised $n_e(z)$ is shown in orange. The grey dashed lines are the $a_0$ thresholds required to ionise the core electrons of nitrogen according to the ADK model \cite{Pak2010}.
			(e) Laser waist $w_0$ (blue) and length $c\tau$ (orange) vs.~$z$ at $n_e \pm (1,2)\sigma_{n_e}$. 
			(f) Normalised slice emittance in $x$ (blue) and $y$ (orange) of II electrons at $n_e \pm (1,2)\sigma_{n_e}$ at $z=1$\,mm. In dashed, solid and dotted green are the corresponding current profiles at $n_e - 2\sigma_{n_e}$, $n_e$ and $n_e + 2\sigma_{n_e}$.
			In higher $n_e$ simulations the $w_0$ and $c\tau$ are smaller, while $a_0$, $Q$ and $\epsilon_n$ are larger.}
		\label{fig:sims}
	\end{figure*}

	From a large dataset it is possible to create meaningful correlations of outputs against measured inputs.
	These can be used to identify which parts of a system should be stabilised to improve its precision \cite{Maier2020}.
	In Fig.~\ref{fig:correlations} the measured charge and peak energies are correlated with the plasma density and laser energy.
	Also shown is the variation in the peak electron energy with charge.
	To negate the effects of long term drifts these plots are of a subset of the first 3600 consecutive shots (5\% of the total).
	They show that the injected charge was particularly sensitive to the plasma density and to a lesser extent the laser energy.
	An increase of electron density from 1.4 to \SI{1.5e19}{\per\cubic\centi\metre} doubled the measured charge.
	The peak electron energy, however, was almost completely unaffected by changes in the plasma density or the laser energy within these ranges.
	Finally, there is a positive correlation between the peak electron energy and the injected charge.
	This is the opposite of what one would naively expect from beam loading.
	Note that this correlation is quite strongly influenced by a small number of exceptionally low charge and peak energy shots.
	If we exclude all shots more than $2\sigma$ below the mean in both charge and peak energy the correlation coefficient reduces from 0.36 to 0.24.
	Overall it is clear that the main limiting factor to the stability of the system was the plasma density fluctuation impacting the amount of accelerated charge per shot.\\
	To further understand these correlations and the physics behind the behaviour of the system a number of particle-in-cell (PIC) simulations were performed using the cylindrical quasi-3D code FBPIC \cite{LEHE2016}.
	The window size was $\SI{40}{\um} \times \SI{30}{\um}$ with 40 and 10 cells/\si{\um} in $z$ and $r$ respectively, and moved in $z$ at the linear group velocity of the laser pulse.
	Three azimuthal modes were used.
	The plasma consisted primarily of a pre-ionised species with a longitudinal profile shown in the inset of Fig.~\ref{fig:setup} and a peak density of \SI{1.47e19}{\per\cubic\centi\metre}, modelled with 108 particles per cell. 
	An additional species modelled the nitrogen component of the plasma, pre-ionised to a charge state of 5+, with 48 particles per cell and the same relative density as in the experiment.
	The laser was modelled as an \SI{800}{\nano\metre} gaussian pulse in space and time, with otherwise the same properties as were used in the experiment: $a_0 = 1.41$, a $1/e^2$ intensity radius of \SI{8.0}{\um} and a FWHM duration of \SI{26.9}{\femto\second}.
	Its vacuum focal position was at the top of the plasma density upramp.\\
	The electron spectrum from the simulation is shown in Fig.~\ref{fig:sims}a, where it is compared to the experimentally measured spectrum.
	Also shown is the simulated spectrum that has been tracked to the electron spectrometer screen.
	The entrance aperture of the CCM had a radius of $\SI{13.3}{\milli\radian}$, so injected electrons in the simulation travelling at angles larger than this were eliminated.
	To produce a spectrum a random sample of 4\,million unique macroparticles that remained from the injected beam were propagated individually in 3D to a plane representing the electron spectrometer screen via the measured magnetic field of the dipole. 
	Since an imaging spectrometer was not used this allows for a better comparison between the simulation and experimental results, i.e.~it includes the effect of the divergence on the measured spectrum.
	An example of this can be seen in Fig.~\ref{fig:sims}a where the spiky `raw' spectrum, caused by periodic ionisation in the oscillating laser field, is washed out via the propagation leaving a smooth spectrum similar to the experiment.
	The propagation algorithm was applied to all further spectral analysis.\\
	To calculate the charge from the simulation, electrons with $\gamma \beta_z < 11 \sim \gamma_{ph}$, the Lorentz factor associated with the wakefield's phase velocity, were removed to eliminate the electrons that make up the plasma wave, as were electrons with a divergence larger than the CCM aperture.
	A \SI{15.0}{\pico\coulomb} bunch remained from the simulation, comprised entirely of II electrons, which is in excellent agreement with the experimental charge measurement.
	The peak energy of the beam was \SI{82.9}{\mega\electronvolt}, close to the mean experimental value of \SI{78}{\mega\electronvolt}.
	Fig.~\ref{fig:sims}a shows that the simulation overestimates the length of the high energy tail and underestimates the amount of charge at low energies.
	Note that the spectra in Fig.~\ref{fig:sims}a are normalised to their maximum values, principally because we do not have a dedicated measurement of the charge between 0 and \SI{37}{\mega\electronvolt} from the experiment.\\
	In parts b, d and e of Fig.~\ref{fig:sims} the dynamics of the interaction are examined to explain the measured beam properties.
	Figs.~\ref{fig:sims}d and e show that the laser quickly self-focuses in the density upramp to a peak $a_0$ of approximately 3 \highlightcorrect{at} a short distance into the flat top density region.
	At this point charge is injected via ionisation injection, which is plotted in Fig.~\ref{fig:sims}c (note that the charge in this figure is the total II charge trapped in the wakefield, not propagated to the screen). 
	The electron beam mean energy and rms energy spread are plotted in Fig.~\ref{fig:sims}b. 
	The rapid initial increase in both energy and energy spread are a result of continuous injection without significant beam-loading.
	After an initial `over-focusing' $a_0$ stabilises at 2.7 due to a combination of slight defocusing and self-compression.
	This drop in $a_0$ causes the rate of charge injection to decrease.
	The electrons that were injected early continue to accelerate while the reduced rate of trapping slows down the increase in the energy spread, beginning the process of reducing the relative energy spread.
	Once the laser reaches the density downramp it rapidly defocuses and no further trapping occurs.
	Simultaneously the bunch reaches dephasing, which further reduces the energy spread via phase space rotation.
	At the end of the simulation there is a broadband electron bunch which is nevertheless peaked, with a relative energy spread of approx.~30\% compared to its initial 100\% spread.
	The lack of correlation of peak energy with density and laser energy in Fig.~\ref{fig:correlations} is attributed to operating close to dephasing, where the beam is located around the zero-crossing of the wakefield at the end of the plasma, such that slight under-acceleration in lower densities and slight over-acceleration at higher densities, relative to the dephasing length, have a minimising effect.\\
	The physics behind the jitter in the output beam parameters will now be elucidated with a range of PIC simulations.
	The laser energy $E_{las}$ and the plasma density $n_e$ were varied independently of each other by 1 and 2 experimental standard deviations ($\sigma_E = 1.3$\% and $\sigma_{n_e} = 3.6$\%).
	The resulting charges and peak energies are overlaid on the experimental correlation data in Fig.~\ref{fig:correlations}a--d and are broadly in good agreement, especially in the case of charge as a function of $n_e$.
	At the highest density there is a rapid increase in charge in the simulation because 5.5\,pC of self-injected (SI) charge passed through the virtual CCM aperture.
	In general there is a strong scaling of II charge ($Q_I$) with $n_e$, increasing by 1.4\,pC (9\%) per $\sigma_{ne}$, 3 times more rapidly than $n_e$. 
	The naive expectation might have been a 1:1 correspondence, assuming there was the same volume of laser ionisation but a higher density of nitrogen atoms.\\
	To explain this observation, the main source of jitter in this experiment, the evolution of $Q_I$ and the laser $a_0$, $w_0$ and duration $c\tau$ are plotted in Fig.~\ref{fig:sims} c-e.
	Black lines are from simulations at density $n_e$, and the results from simulations at $n_e \pm \sigma_{n_e}$ and $n_e \pm 2\sigma_{n_e}$ are bounded by the dark and pale shaded areas respectively.
	In all simulations the initially mismatched laser pulse rapidly self-focuses to a similar $w_0$ and $a_0$, which is above the ionisation threshold for both core electrons of nitrogen.
	Consequently, charge is injected rapidly and at a similar rate for all densities.
	The critical part of the simulation is between $z = 0.6 \ \textrm{and} \ 0.9$\,mm where the $n_e$ fluctuations provide an order 10\% correction to the matched spot size, the spot size that drives a stable blown out wake \cite{Lu2007}, and thus a significant increase in $a_0$ in the high $n_e$ simulations.
	The effect of a higher density, a smaller matched $w_0$ and a higher $a_0$ is to \emph{increase} the area of the laser pulse with $a_0$ above the ionisation thresholds, a proxy for the amount of nitrogen being ionised. 
	This is demonstrated in Fig.~\ref{fig:sims}c, where from $z=0.6$\,mm the area rapidly diverges between the lowest and highest density simulations.
	In all simulations, a higher density resulted in a larger area of the laser pulse with an above-threshold $a_0$.
	This is why the rate of charge injection diverges after $z=0.6$\,mm in Fig.~\ref{fig:sims}c, with high $n_e$ simulations injecting charge at a significantly higher rate than those at low $n_e$.
	Up until $z=1$\,mm there is no large difference in pulse length, and after this point trapping of II charge is terminated as the laser defocuses in the downramp. 
	Changes in the matched spot size are the principal cause of the observed charge jitter.
	The argument that the charge varies according to the volume of the focused pulse above ionisation threshold would also easily explain the correlation between charge and laser energy in Fig.~\ref{fig:correlations}b, although we do not examine this in depth here.\\
	As the volume over which the laser can ionise varies with $n_e$, so do the current profiles and transverse phase spaces of the injected beams.
	Fig.~\ref{fig:sims}f plots the normalised slice emittances $\epsilon_n$ in $x$ (the laser polarisation plane in the simulation) and $y$ as well as the current profile of the beam at $z=1$\,mm, the point at which injection is terminated in all simulations.
	The laser evolution is clearly imprinted on to the bunch.
	The first electrons to be injected are expected to be at the front (positive $(z-ct)$) and the last at the rear due to the phase advance of electrons in the wake over time.
	The longitudinal phase space of the beam is approximately linear and positively chirped in each case, supporting this lemma.
	The front of the beam has the highest current since it was injected first during the over-focusing of the laser, and it is trailed by the electrons injected during the matched spot phase where there is a large variation in current with density.
	While the expected behaviour of particularly $\epsilon_{n,x}$ is harder to predict since it is a function of the ionising volume and the field strength where each electron is ionised, it is clear that substantial differences exist in the slice emittance for relatively small density changes.
	This could, for example, lead to large spot size variations if the electron beam was focused downstream of the plasma.\\
	While the simulations presented here are specific to our experimental set up, the general behaviour is likely to extend to any II based LPA reliant on self-focusing, which are common in the literature.
	With II it is often the case that the laser intensity in the plasma is close to the ionisation threshold of the dopant species, as this is required to ionise in a small volume around the centre of the wakefield so that a good quality beam can be trapped and accelerated.
	Our results suggest that to produce a system based on II with per-cent level shot-to-shot charge fluctuations, the plasma density and laser energy must be controlled at the level of a fraction of a per-cent.
	Otherwise, the nonlinear process of self-focusing and the threshold process of ionisation combine to enhance any fluctuations in the inputs.
	It is likely that this problem can be mitigated by using a guiding channel based on hollow capillary waveguides \cite{Hansson2014}, or dynamically formed plasma waveguides based on laser \cite{Geddes2004a} or discharge ionisation and heating \cite{Karsch2007}.
	Each of these comes at the expense of a more complicated setup and/ or increased risk of damage from the LPA driver, and will contain other sources of jitter which should be evaluated in an LPA context.
	
	\section{Conclusions}
	\label{sec:summary}
	This work has demonstrated that even with simple targetry and a modest laser energy, that can be produced with a table-top laser system, an ionisation injection based LPA producing $\sim 80$\,MeV electrons can be run at repetition rates of multiple Hz and remain stable over 8 hours.
	This is a significant step towards the use of LPAs as workhorse machines requiring minimal user intervention over full day timescales and we envisage wider adoption of these machines outside the traditional national laboratory or university laser lab environments.\\
	72,000 consecutive shots all demonstrated charge injection and acceleration, and the mean charge was robust against drifts (within the measurement error) over the full run. 
	The peak energy dropped by \SI{1.3}{\mega\electronvolt} per 10,000 shots, from an initial \SI{82.3}{\mega\electronvolt}.
	However, the average charge per shot in selected energy bands, important for future applications, was extremely stable.
	For example we found that the system accelerated an average of \SI{2.2}{\pico\coulomb} to energies between 70 and \SI{80}{\mega\electronvolt} with no drift over the full 8 hours.
	The largest shot-to-shot fluctuation was in the total beam charge.
	Correlations of the measured beam properties revealed that the 3.6\% fluctuations in the plasma density were the principal driver of the shot-to-shot charge jitter, however they had no discernible effect on the peak electron energy.
	Particle-in-cell simulations, in agreement with the data, showed that the relative charge fluctuation was $\sim 3$ times the density fluctuation, and was mostly a result of a smaller matched spot size of the laser in the plasma increasing the volume of the (same energy) laser pulse that was able to ionise the core nitrogen electrons.
	This work suggests that to achieve low jitter systems based on ionisation injection very fine control over the intensity of the laser pulse in the plasma is required, which may be most readily achieved by sub-per-cent level reproducibility of the plasma density and laser energy.
	
	\begin{acknowledgments}
		The authors would like to thank M.~Dinter, S.~Karstensen, S.~Kottler, K.~Ludwig, F.~Marutzky, A.~Rahali, V.~Rybnikov, and A.~Schleiermacher for their engineering and technical support. The authors acknowledge funding from Helmholtz ARD, Helmholtz ATHENA, the DESY Strategy Fund, and the BMBF InnovationPool through project PLASMED X. This work was supported by the Maxwell computational resources at DESY.
		
	\end{acknowledgments}
	
	\typeout{}
	\section*{References}

\begin{thebibliography}{67}%
		\makeatletter
		\providecommand \@ifxundefined [1]{%
			\@ifx{#1\undefined}
		}%
		\providecommand \@ifnum [1]{%
			\ifnum #1\expandafter \@firstoftwo
			\else \expandafter \@secondoftwo
			\fi
		}%
		\providecommand \@ifx [1]{%
			\ifx #1\expandafter \@firstoftwo
			\else \expandafter \@secondoftwo
			\fi
		}%
		\providecommand \natexlab [1]{#1}%
		\providecommand \enquote  [1]{``#1''}%
		\providecommand \bibnamefont  [1]{#1}%
		\providecommand \bibfnamefont [1]{#1}%
		\providecommand \citenamefont [1]{#1}%
		\providecommand \href@noop [0]{\@secondoftwo}%
		\providecommand \href [0]{\begingroup \@sanitize@url \@href}%
		\providecommand \@href[1]{\@@startlink{#1}\@@href}%
		\providecommand \@@href[1]{\endgroup#1\@@endlink}%
		\providecommand \@sanitize@url [0]{\catcode `\\12\catcode `\$12\catcode
			`\&12\catcode `\#12\catcode `\^12\catcode `\_12\catcode `\%12\relax}%
		\providecommand \@@startlink[1]{}%
		\providecommand \@@endlink[0]{}%
		\providecommand \url  [0]{\begingroup\@sanitize@url \@url }%
		\providecommand \@url [1]{\endgroup\@href {#1}{\urlprefix }}%
		\providecommand \urlprefix  [0]{URL }%
		\providecommand \Eprint [0]{\href }%
		\providecommand \doibase [0]{http://dx.doi.org/}%
		\providecommand \selectlanguage [0]{\@gobble}%
		\providecommand \bibinfo  [0]{\@secondoftwo}%
		\providecommand \bibfield  [0]{\@secondoftwo}%
		\providecommand \translation [1]{[#1]}%
		\providecommand \BibitemOpen [0]{}%
		\providecommand \bibitemStop [0]{}%
		\providecommand \bibitemNoStop [0]{.\EOS\space}%
		\providecommand \EOS [0]{\spacefactor3000\relax}%
		\providecommand \BibitemShut  [1]{\csname bibitem#1\endcsname}%
		\let\auto@bib@innerbib\@empty
		\bibitem [{\citenamefont {Tajima}\ and\ \citenamefont
			{Dawson}(1979)}]{Tajima1979}%
		\BibitemOpen
		\bibfield  {author} {\bibinfo {author} {\bibfnamefont {T.}~\bibnamefont
				{Tajima}}\ and\ \bibinfo {author} {\bibfnamefont {J.~M.}\ \bibnamefont
				{Dawson}},\ }\bibfield  {title} {\enquote {\bibinfo {title} {Laser electron
					accelerator},}\ }\href {\doibase 10.1103/PhysRevLett.43.267} {\bibfield
			{journal} {\bibinfo  {journal} {Phys. Rev. Lett.}\ }\textbf {\bibinfo
				{volume} {43}},\ \bibinfo {pages} {267--270} (\bibinfo {year}
			{1979})}\BibitemShut {NoStop}%
		\bibitem [{\citenamefont {Fuchs}\ \emph {et~al.}(2009)\citenamefont {Fuchs},
			\citenamefont {Weingartner}, \citenamefont {Popp}, \citenamefont {Major},
			\citenamefont {Becker}, \citenamefont {Osterhoff}, \citenamefont {Cortrie},
			\citenamefont {Zeitler}, \citenamefont {H{\"o}rlein}, \citenamefont
			{Tsakiris}, \citenamefont {Schramm}, \citenamefont {Rowlands-Rees},
			\citenamefont {Hooker}, \citenamefont {Habs}, \citenamefont {Krausz},
			\citenamefont {Karsch},\ and\ \citenamefont {Gr{\"u}ner}}]{Fuchs2009}%
		\BibitemOpen
		\bibfield  {author} {\bibinfo {author} {\bibfnamefont {M.}~\bibnamefont
				{Fuchs}}, \bibinfo {author} {\bibfnamefont {R.}~\bibnamefont {Weingartner}},
			\bibinfo {author} {\bibfnamefont {A.}~\bibnamefont {Popp}}, \bibinfo {author}
			{\bibfnamefont {Z.}~\bibnamefont {Major}}, \bibinfo {author} {\bibfnamefont
				{S.}~\bibnamefont {Becker}}, \bibinfo {author} {\bibfnamefont
				{J.}~\bibnamefont {Osterhoff}}, \bibinfo {author} {\bibfnamefont
				{I.}~\bibnamefont {Cortrie}}, \bibinfo {author} {\bibfnamefont
				{B.}~\bibnamefont {Zeitler}}, \bibinfo {author} {\bibfnamefont
				{R.}~\bibnamefont {H{\"o}rlein}}, \bibinfo {author} {\bibfnamefont {G.~D.}\
				\bibnamefont {Tsakiris}}, \bibinfo {author} {\bibfnamefont {U.}~\bibnamefont
				{Schramm}}, \bibinfo {author} {\bibfnamefont {T.~P.}\ \bibnamefont
				{Rowlands-Rees}}, \bibinfo {author} {\bibfnamefont {S.~M.}\ \bibnamefont
				{Hooker}}, \bibinfo {author} {\bibfnamefont {D.}~\bibnamefont {Habs}},
			\bibinfo {author} {\bibfnamefont {F.}~\bibnamefont {Krausz}}, \bibinfo
			{author} {\bibfnamefont {S.}~\bibnamefont {Karsch}}, \ and\ \bibinfo {author}
			{\bibfnamefont {F.}~\bibnamefont {Gr{\"u}ner}},\ }\bibfield  {title}
		{\enquote {\bibinfo {title} {Laser-driven soft-x-ray undulator source},}\
		}\href {\doibase https://doi.org/10.1038/nphys1404} {\bibfield  {journal}
			{\bibinfo  {journal} {Nature Physics}\ }\textbf {\bibinfo {volume} {5}},\
			\bibinfo {pages} {826--829} (\bibinfo {year} {2009})}\BibitemShut {NoStop}%
		\bibitem [{\citenamefont {Compton}(1923)}]{Compton1923}%
		\BibitemOpen
		\bibfield  {author} {\bibinfo {author} {\bibfnamefont {A.~H.}\ \bibnamefont
				{Compton}},\ }\bibfield  {title} {\enquote {\bibinfo {title} {A quantum
					theory of the scattering of x-rays by light elements},}\ }\href {\doibase
			10.1103/PhysRev.21.483} {\bibfield  {journal} {\bibinfo  {journal} {Phys.
					Rev.}\ }\textbf {\bibinfo {volume} {21}},\ \bibinfo {pages} {483--502}
			(\bibinfo {year} {1923})}\BibitemShut {NoStop}%
		\bibitem [{\citenamefont {Catravas}, \citenamefont {Esarey},\ and\
			\citenamefont {Leemans}(2001)}]{Catravas_2001}%
		\BibitemOpen
		\bibfield  {author} {\bibinfo {author} {\bibfnamefont {P.}~\bibnamefont
				{Catravas}}, \bibinfo {author} {\bibfnamefont {E.}~\bibnamefont {Esarey}}, \
			and\ \bibinfo {author} {\bibfnamefont {W.~P.}\ \bibnamefont {Leemans}},\
		}\bibfield  {title} {\enquote {\bibinfo {title} {Femtosecond x-rays from
					thomson scattering using laser wakefield accelerators},}\ }\href {\doibase
			10.1088/0957-0233/12/11/310} {\bibfield  {journal} {\bibinfo  {journal}
				{Measurement Science and Technology}\ }\textbf {\bibinfo {volume} {12}},\
			\bibinfo {pages} {1828--1834} (\bibinfo {year} {2001})}\BibitemShut {NoStop}%
		\bibitem [{\citenamefont {Schwoerer}\ \emph {et~al.}(2006)\citenamefont
			{Schwoerer}, \citenamefont {Liesfeld}, \citenamefont {Schlenvoigt},
			\citenamefont {Amthor},\ and\ \citenamefont {Sauerbrey}}]{Schwoerer2006}%
		\BibitemOpen
		\bibfield  {author} {\bibinfo {author} {\bibfnamefont {H.}~\bibnamefont
				{Schwoerer}}, \bibinfo {author} {\bibfnamefont {B.}~\bibnamefont {Liesfeld}},
			\bibinfo {author} {\bibfnamefont {H.-P.}\ \bibnamefont {Schlenvoigt}},
			\bibinfo {author} {\bibfnamefont {K.-U.}\ \bibnamefont {Amthor}}, \ and\
			\bibinfo {author} {\bibfnamefont {R.}~\bibnamefont {Sauerbrey}},\ }\bibfield
		{title} {\enquote {\bibinfo {title} {Thomson-backscattered x rays from
					laser-accelerated electrons},}\ }\href {\doibase
			10.1103/PhysRevLett.96.014802} {\bibfield  {journal} {\bibinfo  {journal}
				{Phys. Rev. Lett.}\ }\textbf {\bibinfo {volume} {96}},\ \bibinfo {pages}
			{014802} (\bibinfo {year} {2006})}\BibitemShut {NoStop}%
		\bibitem [{\citenamefont {Ta~Phuoc}\ \emph {et~al.}(2012)\citenamefont
			{Ta~Phuoc}, \citenamefont {Corde}, \citenamefont {Thaury}, \citenamefont
			{Malka}, \citenamefont {Tafzi}, \citenamefont {Goddet}, \citenamefont {Shah},
			\citenamefont {Sebban},\ and\ \citenamefont {Rousse}}]{TaPhuoc2012}%
		\BibitemOpen
		\bibfield  {author} {\bibinfo {author} {\bibfnamefont {K.}~\bibnamefont
				{Ta~Phuoc}}, \bibinfo {author} {\bibfnamefont {S.}~\bibnamefont {Corde}},
			\bibinfo {author} {\bibfnamefont {C.}~\bibnamefont {Thaury}}, \bibinfo
			{author} {\bibfnamefont {V.}~\bibnamefont {Malka}}, \bibinfo {author}
			{\bibfnamefont {A.}~\bibnamefont {Tafzi}}, \bibinfo {author} {\bibfnamefont
				{J.~P.}\ \bibnamefont {Goddet}}, \bibinfo {author} {\bibfnamefont {R.~C.}\
				\bibnamefont {Shah}}, \bibinfo {author} {\bibfnamefont {S.}~\bibnamefont
				{Sebban}}, \ and\ \bibinfo {author} {\bibfnamefont {A.}~\bibnamefont
				{Rousse}},\ }\bibfield  {title} {\enquote {\bibinfo {title} {All-optical
					compton gamma-ray source},}\ }\href {\doibase
			https://doi.org/10.1038/nphoton.2012.82} {\bibfield  {journal} {\bibinfo
				{journal} {Nature Photonics}\ }\textbf {\bibinfo {volume} {6}},\ \bibinfo
			{pages} {308--311} (\bibinfo {year} {2012})}\BibitemShut {NoStop}%
		\bibitem [{\citenamefont {Bethe}\ and\ \citenamefont
			{Maximon}(1954)}]{Bethe1954}%
		\BibitemOpen
		\bibfield  {author} {\bibinfo {author} {\bibfnamefont {H.~A.}\ \bibnamefont
				{Bethe}}\ and\ \bibinfo {author} {\bibfnamefont {L.~C.}\ \bibnamefont
				{Maximon}},\ }\bibfield  {title} {\enquote {\bibinfo {title} {Theory of
					bremsstrahlung and pair production. i. differential cross section},}\ }\href
		{\doibase 10.1103/PhysRev.93.768} {\bibfield  {journal} {\bibinfo  {journal}
				{Phys. Rev.}\ }\textbf {\bibinfo {volume} {93}},\ \bibinfo {pages} {768--784}
			(\bibinfo {year} {1954})}\BibitemShut {NoStop}%
		\bibitem [{\citenamefont {Edwards}\ \emph {et~al.}(2002)\citenamefont
			{Edwards}, \citenamefont {Sinclair}, \citenamefont {Goldsack}, \citenamefont
			{Krushelnick}, \citenamefont {Beg}, \citenamefont {Clark}, \citenamefont
			{Dangor}, \citenamefont {Najmudin}, \citenamefont {Tatarakis}, \citenamefont
			{Walton}, \citenamefont {Zepf}, \citenamefont {Ledingham}, \citenamefont
			{Spencer}, \citenamefont {Norreys}, \citenamefont {Clarke}, \citenamefont
			{Kodama}, \citenamefont {Toyama},\ and\ \citenamefont {Tampo}}]{Edwards2002}%
		\BibitemOpen
		\bibfield  {author} {\bibinfo {author} {\bibfnamefont {R.~D.}\ \bibnamefont
				{Edwards}}, \bibinfo {author} {\bibfnamefont {M.~A.}\ \bibnamefont
				{Sinclair}}, \bibinfo {author} {\bibfnamefont {T.~J.}\ \bibnamefont
				{Goldsack}}, \bibinfo {author} {\bibfnamefont {K.}~\bibnamefont
				{Krushelnick}}, \bibinfo {author} {\bibfnamefont {F.~N.}\ \bibnamefont
				{Beg}}, \bibinfo {author} {\bibfnamefont {E.~L.}\ \bibnamefont {Clark}},
			\bibinfo {author} {\bibfnamefont {A.~E.}\ \bibnamefont {Dangor}}, \bibinfo
			{author} {\bibfnamefont {Z.}~\bibnamefont {Najmudin}}, \bibinfo {author}
			{\bibfnamefont {M.}~\bibnamefont {Tatarakis}}, \bibinfo {author}
			{\bibfnamefont {B.}~\bibnamefont {Walton}}, \bibinfo {author} {\bibfnamefont
				{M.}~\bibnamefont {Zepf}}, \bibinfo {author} {\bibfnamefont {K.~W.~D.}\
				\bibnamefont {Ledingham}}, \bibinfo {author} {\bibfnamefont {I.}~\bibnamefont
				{Spencer}}, \bibinfo {author} {\bibfnamefont {P.~A.}\ \bibnamefont
				{Norreys}}, \bibinfo {author} {\bibfnamefont {R.~J.}\ \bibnamefont {Clarke}},
			\bibinfo {author} {\bibfnamefont {R.}~\bibnamefont {Kodama}}, \bibinfo
			{author} {\bibfnamefont {Y.}~\bibnamefont {Toyama}}, \ and\ \bibinfo {author}
			{\bibfnamefont {M.}~\bibnamefont {Tampo}},\ }\bibfield  {title} {\enquote
			{\bibinfo {title} {Characterization of a gamma-ray source based on a
					laser-plasma accelerator with applications to radiography},}\ }\href
		{\doibase 10.1063/1.1464221} {\bibfield  {journal} {\bibinfo  {journal}
				{Applied Physics Letters}\ }\textbf {\bibinfo {volume} {80}},\ \bibinfo
			{pages} {2129--2131} (\bibinfo {year} {2002})},\ \Eprint
		{http://arxiv.org/abs/https://doi.org/10.1063/1.1464221}
		{https://doi.org/10.1063/1.1464221} \BibitemShut {NoStop}%
		\bibitem [{\citenamefont {Glinec}\ \emph {et~al.}(2005)\citenamefont {Glinec},
			\citenamefont {Faure}, \citenamefont {LeDain}, \citenamefont {Darbon},
			\citenamefont {Hosokai}, \citenamefont {Santos}, \citenamefont {Lefebvre},
			\citenamefont {Rousseau}, \citenamefont {Burgy}, \citenamefont {Mercier},\
			and\ \citenamefont {Malka}}]{Glinec2005}%
		\BibitemOpen
		\bibfield  {author} {\bibinfo {author} {\bibfnamefont {Y.}~\bibnamefont
				{Glinec}}, \bibinfo {author} {\bibfnamefont {J.}~\bibnamefont {Faure}},
			\bibinfo {author} {\bibfnamefont {L.}\ \bibnamefont {LeDain}}, \bibinfo
			{author} {\bibfnamefont {S.}~\bibnamefont {Darbon}}, \bibinfo {author}
			{\bibfnamefont {T.}~\bibnamefont {Hosokai}}, \bibinfo {author} {\bibfnamefont
				{J.~J.}\ \bibnamefont {Santos}}, \bibinfo {author} {\bibfnamefont
				{E.}~\bibnamefont {Lefebvre}}, \bibinfo {author} {\bibfnamefont {J.~P.}\
				\bibnamefont {Rousseau}}, \bibinfo {author} {\bibfnamefont {F.}~\bibnamefont
				{Burgy}}, \bibinfo {author} {\bibfnamefont {B.}~\bibnamefont {Mercier}}, \
			and\ \bibinfo {author} {\bibfnamefont {V.}~\bibnamefont {Malka}},\ }\bibfield
		{title} {\enquote {\bibinfo {title} {High-resolution
					$\ensuremath{\gamma}$-ray radiography produced by a laser-plasma driven
					electron source},}\ }\href {\doibase 10.1103/PhysRevLett.94.025003}
		{\bibfield  {journal} {\bibinfo  {journal} {Phys. Rev. Lett.}\ }\textbf
			{\bibinfo {volume} {94}},\ \bibinfo {pages} {025003} (\bibinfo {year}
			{2005})}\BibitemShut {NoStop}%
		\bibitem [{\citenamefont {Döpp}\ \emph {et~al.}(2016)\citenamefont {Döpp},
			\citenamefont {Guillaume}, \citenamefont {Thaury}, \citenamefont {Lifschitz},
			\citenamefont {Sylla}, \citenamefont {Goddet}, \citenamefont {Tafzi},
			\citenamefont {Iaquanello}, \citenamefont {Lefrou}, \citenamefont {Rousseau},
			\citenamefont {Conejero}, \citenamefont {Ruiz}, \citenamefont {Phuoc},\ and\
			\citenamefont {Malka}}]{DOPP2016515}%
		\BibitemOpen
		\bibfield  {author} {\bibinfo {author} {\bibfnamefont {A.}~\bibnamefont
				{Döpp}}, \bibinfo {author} {\bibfnamefont {E.}~\bibnamefont {Guillaume}},
			\bibinfo {author} {\bibfnamefont {C.}~\bibnamefont {Thaury}}, \bibinfo
			{author} {\bibfnamefont {A.}~\bibnamefont {Lifschitz}}, \bibinfo {author}
			{\bibfnamefont {F.}~\bibnamefont {Sylla}}, \bibinfo {author} {\bibfnamefont
				{J.-P.}\ \bibnamefont {Goddet}}, \bibinfo {author} {\bibfnamefont
				{A.}~\bibnamefont {Tafzi}}, \bibinfo {author} {\bibfnamefont
				{G.}~\bibnamefont {Iaquanello}}, \bibinfo {author} {\bibfnamefont
				{T.}~\bibnamefont {Lefrou}}, \bibinfo {author} {\bibfnamefont
				{P.}~\bibnamefont {Rousseau}}, \bibinfo {author} {\bibfnamefont
				{E.}~\bibnamefont {Conejero}}, \bibinfo {author} {\bibfnamefont
				{C.}~\bibnamefont {Ruiz}}, \bibinfo {author} {\bibfnamefont {K.~T.}\
				\bibnamefont {Phuoc}}, \ and\ \bibinfo {author} {\bibfnamefont
				{V.}~\bibnamefont {Malka}},\ }\bibfield  {title} {\enquote {\bibinfo {title}
				{A bremsstrahlung gamma-ray source based on stable ionization injection of
					electrons into a laser wakefield accelerator},}\ }\href {\doibase
			https://doi.org/10.1016/j.nima.2016.01.086} {\bibfield  {journal} {\bibinfo
				{journal} {Nuclear Instruments and Methods in Physics Research Section A:
					Accelerators, Spectrometers, Detectors and Associated Equipment}\ }\textbf
			{\bibinfo {volume} {830}},\ \bibinfo {pages} {515 -- 519} (\bibinfo {year}
			{2016})}\BibitemShut {NoStop}%
		\bibitem [{\citenamefont {Cole}\ \emph {et~al.}(2015)\citenamefont {Cole},
			\citenamefont {Wood}, \citenamefont {Lopes}, \citenamefont {Poder},
			\citenamefont {Abel}, \citenamefont {Alatabi}, \citenamefont {Bryant},
			\citenamefont {Jin}, \citenamefont {Kneip}, \citenamefont {Mecseki},
			\citenamefont {Symes}, \citenamefont {Mangles},\ and\ \citenamefont
			{Najmudin}}]{Cole2015}%
		\BibitemOpen
		\bibfield  {author} {\bibinfo {author} {\bibfnamefont {J.~M.}\ \bibnamefont
				{Cole}}, \bibinfo {author} {\bibfnamefont {J.~C.}\ \bibnamefont {Wood}},
			\bibinfo {author} {\bibfnamefont {N.~C.}\ \bibnamefont {Lopes}}, \bibinfo
			{author} {\bibfnamefont {K.}~\bibnamefont {Poder}}, \bibinfo {author}
			{\bibfnamefont {R.~L.}\ \bibnamefont {Abel}}, \bibinfo {author}
			{\bibfnamefont {S.}~\bibnamefont {Alatabi}}, \bibinfo {author} {\bibfnamefont
				{J.~S.~J.}\ \bibnamefont {Bryant}}, \bibinfo {author} {\bibfnamefont
				{A.}~\bibnamefont {Jin}}, \bibinfo {author} {\bibfnamefont {S.}~\bibnamefont
				{Kneip}}, \bibinfo {author} {\bibfnamefont {K.}~\bibnamefont {Mecseki}},
			\bibinfo {author} {\bibfnamefont {D.~R.}\ \bibnamefont {Symes}}, \bibinfo
			{author} {\bibfnamefont {S.~P.~D.}\ \bibnamefont {Mangles}}, \ and\ \bibinfo
			{author} {\bibfnamefont {Z.}~\bibnamefont {Najmudin}},\ }\bibfield  {title}
		{\enquote {\bibinfo {title} {Laser-wakefield accelerators as hard x-ray
					sources for 3d medical imaging of human bone},}\ }\href@noop {} {\bibfield
			{journal} {\bibinfo  {journal} {Scientific Reports}\ }\textbf {\bibinfo
				{volume} {5}},\ \bibinfo {pages} {13244} (\bibinfo {year}
			{2015})}\BibitemShut {NoStop}%
		\bibitem [{\citenamefont {Wenz}\ \emph {et~al.}(2015)\citenamefont {Wenz},
			\citenamefont {Schleede}, \citenamefont {Khrennikov}, \citenamefont {Bech},
			\citenamefont {Thibault}, \citenamefont {Heigoldt}, \citenamefont
			{Pfeiffer},\ and\ \citenamefont {Karsch}}]{Wenz2015}%
		\BibitemOpen
		\bibfield  {author} {\bibinfo {author} {\bibfnamefont {J.}~\bibnamefont
				{Wenz}}, \bibinfo {author} {\bibfnamefont {S.}~\bibnamefont {Schleede}},
			\bibinfo {author} {\bibfnamefont {K.}~\bibnamefont {Khrennikov}}, \bibinfo
			{author} {\bibfnamefont {M.}~\bibnamefont {Bech}}, \bibinfo {author}
			{\bibfnamefont {P.}~\bibnamefont {Thibault}}, \bibinfo {author}
			{\bibfnamefont {M.}~\bibnamefont {Heigoldt}}, \bibinfo {author}
			{\bibfnamefont {F.}~\bibnamefont {Pfeiffer}}, \ and\ \bibinfo {author}
			{\bibfnamefont {S.}~\bibnamefont {Karsch}},\ }\bibfield  {title} {\enquote
			{\bibinfo {title} {Quantitative x-ray phase-contrast microtomography from a
					compact laser-driven betatron source},}\ }\href@noop {} {\bibfield  {journal}
			{\bibinfo  {journal} {Nat. Communic.}\ }\textbf {\bibinfo {volume}
				{6}} \bibinfo {pages} {7568} (\bibinfo {year} {2015})}\BibitemShut {NoStop}%
		\bibitem [{\citenamefont {Cole}\ \emph
			{et~al.}(2018{\natexlab{a}})\citenamefont {Cole}, \citenamefont {Symes},
			\citenamefont {Lopes}, \citenamefont {Wood}, \citenamefont {Poder},
			\citenamefont {Alatabi}, \citenamefont {Botchway}, \citenamefont {Foster},
			\citenamefont {Gratton}, \citenamefont {Johnson}, \citenamefont {Kamperidis},
			\citenamefont {Kononenko}, \citenamefont {De~Lazzari}, \citenamefont
			{Palmer}, \citenamefont {Rusby}, \citenamefont {Sanderson}, \citenamefont
			{Sandholzer}, \citenamefont {Sarri}, \citenamefont {Szoke-Kovacs},
			\citenamefont {Teboul}, \citenamefont {Thompson}, \citenamefont {Warwick},
			\citenamefont {Westerberg}, \citenamefont {Hill}, \citenamefont {Norris},
			\citenamefont {Mangles},\ and\ \citenamefont {Najmudin}}]{Cole6335}%
		\BibitemOpen
		\bibfield  {author} {\bibinfo {author} {\bibfnamefont {J.~M.}\ \bibnamefont
				{Cole}}, \bibinfo {author} {\bibfnamefont {D.~R.}\ \bibnamefont {Symes}},
			\bibinfo {author} {\bibfnamefont {N.~C.}\ \bibnamefont {Lopes}}, \bibinfo
			{author} {\bibfnamefont {J.~C.}\ \bibnamefont {Wood}}, \bibinfo {author}
			{\bibfnamefont {K.}~\bibnamefont {Poder}}, \bibinfo {author} {\bibfnamefont
				{S.}~\bibnamefont {Alatabi}}, \bibinfo {author} {\bibfnamefont {S.~W.}\
				\bibnamefont {Botchway}}, \bibinfo {author} {\bibfnamefont {P.~S.}\
				\bibnamefont {Foster}}, \bibinfo {author} {\bibfnamefont {S.}~\bibnamefont
				{Gratton}}, \bibinfo {author} {\bibfnamefont {S.}~\bibnamefont {Johnson}},
			\bibinfo {author} {\bibfnamefont {C.}~\bibnamefont {Kamperidis}}, \bibinfo
			{author} {\bibfnamefont {O.}~\bibnamefont {Kononenko}}, \bibinfo {author}
			{\bibfnamefont {M.}~\bibnamefont {De~Lazzari}}, \bibinfo {author}
			{\bibfnamefont {C.~A.~J.}\ \bibnamefont {Palmer}}, \bibinfo {author}
			{\bibfnamefont {D.}~\bibnamefont {Rusby}}, \bibinfo {author} {\bibfnamefont
				{J.}~\bibnamefont {Sanderson}}, \bibinfo {author} {\bibfnamefont
				{M.}~\bibnamefont {Sandholzer}}, \bibinfo {author} {\bibfnamefont
				{G.}~\bibnamefont {Sarri}}, \bibinfo {author} {\bibfnamefont
				{Z.}~\bibnamefont {Szoke-Kovacs}}, \bibinfo {author} {\bibfnamefont
				{L.}~\bibnamefont {Teboul}}, \bibinfo {author} {\bibfnamefont {J.~M.}\
				\bibnamefont {Thompson}}, \bibinfo {author} {\bibfnamefont {J.~R.}\
				\bibnamefont {Warwick}}, \bibinfo {author} {\bibfnamefont {H.}~\bibnamefont
				{Westerberg}}, \bibinfo {author} {\bibfnamefont {M.~A.}\ \bibnamefont
				{Hill}}, \bibinfo {author} {\bibfnamefont {D.~P.}\ \bibnamefont {Norris}},
			\bibinfo {author} {\bibfnamefont {S.~P.~D.}\ \bibnamefont {Mangles}}, \ and\
			\bibinfo {author} {\bibfnamefont {Z.}~\bibnamefont {Najmudin}},\ }\bibfield
		{title} {\enquote {\bibinfo {title} {High-resolution $\mu$ct of a mouse
					embryo using a compact laser-driven x-ray betatron source},}\ }\href
		{\doibase 10.1073/pnas.1802314115} {\bibfield  {journal} {\bibinfo  {journal}
				{Proceedings of the National Academy of Sciences}\ }\textbf {\bibinfo
				{volume} {115}},\ \bibinfo {pages} {6335--6340} (\bibinfo {year}
			{2018}{\natexlab{a}})},\ \Eprint
		{http://arxiv.org/abs/https://www.pnas.org/content/115/25/6335.full.pdf}
		{https://www.pnas.org/content/115/25/6335.full.pdf} \BibitemShut {NoStop}%
		\bibitem [{\citenamefont {Rousse}\ \emph {et~al.}(2004)\citenamefont {Rousse},
			\citenamefont {Phuoc}, \citenamefont {Shah}, \citenamefont {Pukhov},
			\citenamefont {Lefebvre}, \citenamefont {Malka}, \citenamefont {Kiselev},
			\citenamefont {Burgy}, \citenamefont {Rousseau}, \citenamefont {Umstadter},\
			and\ \citenamefont {Hulin}}]{Rousse2004}%
		\BibitemOpen
		\bibfield  {author} {\bibinfo {author} {\bibfnamefont {A.}~\bibnamefont
				{Rousse}}, \bibinfo {author} {\bibfnamefont {K.~T.}\ \bibnamefont {Phuoc}},
			\bibinfo {author} {\bibfnamefont {R.}~\bibnamefont {Shah}}, \bibinfo {author}
			{\bibfnamefont {A.}~\bibnamefont {Pukhov}}, \bibinfo {author} {\bibfnamefont
				{E.}~\bibnamefont {Lefebvre}}, \bibinfo {author} {\bibfnamefont
				{V.}~\bibnamefont {Malka}}, \bibinfo {author} {\bibfnamefont
				{S.}~\bibnamefont {Kiselev}}, \bibinfo {author} {\bibfnamefont
				{F.}~\bibnamefont {Burgy}}, \bibinfo {author} {\bibfnamefont {J.-P.}\
				\bibnamefont {Rousseau}}, \bibinfo {author} {\bibfnamefont {D.}~\bibnamefont
				{Umstadter}}, \ and\ \bibinfo {author} {\bibfnamefont {D.}~\bibnamefont
				{Hulin}},\ }\bibfield  {title} {\enquote {\bibinfo {title} {Production of a
					kev x-ray beam from synchrotron radiation in relativistic laser-plasma
					interaction},}\ }\href {\doibase 10.1103/PhysRevLett.93.135005} {\bibfield
			{journal} {\bibinfo  {journal} {Phys. Rev. Lett.}\ }\textbf {\bibinfo
				{volume} {93}},\ \bibinfo {pages} {135005} (\bibinfo {year}
			{2004})}\BibitemShut {NoStop}%
		\bibitem [{\citenamefont {Kneip}\ \emph {et~al.}(2010)\citenamefont {Kneip},
			\citenamefont {McGuffey}, \citenamefont {Martins}, \citenamefont {Martins},
			\citenamefont {Bellei}, \citenamefont {Chvykov}, \citenamefont {Dollar},
			\citenamefont {Fonseca}, \citenamefont {Huntington}, \citenamefont
			{Kalintchenko}, \citenamefont {Maksimchuk}, \citenamefont {Mangles},
			\citenamefont {Matsuoka}, \citenamefont {Nagel}, \citenamefont {Palmer},
			\citenamefont {Schreiber}, \citenamefont {Ta~Phuoc}, \citenamefont {Thomas},
			\citenamefont {Yanovsky}, \citenamefont {Silva}, \citenamefont
			{Krushelnick},\ and\ \citenamefont {Najmudin}}]{Kneip2010}%
		\BibitemOpen
		\bibfield  {author} {\bibinfo {author} {\bibfnamefont {S.}~\bibnamefont
				{Kneip}}, \bibinfo {author} {\bibfnamefont {C.}~\bibnamefont {McGuffey}},
			\bibinfo {author} {\bibfnamefont {J.~L.}\ \bibnamefont {Martins}}, \bibinfo
			{author} {\bibfnamefont {S.~F.}\ \bibnamefont {Martins}}, \bibinfo {author}
			{\bibfnamefont {C.}~\bibnamefont {Bellei}}, \bibinfo {author} {\bibfnamefont
				{V.}~\bibnamefont {Chvykov}}, \bibinfo {author} {\bibfnamefont
				{F.}~\bibnamefont {Dollar}}, \bibinfo {author} {\bibfnamefont
				{R.}~\bibnamefont {Fonseca}}, \bibinfo {author} {\bibfnamefont
				{C.}~\bibnamefont {Huntington}}, \bibinfo {author} {\bibfnamefont
				{G.}~\bibnamefont {Kalintchenko}}, \bibinfo {author} {\bibfnamefont
				{A.}~\bibnamefont {Maksimchuk}}, \bibinfo {author} {\bibfnamefont {S.~P.~D.}\
				\bibnamefont {Mangles}}, \bibinfo {author} {\bibfnamefont {T.}~\bibnamefont
				{Matsuoka}}, \bibinfo {author} {\bibfnamefont {S.~R.}\ \bibnamefont {Nagel}},
			\bibinfo {author} {\bibfnamefont {C.~A.~J.}\ \bibnamefont {Palmer}}, \bibinfo
			{author} {\bibfnamefont {J.}~\bibnamefont {Schreiber}}, \bibinfo {author}
			{\bibfnamefont {K.}~\bibnamefont {Ta~Phuoc}}, \bibinfo {author}
			{\bibfnamefont {A.~G.~R.}\ \bibnamefont {Thomas}}, \bibinfo {author}
			{\bibfnamefont {V.}~\bibnamefont {Yanovsky}}, \bibinfo {author}
			{\bibfnamefont {L.~O.}\ \bibnamefont {Silva}}, \bibinfo {author}
			{\bibfnamefont {K.}~\bibnamefont {Krushelnick}}, \ and\ \bibinfo {author}
			{\bibfnamefont {Z.}~\bibnamefont {Najmudin}},\ }\bibfield  {title} {\enquote
			{\bibinfo {title} {Bright spatially coherent synchrotron x-rays from a
					table-top source},}\ }\href {\doibase https://doi.org/10.1038/nphys1789}
		{\bibfield  {journal} {\bibinfo  {journal} {Nature Physics}\ }\textbf
			{\bibinfo {volume} {6}},\ \bibinfo {pages} {980--983} (\bibinfo {year}
			{2010})}\BibitemShut {NoStop}%
		\bibitem [{\citenamefont {Wang}\ \emph {et~al.}(2021)\citenamefont {Wang},
			\citenamefont {Feng}, \citenamefont {Ke}, \citenamefont {Yu}, \citenamefont
			{Xu}, \citenamefont {Qi}, \citenamefont {Chen}, \citenamefont {Qin},
			\citenamefont {Zhang}, \citenamefont {Fang}, \citenamefont {Liu},
			\citenamefont {Jiang}, \citenamefont {Wang}, \citenamefont {Wang},
			\citenamefont {Yang}, \citenamefont {Wu}, \citenamefont {Leng}, \citenamefont
			{Liu}, \citenamefont {Li},\ and\ \citenamefont {Xu}}]{Wang2021}%
		\BibitemOpen
		\bibfield  {author} {\bibinfo {author} {\bibfnamefont {W.}~\bibnamefont
				{Wang}}, \bibinfo {author} {\bibfnamefont {K.}~\bibnamefont {Feng}}, \bibinfo
			{author} {\bibfnamefont {L.}~\bibnamefont {Ke}}, \bibinfo {author}
			{\bibfnamefont {C.}~\bibnamefont {Yu}}, \bibinfo {author} {\bibfnamefont
				{Y.}~\bibnamefont {Xu}}, \bibinfo {author} {\bibfnamefont {R.}~\bibnamefont
				{Qi}}, \bibinfo {author} {\bibfnamefont {Y.}~\bibnamefont {Chen}}, \bibinfo
			{author} {\bibfnamefont {Z.}~\bibnamefont {Qin}}, \bibinfo {author}
			{\bibfnamefont {Z.}~\bibnamefont {Zhang}}, \bibinfo {author} {\bibfnamefont
				{M.}~\bibnamefont {Fang}}, \bibinfo {author} {\bibfnamefont {J.}~\bibnamefont
				{Liu}}, \bibinfo {author} {\bibfnamefont {K.}~\bibnamefont {Jiang}}, \bibinfo
			{author} {\bibfnamefont {H.}~\bibnamefont {Wang}}, \bibinfo {author}
			{\bibfnamefont {C.}~\bibnamefont {Wang}}, \bibinfo {author} {\bibfnamefont
				{X.}~\bibnamefont {Yang}}, \bibinfo {author} {\bibfnamefont {F.}~\bibnamefont
				{Wu}}, \bibinfo {author} {\bibfnamefont {Y.}~\bibnamefont {Leng}}, \bibinfo
			{author} {\bibfnamefont {J.}~\bibnamefont {Liu}}, \bibinfo {author}
			{\bibfnamefont {R.}~\bibnamefont {Li}}, \ and\ \bibinfo {author}
			{\bibfnamefont {Z.}~\bibnamefont {Xu}},\ }\bibfield  {title} {\enquote
			{\bibinfo {title} {Free-electron lasing at 27 nanometres based on a laser
					wakefield accelerator},}\ }\href {\doibase 10.1038/s41586-021-03678-x}
		{\bibfield  {journal} {\bibinfo  {journal} {Nature}\ }\textbf {\bibinfo
				{volume} {595}} (\bibinfo {year} {2021}),\
			10.1038/s41586-021-03678-x}\BibitemShut {NoStop}%
		\bibitem [{\citenamefont {Mangles}\ \emph {et~al.}(2004)\citenamefont
			{Mangles}, \citenamefont {Murphy}, \citenamefont {Najmudin}, \citenamefont
			{Thomas}, \citenamefont {Collier}, \citenamefont {Dangor}, \citenamefont
			{Divall}, \citenamefont {Foster}, \citenamefont {Gallacher}, \citenamefont
			{Hooker}, \citenamefont {Jaroszynski}, \citenamefont {Langley}, \citenamefont
			{Mori}, \citenamefont {Norreys}, \citenamefont {Tsung}, \citenamefont
			{Viskup}, \citenamefont {Walton},\ and\ \citenamefont
			{Krushelnick}}]{Mangles2004a}%
		\BibitemOpen
		\bibfield  {author} {\bibinfo {author} {\bibfnamefont {S.~P.~D.}\
				\bibnamefont {Mangles}}, \bibinfo {author} {\bibfnamefont {C.~D.}\
				\bibnamefont {Murphy}}, \bibinfo {author} {\bibfnamefont {Z.}~\bibnamefont
				{Najmudin}}, \bibinfo {author} {\bibfnamefont {A.~G.~R.}\ \bibnamefont
				{Thomas}}, \bibinfo {author} {\bibfnamefont {J.~L.}\ \bibnamefont {Collier}},
			\bibinfo {author} {\bibfnamefont {A.~E.}\ \bibnamefont {Dangor}}, \bibinfo
			{author} {\bibfnamefont {E.~J.}\ \bibnamefont {Divall}}, \bibinfo {author}
			{\bibfnamefont {P.~S.}\ \bibnamefont {Foster}}, \bibinfo {author}
			{\bibfnamefont {J.~G.}\ \bibnamefont {Gallacher}}, \bibinfo {author}
			{\bibfnamefont {C.~J.}\ \bibnamefont {Hooker}}, \bibinfo {author}
			{\bibfnamefont {D.~A.}\ \bibnamefont {Jaroszynski}}, \bibinfo {author}
			{\bibfnamefont {A.~J.}\ \bibnamefont {Langley}}, \bibinfo {author}
			{\bibfnamefont {W.~B.}\ \bibnamefont {Mori}}, \bibinfo {author}
			{\bibfnamefont {P.~A.}\ \bibnamefont {Norreys}}, \bibinfo {author}
			{\bibfnamefont {F.~S.}\ \bibnamefont {Tsung}}, \bibinfo {author}
			{\bibfnamefont {R.}~\bibnamefont {Viskup}}, \bibinfo {author} {\bibfnamefont
				{B.~R.}\ \bibnamefont {Walton}}, \ and\ \bibinfo {author} {\bibfnamefont
				{K.}~\bibnamefont {Krushelnick}},\ }\bibfield  {title} {\enquote {\bibinfo
				{title} {{Monoenergetic beams of relativistic electrons from intense laser
						plasma interactions}},}\ }\href {\doibase 10.1038/nature02930.1.} {\bibfield
			{journal} {\bibinfo  {journal} {Nature}\ }\textbf {\bibinfo {volume} {431}},\
			\bibinfo {pages} {535--538} (\bibinfo {year} {2004})}\BibitemShut {NoStop}%
		\bibitem [{\citenamefont {Geddes}\ \emph {et~al.}(2004)\citenamefont {Geddes},
			\citenamefont {Toth}, \citenamefont {Tilborg}, \citenamefont {Esarey},
			\citenamefont {Schroeder}, \citenamefont {Bruhwiler}, \citenamefont {Nieter},
			\citenamefont {Cary},\ and\ \citenamefont {Leemans}}]{Geddes2004a}%
		\BibitemOpen
		\bibfield  {author} {\bibinfo {author} {\bibfnamefont {C.}~\bibnamefont
				{Geddes}}, \bibinfo {author} {\bibfnamefont {C.}~\bibnamefont {Toth}},
			\bibinfo {author} {\bibfnamefont {J.~V.}\ \bibnamefont {Tilborg}}, \bibinfo
			{author} {\bibfnamefont {E.}~\bibnamefont {Esarey}}, \bibinfo {author}
			{\bibfnamefont {C.~B.}\ \bibnamefont {Schroeder}}, \bibinfo {author}
			{\bibfnamefont {D.}~\bibnamefont {Bruhwiler}}, \bibinfo {author}
			{\bibfnamefont {C.}~\bibnamefont {Nieter}}, \bibinfo {author} {\bibfnamefont
				{J.}~\bibnamefont {Cary}}, \ and\ \bibinfo {author} {\bibfnamefont {W.~P.}\
				\bibnamefont {Leemans}},\ }\bibfield  {title} {\enquote {\bibinfo {title}
				{{High-quality electron beams from a laser wakefield accelerator using
						plasma-channel guiding}},}\ }\href {\doibase 10.1038/nature02939.1.}
		{\bibfield  {journal} {\bibinfo  {journal} {Nature}\ }\textbf {\bibinfo
				{volume} {431}},\ \bibinfo {pages} {538--541} (\bibinfo {year}
			{2004})}\BibitemShut {NoStop}%
		\bibitem [{\citenamefont {Faure}\ \emph {et~al.}(2004)\citenamefont {Faure},
			\citenamefont {Glinec}, \citenamefont {Pukhov}, \citenamefont {Kiselev},
			\citenamefont {Gordienko}, \citenamefont {Lefebvre}, \citenamefont
			{Rousseau}, \citenamefont {Burgy},\ and\ \citenamefont {Malka}}]{Faure2004a}%
		\BibitemOpen
		\bibfield  {author} {\bibinfo {author} {\bibfnamefont {J.}~\bibnamefont
				{Faure}}, \bibinfo {author} {\bibfnamefont {Y.}~\bibnamefont {Glinec}},
			\bibinfo {author} {\bibfnamefont {A.}~\bibnamefont {Pukhov}}, \bibinfo
			{author} {\bibfnamefont {S.}~\bibnamefont {Kiselev}}, \bibinfo {author}
			{\bibfnamefont {S.}~\bibnamefont {Gordienko}}, \bibinfo {author}
			{\bibfnamefont {E.}~\bibnamefont {Lefebvre}}, \bibinfo {author}
			{\bibfnamefont {J.-P.}\ \bibnamefont {Rousseau}}, \bibinfo {author}
			{\bibfnamefont {F.}~\bibnamefont {Burgy}}, \ and\ \bibinfo {author}
			{\bibfnamefont {V.}~\bibnamefont {Malka}},\ }\bibfield  {title} {\enquote
			{\bibinfo {title} {{A laser plasma accelerator producing monoenergetic
						electron beams}},}\ }\href {\doibase 10.1038/nature02900.1.} {\bibfield
			{journal} {\bibinfo  {journal} {Nature}\ }\textbf {\bibinfo {volume} {431}},\
			\bibinfo {pages} {541--544} (\bibinfo {year} {2004})}\BibitemShut {NoStop}%
		\bibitem [{\citenamefont {Wang}\ \emph {et~al.}(2016)\citenamefont {Wang},
			\citenamefont {Li}, \citenamefont {Liu}, \citenamefont {Zhang}, \citenamefont
			{Qi}, \citenamefont {Yu}, \citenamefont {Liu}, \citenamefont {Fang},
			\citenamefont {Qin}, \citenamefont {Wang}, \citenamefont {Xu}, \citenamefont
			{Wu}, \citenamefont {Leng}, \citenamefont {Li},\ and\ \citenamefont
			{Xu}}]{Wang2016}%
		\BibitemOpen
		\bibfield  {author} {\bibinfo {author} {\bibfnamefont {W.~T.}\ \bibnamefont
				{Wang}}, \bibinfo {author} {\bibfnamefont {W.~T.}\ \bibnamefont {Li}},
			\bibinfo {author} {\bibfnamefont {J.~S.}\ \bibnamefont {Liu}}, \bibinfo
			{author} {\bibfnamefont {Z.~J.}\ \bibnamefont {Zhang}}, \bibinfo {author}
			{\bibfnamefont {R.}~\bibnamefont {Qi}}, \bibinfo {author} {\bibfnamefont
				{C.~H.}\ \bibnamefont {Yu}}, \bibinfo {author} {\bibfnamefont {J.~Q.}\
				\bibnamefont {Liu}}, \bibinfo {author} {\bibfnamefont {M.}~\bibnamefont
				{Fang}}, \bibinfo {author} {\bibfnamefont {Z.~Y.}\ \bibnamefont {Qin}},
			\bibinfo {author} {\bibfnamefont {C.}~\bibnamefont {Wang}}, \bibinfo {author}
			{\bibfnamefont {Y.}~\bibnamefont {Xu}}, \bibinfo {author} {\bibfnamefont
				{F.~X.}\ \bibnamefont {Wu}}, \bibinfo {author} {\bibfnamefont {Y.~X.}\
				\bibnamefont {Leng}}, \bibinfo {author} {\bibfnamefont {R.~X.}\ \bibnamefont
				{Li}}, \ and\ \bibinfo {author} {\bibfnamefont {Z.~Z.}\ \bibnamefont {Xu}},\
		}\bibfield  {title} {\enquote {\bibinfo {title} {High-brightness high-energy
					electron beams from a laser wakefield accelerator via energy chirp
					control},}\ }\href {\doibase 10.1103/PhysRevLett.117.124801} {\bibfield
			{journal} {\bibinfo  {journal} {Phys. Rev. Lett.}\ }\textbf {\bibinfo
				{volume} {117}},\ \bibinfo {pages} {124801} (\bibinfo {year}
			{2016})}\BibitemShut {NoStop}%
		\bibitem [{\citenamefont {Wang}\ \emph {et~al.}(2013)\citenamefont {Wang},
			\citenamefont {Zgadzaj}, \citenamefont {Fazel}, \citenamefont {Li},
			\citenamefont {Yi}, \citenamefont {Zhang}, \citenamefont {Henderson},
			\citenamefont {Chang}, \citenamefont {Korzekwa}, \citenamefont {Tsai},
			\citenamefont {Pai}, \citenamefont {Quevedo}, \citenamefont {Dyer},
			\citenamefont {Gaul}, \citenamefont {Martinez}, \citenamefont {Bernstein},
			\citenamefont {Borger}, \citenamefont {Spinks}, \citenamefont {Donovan},
			\citenamefont {Khudik}, \citenamefont {Shvets}, \citenamefont {Ditmire},\
			and\ \citenamefont {Downer}}]{Wang2013}%
		\BibitemOpen
		\bibfield  {author} {\bibinfo {author} {\bibfnamefont {X.}~\bibnamefont
				{Wang}}, \bibinfo {author} {\bibfnamefont {R.}~\bibnamefont {Zgadzaj}},
			\bibinfo {author} {\bibfnamefont {N.}~\bibnamefont {Fazel}}, \bibinfo
			{author} {\bibfnamefont {Z.}~\bibnamefont {Li}}, \bibinfo {author}
			{\bibfnamefont {S.~A.}\ \bibnamefont {Yi}}, \bibinfo {author} {\bibfnamefont
				{X.}~\bibnamefont {Zhang}}, \bibinfo {author} {\bibfnamefont
				{W.}~\bibnamefont {Henderson}}, \bibinfo {author} {\bibfnamefont {Y.-Y.}\
				\bibnamefont {Chang}}, \bibinfo {author} {\bibfnamefont {R.}~\bibnamefont
				{Korzekwa}}, \bibinfo {author} {\bibfnamefont {H.-E.}\ \bibnamefont {Tsai}},
			\bibinfo {author} {\bibfnamefont {C.~H.}\ \bibnamefont {Pai}}, \bibinfo
			{author} {\bibfnamefont {H.}~\bibnamefont {Quevedo}}, \bibinfo {author}
			{\bibfnamefont {G.}~\bibnamefont {Dyer}}, \bibinfo {author} {\bibfnamefont
				{E.}~\bibnamefont {Gaul}}, \bibinfo {author} {\bibfnamefont {M.}~\bibnamefont
				{Martinez}}, \bibinfo {author} {\bibfnamefont {A.~C.}\ \bibnamefont
				{Bernstein}}, \bibinfo {author} {\bibfnamefont {T.}~\bibnamefont {Borger}},
			\bibinfo {author} {\bibfnamefont {M.}~\bibnamefont {Spinks}}, \bibinfo
			{author} {\bibfnamefont {M.}~\bibnamefont {Donovan}}, \bibinfo {author}
			{\bibfnamefont {V.}~\bibnamefont {Khudik}}, \bibinfo {author} {\bibfnamefont
				{G.}~\bibnamefont {Shvets}}, \bibinfo {author} {\bibfnamefont
				{T.}~\bibnamefont {Ditmire}}, \ and\ \bibinfo {author} {\bibfnamefont
				{M.~C.}\ \bibnamefont {Downer}},\ }\bibfield  {title} {\enquote {\bibinfo
				{title} {{Quasi monoenergetic laser plasma acceleration of electrons to 2
						GeV}},}\ }\href {\doibase 10.1038/ncomms2988} {\bibfield  {journal} {\bibinfo
				{journal} {Nature Communications}\ }\textbf {\bibinfo {volume} {4}}
			(\bibinfo {year} {2013}),\ 10.1038/ncomms2988}\BibitemShut {NoStop}%
		\bibitem [{\citenamefont {Kim}\ \emph {et~al.}(2013)\citenamefont {Kim},
			\citenamefont {Pae}, \citenamefont {Cha}, \citenamefont {Kim}, \citenamefont
			{Yu}, \citenamefont {Sung}, \citenamefont {Lee}, \citenamefont {Jeong},\ and\
			\citenamefont {Lee}}]{Kim2013}%
		\BibitemOpen
		\bibfield  {author} {\bibinfo {author} {\bibfnamefont {H.~T.}\ \bibnamefont
				{Kim}}, \bibinfo {author} {\bibfnamefont {K.~H.}\ \bibnamefont {Pae}},
			\bibinfo {author} {\bibfnamefont {H.~J.}\ \bibnamefont {Cha}}, \bibinfo
			{author} {\bibfnamefont {I.~J.}\ \bibnamefont {Kim}}, \bibinfo {author}
			{\bibfnamefont {T.~J.}\ \bibnamefont {Yu}}, \bibinfo {author} {\bibfnamefont
				{J.~H.}\ \bibnamefont {Sung}}, \bibinfo {author} {\bibfnamefont {S.~K.}\
				\bibnamefont {Lee}}, \bibinfo {author} {\bibfnamefont {T.~M.}\ \bibnamefont
				{Jeong}}, \ and\ \bibinfo {author} {\bibfnamefont {J.}~\bibnamefont {Lee}},\
		}\bibfield  {title} {\enquote {\bibinfo {title} {Enhancement of electron
					energy to the multi-gev regime by a dual-stage laser-wakefield accelerator
					pumped by petawatt laser pulses},}\ }\href {\doibase
			10.1103/PhysRevLett.111.165002} {\bibfield  {journal} {\bibinfo  {journal}
				{Phys. Rev. Lett.}\ }\textbf {\bibinfo {volume} {111}},\ \bibinfo {pages}
			{165002} (\bibinfo {year} {2013})}\BibitemShut {NoStop}%
		\bibitem [{\citenamefont {Gonsalves}\ \emph {et~al.}(2019)\citenamefont
			{Gonsalves}, \citenamefont {Nakamura}, \citenamefont {Daniels}, \citenamefont
			{Benedetti}, \citenamefont {Pieronek}, \citenamefont {de~Raadt},
			\citenamefont {Steinke}, \citenamefont {Bin}, \citenamefont {Bulanov},
			\citenamefont {van Tilborg}, \citenamefont {Geddes}, \citenamefont
			{Schroeder}, \citenamefont {T\'oth}, \citenamefont {Esarey}, \citenamefont
			{Swanson}, \citenamefont {Fan-Chiang}, \citenamefont {Bagdasarov},
			\citenamefont {Bobrova}, \citenamefont {Gasilov}, \citenamefont {Korn},
			\citenamefont {Sasorov},\ and\ \citenamefont {Leemans}}]{Gonsalves2019}%
		\BibitemOpen
		\bibfield  {author} {\bibinfo {author} {\bibfnamefont {A.~J.}\ \bibnamefont
				{Gonsalves}}, \bibinfo {author} {\bibfnamefont {K.}~\bibnamefont {Nakamura}},
			\bibinfo {author} {\bibfnamefont {J.}~\bibnamefont {Daniels}}, \bibinfo
			{author} {\bibfnamefont {C.}~\bibnamefont {Benedetti}}, \bibinfo {author}
			{\bibfnamefont {C.}~\bibnamefont {Pieronek}}, \bibinfo {author}
			{\bibfnamefont {T.~C.~H.}\ \bibnamefont {de~Raadt}}, \bibinfo {author}
			{\bibfnamefont {S.}~\bibnamefont {Steinke}}, \bibinfo {author} {\bibfnamefont
				{J.~H.}\ \bibnamefont {Bin}}, \bibinfo {author} {\bibfnamefont {S.~S.}\
				\bibnamefont {Bulanov}}, \bibinfo {author} {\bibfnamefont {J.}~\bibnamefont
				{van Tilborg}}, \bibinfo {author} {\bibfnamefont {C.~G.~R.}\ \bibnamefont
				{Geddes}}, \bibinfo {author} {\bibfnamefont {C.~B.}\ \bibnamefont
				{Schroeder}}, \bibinfo {author} {\bibfnamefont {C.}~\bibnamefont {T\'oth}},
			\bibinfo {author} {\bibfnamefont {E.}~\bibnamefont {Esarey}}, \bibinfo
			{author} {\bibfnamefont {K.}~\bibnamefont {Swanson}}, \bibinfo {author}
			{\bibfnamefont {L.}~\bibnamefont {Fan-Chiang}}, \bibinfo {author}
			{\bibfnamefont {G.}~\bibnamefont {Bagdasarov}}, \bibinfo {author}
			{\bibfnamefont {N.}~\bibnamefont {Bobrova}}, \bibinfo {author} {\bibfnamefont
				{V.}~\bibnamefont {Gasilov}}, \bibinfo {author} {\bibfnamefont
				{G.}~\bibnamefont {Korn}}, \bibinfo {author} {\bibfnamefont {P.}~\bibnamefont
				{Sasorov}}, \ and\ \bibinfo {author} {\bibfnamefont {W.~P.}\ \bibnamefont
				{Leemans}},\ }\bibfield  {title} {\enquote {\bibinfo {title} {Petawatt laser
					guiding and electron beam acceleration to 8 gev in a laser-heated capillary
					discharge waveguide},}\ }\href {\doibase 10.1103/PhysRevLett.122.084801}
		{\bibfield  {journal} {\bibinfo  {journal} {Phys. Rev. Lett.}\ }\textbf
			{\bibinfo {volume} {122}},\ \bibinfo {pages} {084801} (\bibinfo {year}
			{2019})}\BibitemShut {NoStop}%
		\bibitem [{\citenamefont {Couperus}\ \emph {et~al.}(2017)\citenamefont
			{Couperus}, \citenamefont {Pausch}, \citenamefont {K{\"o}hler}, \citenamefont
			{Zarini}, \citenamefont {Kr{\"a}mer}, \citenamefont {Garten}, \citenamefont
			{Huebl}, \citenamefont {Gebhardt}, \citenamefont {Helbig}, \citenamefont
			{Bock}, \citenamefont {Zeil}, \citenamefont {Debus}, \citenamefont
			{Bussmann}, \citenamefont {Schramm},\ and\ \citenamefont
			{Irman}}]{Couperus2017}%
		\BibitemOpen
		\bibfield  {author} {\bibinfo {author} {\bibfnamefont {J.~P.}\ \bibnamefont
				{Couperus}}, \bibinfo {author} {\bibfnamefont {R.}~\bibnamefont {Pausch}},
			\bibinfo {author} {\bibfnamefont {A.}~\bibnamefont {K{\"o}hler}}, \bibinfo
			{author} {\bibfnamefont {O.}~\bibnamefont {Zarini}}, \bibinfo {author}
			{\bibfnamefont {J.~M.}\ \bibnamefont {Kr{\"a}mer}}, \bibinfo {author}
			{\bibfnamefont {M.}~\bibnamefont {Garten}}, \bibinfo {author} {\bibfnamefont
				{A.}~\bibnamefont {Huebl}}, \bibinfo {author} {\bibfnamefont
				{R.}~\bibnamefont {Gebhardt}}, \bibinfo {author} {\bibfnamefont
				{U.}~\bibnamefont {Helbig}}, \bibinfo {author} {\bibfnamefont
				{S.}~\bibnamefont {Bock}}, \bibinfo {author} {\bibfnamefont {K.}~\bibnamefont
				{Zeil}}, \bibinfo {author} {\bibfnamefont {A.}~\bibnamefont {Debus}},
			\bibinfo {author} {\bibfnamefont {M.}~\bibnamefont {Bussmann}}, \bibinfo
			{author} {\bibfnamefont {U.}~\bibnamefont {Schramm}}, \ and\ \bibinfo
			{author} {\bibfnamefont {A.}~\bibnamefont {Irman}},\ }\bibfield  {title}
		{\enquote {\bibinfo {title} {{Demonstration of a beam loaded
						nanocoulomb-class laser wakefield accelerator}},}\ }\href {\doibase
			10.1038/s41467-017-00592-7} {\bibfield  {journal} {\bibinfo  {journal}
				{Nature Communications}\ }\textbf {\bibinfo {volume} {8}} (\bibinfo {year}
			{2017}),\ 10.1038/s41467-017-00592-7}\BibitemShut {NoStop}%
		\bibitem [{\citenamefont {Shaw}\ \emph {et~al.}(2021)\citenamefont {Shaw},
			\citenamefont {Romo-Gonzalez}, \citenamefont {Lemos}, \citenamefont {King},
			\citenamefont {Bruhaug}, \citenamefont {Miller}, \citenamefont {Dorrer},
			\citenamefont {Kruschwitz}, \citenamefont {Waxer}, \citenamefont {Williams},
			\citenamefont {Ambat}, \citenamefont {McKie}, \citenamefont {Sinclair},
			\citenamefont {Mori}, \citenamefont {Joshi}, \citenamefont {Chen},
			\citenamefont {Palastro}, \citenamefont {Albert},\ and\ \citenamefont
			{Froula}}]{Shaw2021}%
		\BibitemOpen
		\bibfield  {author} {\bibinfo {author} {\bibfnamefont {J.~L.}\ \bibnamefont
				{Shaw}}, \bibinfo {author} {\bibfnamefont {M.~A.}\ \bibnamefont
				{Romo-Gonzalez}}, \bibinfo {author} {\bibfnamefont {N.}~\bibnamefont
				{Lemos}}, \bibinfo {author} {\bibfnamefont {P.~M.}\ \bibnamefont {King}},
			\bibinfo {author} {\bibfnamefont {G.}~\bibnamefont {Bruhaug}}, \bibinfo
			{author} {\bibfnamefont {K.~G.}\ \bibnamefont {Miller}}, \bibinfo {author}
			{\bibfnamefont {C.}~\bibnamefont {Dorrer}}, \bibinfo {author} {\bibfnamefont
				{B.}~\bibnamefont {Kruschwitz}}, \bibinfo {author} {\bibfnamefont
				{L.}~\bibnamefont {Waxer}}, \bibinfo {author} {\bibfnamefont {G.~J.}\
				\bibnamefont {Williams}}, \bibinfo {author} {\bibfnamefont {M.~V.}\
				\bibnamefont {Ambat}}, \bibinfo {author} {\bibfnamefont {M.~M.}\ \bibnamefont
				{McKie}}, \bibinfo {author} {\bibfnamefont {M.~D.}\ \bibnamefont {Sinclair}},
			\bibinfo {author} {\bibfnamefont {W.~B.}\ \bibnamefont {Mori}}, \bibinfo
			{author} {\bibfnamefont {C.}~\bibnamefont {Joshi}}, \bibinfo {author}
			{\bibfnamefont {H.}~\bibnamefont {Chen}}, \bibinfo {author} {\bibfnamefont
				{J.~P.}\ \bibnamefont {Palastro}}, \bibinfo {author} {\bibfnamefont
				{F.}~\bibnamefont {Albert}}, \ and\ \bibinfo {author} {\bibfnamefont
				{D.}~\bibnamefont {Froula}},\ }\bibfield  {title} {\enquote {\bibinfo {title}
				{Microcoulomb $(0.7 \pm \frac{0.4}{0.2} \mu \mathrm{C})$ laser plasma
					accelerator on omega ep},}\ }\href {\doibase
			https://doi.org/10.1038/s41598-021-86523-5} {\bibfield  {journal} {\bibinfo
				{journal} {Scientific Reports}\ }\textbf {\bibinfo {volume} {11}} (\bibinfo
			{year} {2021}),\ https://doi.org/10.1038/s41598-021-86523-5}\BibitemShut
		{NoStop}%
		\bibitem [{\citenamefont {Weingartner}\ \emph {et~al.}(2012)\citenamefont
			{Weingartner}, \citenamefont {Raith}, \citenamefont {Popp}, \citenamefont
			{Chou}, \citenamefont {Wenz}, \citenamefont {Khrennikov}, \citenamefont
			{Heigoldt}, \citenamefont {Maier}, \citenamefont {Kajumba}, \citenamefont
			{Fuchs}, \citenamefont {Zeitler}, \citenamefont {Krausz}, \citenamefont
			{Karsch},\ and\ \citenamefont {Gr\"uner}}]{Weingartner2012}%
		\BibitemOpen
		\bibfield  {author} {\bibinfo {author} {\bibfnamefont {R.}~\bibnamefont
				{Weingartner}}, \bibinfo {author} {\bibfnamefont {S.}~\bibnamefont {Raith}},
			\bibinfo {author} {\bibfnamefont {A.}~\bibnamefont {Popp}}, \bibinfo {author}
			{\bibfnamefont {S.}~\bibnamefont {Chou}}, \bibinfo {author} {\bibfnamefont
				{J.}~\bibnamefont {Wenz}}, \bibinfo {author} {\bibfnamefont {K.}~\bibnamefont
				{Khrennikov}}, \bibinfo {author} {\bibfnamefont {M.}~\bibnamefont
				{Heigoldt}}, \bibinfo {author} {\bibfnamefont {A.~R.}\ \bibnamefont {Maier}},
			\bibinfo {author} {\bibfnamefont {N.}~\bibnamefont {Kajumba}}, \bibinfo
			{author} {\bibfnamefont {M.}~\bibnamefont {Fuchs}}, \bibinfo {author}
			{\bibfnamefont {B.}~\bibnamefont {Zeitler}}, \bibinfo {author} {\bibfnamefont
				{F.}~\bibnamefont {Krausz}}, \bibinfo {author} {\bibfnamefont
				{S.}~\bibnamefont {Karsch}}, \ and\ \bibinfo {author} {\bibfnamefont
				{F.}~\bibnamefont {Gr\"uner}},\ }\bibfield  {title} {\enquote {\bibinfo
				{title} {Ultralow emittance electron beams from a laser-wakefield
					accelerator},}\ }\href {\doibase 10.1103/PhysRevSTAB.15.111302} {\bibfield
			{journal} {\bibinfo  {journal} {Phys. Rev. ST Accel. Beams}\ }\textbf
			{\bibinfo {volume} {15}},\ \bibinfo {pages} {111302} (\bibinfo {year}
			{2012})}\BibitemShut {NoStop}%
		\bibitem [{\citenamefont {Mangles}\ \emph {et~al.}(2006)\citenamefont
			{Mangles}, \citenamefont {Thomas}, \citenamefont {Kaluza}, \citenamefont
			{Lundh}, \citenamefont {Lindau}, \citenamefont {Persson}, \citenamefont
			{Tsung}, \citenamefont {Najmudin}, \citenamefont {Mori}, \citenamefont
			{Wahlstr\"om},\ and\ \citenamefont {Krushelnick}}]{Mangles2006}%
		\BibitemOpen
		\bibfield  {author} {\bibinfo {author} {\bibfnamefont {S.~P.~D.}\
				\bibnamefont {Mangles}}, \bibinfo {author} {\bibfnamefont {A.~G.~R.}\
				\bibnamefont {Thomas}}, \bibinfo {author} {\bibfnamefont {M.~C.}\
				\bibnamefont {Kaluza}}, \bibinfo {author} {\bibfnamefont {O.}~\bibnamefont
				{Lundh}}, \bibinfo {author} {\bibfnamefont {F.}~\bibnamefont {Lindau}},
			\bibinfo {author} {\bibfnamefont {A.}~\bibnamefont {Persson}}, \bibinfo
			{author} {\bibfnamefont {F.~S.}\ \bibnamefont {Tsung}}, \bibinfo {author}
			{\bibfnamefont {Z.}~\bibnamefont {Najmudin}}, \bibinfo {author}
			{\bibfnamefont {W.~B.}\ \bibnamefont {Mori}}, \bibinfo {author}
			{\bibfnamefont {C.-G.}\ \bibnamefont {Wahlstr\"om}}, \ and\ \bibinfo {author}
			{\bibfnamefont {K.}~\bibnamefont {Krushelnick}},\ }\bibfield  {title}
		{\enquote {\bibinfo {title} {Laser-wakefield acceleration of monoenergetic
					electron beams in the first plasma-wave period},}\ }\href {\doibase
			10.1103/PhysRevLett.96.215001} {\bibfield  {journal} {\bibinfo  {journal}
				{Phys. Rev. Lett.}\ }\textbf {\bibinfo {volume} {96}},\ \bibinfo {pages}
			{215001} (\bibinfo {year} {2006})}\BibitemShut {NoStop}%
		\bibitem [{\citenamefont {Lundh}\ \emph {et~al.}(2011)\citenamefont {Lundh},
			\citenamefont {Lim}, \citenamefont {Rechatin}, \citenamefont {Ammoura},
			\citenamefont {Ben-Isma\"{i}l}, \citenamefont {Davoine}, \citenamefont
			{Gallot}, \citenamefont {Goddet}, \citenamefont {Lefebvre}, \citenamefont
			{Malka},\ and\ \citenamefont {Faure}}]{Lundh2011}%
		\BibitemOpen
		\bibfield  {author} {\bibinfo {author} {\bibfnamefont {O.}~\bibnamefont
				{Lundh}}, \bibinfo {author} {\bibfnamefont {J.}~\bibnamefont {Lim}}, \bibinfo
			{author} {\bibfnamefont {C.}~\bibnamefont {Rechatin}}, \bibinfo {author}
			{\bibfnamefont {L.}~\bibnamefont {Ammoura}}, \bibinfo {author} {\bibfnamefont
				{A.}~\bibnamefont {Ben-Isma\"{i}l}}, \bibinfo {author} {\bibfnamefont
				{X.}~\bibnamefont {Davoine}}, \bibinfo {author} {\bibfnamefont
				{G.}~\bibnamefont {Gallot}}, \bibinfo {author} {\bibfnamefont {J.-P.}\
				\bibnamefont {Goddet}}, \bibinfo {author} {\bibfnamefont {E.}~\bibnamefont
				{Lefebvre}}, \bibinfo {author} {\bibfnamefont {V.}~\bibnamefont {Malka}}, \
			and\ \bibinfo {author} {\bibfnamefont {J.}~\bibnamefont {Faure}},\ }\bibfield
		{title} {\enquote {\bibinfo {title} {Few femtosecond, few kiloampere
					electron bunch produced by a laser–plasma accelerator},}\ }\href {\doibase
			https://doi.org/10.1038/nphys1872} {\bibfield  {journal} {\bibinfo  {journal}
				{Nature Phys}\ }\textbf {\bibinfo {volume} {7}},\ \bibinfo {pages} {219--222}
			(\bibinfo {year} {2011})}\BibitemShut {NoStop}%
		\bibitem [{\citenamefont {Albert}\ \emph {et~al.}(2021)\citenamefont {Albert},
			\citenamefont {Couprie}, \citenamefont {Debus}, \citenamefont {Downder},
			\citenamefont {Faure}, \citenamefont {Flacco}, \citenamefont {Gizzi},
			\citenamefont {Grismayer}, \citenamefont {Huebl}, \citenamefont {Joshi},
			\citenamefont {Labat}, \citenamefont {Leemans}, \citenamefont {Maier},
			\citenamefont {Mangles}, \citenamefont {Mason}, \citenamefont {Mathieu},
			\citenamefont {Muggli}, \citenamefont {Nishiuchi}, \citenamefont {Osterhoff},
			\citenamefont {Rajeev}, \citenamefont {Schramm}, \citenamefont {Schreiber},
			\citenamefont {Thomas}, \citenamefont {Vay}, \citenamefont {Vranic},\ and\
			\citenamefont {Zeil}}]{Albert2020}%
		\BibitemOpen
		\bibfield  {author} {\bibinfo {author} {\bibfnamefont {F.}~\bibnamefont
				{Albert}}, \bibinfo {author} {\bibfnamefont {M.~E.}\ \bibnamefont {Couprie}},
			\bibinfo {author} {\bibfnamefont {A.}~\bibnamefont {Debus}}, \bibinfo
			{author} {\bibfnamefont {M.~C.}\ \bibnamefont {Downder}}, \bibinfo {author}
			{\bibfnamefont {J.}~\bibnamefont {Faure}}, \bibinfo {author} {\bibfnamefont
				{A.}~\bibnamefont {Flacco}}, \bibinfo {author} {\bibfnamefont {L.~A.}\
				\bibnamefont {Gizzi}}, \bibinfo {author} {\bibfnamefont {T.}~\bibnamefont
				{Grismayer}}, \bibinfo {author} {\bibfnamefont {A.}~\bibnamefont {Huebl}},
			\bibinfo {author} {\bibfnamefont {C.}~\bibnamefont {Joshi}}, \bibinfo
			{author} {\bibfnamefont {M.}~\bibnamefont {Labat}}, \bibinfo {author}
			{\bibfnamefont {W.~P.}\ \bibnamefont {Leemans}}, \bibinfo {author}
			{\bibfnamefont {A.~R.}\ \bibnamefont {Maier}}, \bibinfo {author}
			{\bibfnamefont {S.~P.~D.}\ \bibnamefont {Mangles}}, \bibinfo {author}
			{\bibfnamefont {P.}~\bibnamefont {Mason}}, \bibinfo {author} {\bibfnamefont
				{F.}~\bibnamefont {Mathieu}}, \bibinfo {author} {\bibfnamefont
				{P.}~\bibnamefont {Muggli}}, \bibinfo {author} {\bibfnamefont
				{M.}~\bibnamefont {Nishiuchi}}, \bibinfo {author} {\bibfnamefont
				{J.}~\bibnamefont {Osterhoff}}, \bibinfo {author} {\bibfnamefont {P.~P.}\
				\bibnamefont {Rajeev}}, \bibinfo {author} {\bibfnamefont {U.}~\bibnamefont
				{Schramm}}, \bibinfo {author} {\bibfnamefont {J.}~\bibnamefont {Schreiber}},
			\bibinfo {author} {\bibfnamefont {A.~G.~R.}\ \bibnamefont {Thomas}}, \bibinfo
			{author} {\bibfnamefont {J.-L.}\ \bibnamefont {Vay}}, \bibinfo {author}
			{\bibfnamefont {M.}~\bibnamefont {Vranic}}, \ and\ \bibinfo {author}
			{\bibfnamefont {K.}~\bibnamefont {Zeil}},\ }\bibfield  {title} {\enquote
			{\bibinfo {title} {2020 roadmap on plasma accelerators},}\ }\href@noop {}
		{\bibfield  {journal} {\bibinfo  {journal} {New Journal of Physics}\ }\textbf
			{\bibinfo {volume} {23}},\ \bibinfo {pages} {031101} (\bibinfo {year}
			{2021})}\BibitemShut {NoStop}%
		\bibitem [{\citenamefont {Maier}\ \emph {et~al.}(2020)\citenamefont {Maier},
			\citenamefont {Delbos}, \citenamefont {Eichner}, \citenamefont {H\"ubner},
			\citenamefont {Jalas}, \citenamefont {Jeppe}, \citenamefont {Jolly},
			\citenamefont {Kirchen}, \citenamefont {Leroux}, \citenamefont {Messner},
			\citenamefont {Schnepp}, \citenamefont {Trunk}, \citenamefont {Walker},
			\citenamefont {Werle},\ and\ \citenamefont {Winkler}}]{Maier2020}%
		\BibitemOpen
		\bibfield  {author} {\bibinfo {author} {\bibfnamefont {A.~R.}\ \bibnamefont
				{Maier}}, \bibinfo {author} {\bibfnamefont {N.~M.}\ \bibnamefont {Delbos}},
			\bibinfo {author} {\bibfnamefont {T.}~\bibnamefont {Eichner}}, \bibinfo
			{author} {\bibfnamefont {L.}~\bibnamefont {H\"ubner}}, \bibinfo {author}
			{\bibfnamefont {S.}~\bibnamefont {Jalas}}, \bibinfo {author} {\bibfnamefont
				{L.}~\bibnamefont {Jeppe}}, \bibinfo {author} {\bibfnamefont {S.~W.}\
				\bibnamefont {Jolly}}, \bibinfo {author} {\bibfnamefont {M.}~\bibnamefont
				{Kirchen}}, \bibinfo {author} {\bibfnamefont {V.}~\bibnamefont {Leroux}},
			\bibinfo {author} {\bibfnamefont {P.}~\bibnamefont {Messner}}, \bibinfo
			{author} {\bibfnamefont {M.}~\bibnamefont {Schnepp}}, \bibinfo {author}
			{\bibfnamefont {M.}~\bibnamefont {Trunk}}, \bibinfo {author} {\bibfnamefont
				{P.~A.}\ \bibnamefont {Walker}}, \bibinfo {author} {\bibfnamefont
				{C.}~\bibnamefont {Werle}}, \ and\ \bibinfo {author} {\bibfnamefont
				{P.}~\bibnamefont {Winkler}},\ }\bibfield  {title} {\enquote {\bibinfo
				{title} {Decoding sources of energy variability in a laser-plasma
					accelerator},}\ }\href {\doibase 10.1103/PhysRevX.10.031039} {\bibfield
			{journal} {\bibinfo  {journal} {Phys. Rev. X}\ }\textbf {\bibinfo {volume}
				{10}},\ \bibinfo {pages} {031039} (\bibinfo {year} {2020})}\BibitemShut
		{NoStop}%
		\bibitem [{\citenamefont {Dann}\ \emph {et~al.}(2019)\citenamefont {Dann},
			\citenamefont {Baird}, \citenamefont {Bourgeois}, \citenamefont {Chekhlov},
			\citenamefont {Eardley}, \citenamefont {Gregory}, \citenamefont {Gruse},
			\citenamefont {Hah}, \citenamefont {Hazra}, \citenamefont {Hawkes},
			\citenamefont {Hooker}, \citenamefont {Krushelnick}, \citenamefont {Mangles},
			\citenamefont {Marshall}, \citenamefont {Murphy}, \citenamefont {Najmudin},
			\citenamefont {Nees}, \citenamefont {Osterhoff}, \citenamefont {Parry},
			\citenamefont {Pourmoussavi}, \citenamefont {Rahul}, \citenamefont {Rajeev},
			\citenamefont {Rozario}, \citenamefont {Scott}, \citenamefont {Smith},
			\citenamefont {Springate}, \citenamefont {Tang}, \citenamefont {Tata},
			\citenamefont {Thomas}, \citenamefont {Thornton}, \citenamefont {Symes},\
			and\ \citenamefont {Streeter}}]{Dann2019}%
		\BibitemOpen
		\bibfield  {author} {\bibinfo {author} {\bibfnamefont {S.~J.~D.}\
				\bibnamefont {Dann}}, \bibinfo {author} {\bibfnamefont {C.~D.}\ \bibnamefont
				{Baird}}, \bibinfo {author} {\bibfnamefont {N.}~\bibnamefont {Bourgeois}},
			\bibinfo {author} {\bibfnamefont {O.}~\bibnamefont {Chekhlov}}, \bibinfo
			{author} {\bibfnamefont {S.}~\bibnamefont {Eardley}}, \bibinfo {author}
			{\bibfnamefont {C.~D.}\ \bibnamefont {Gregory}}, \bibinfo {author}
			{\bibfnamefont {J.-N.}\ \bibnamefont {Gruse}}, \bibinfo {author}
			{\bibfnamefont {J.}~\bibnamefont {Hah}}, \bibinfo {author} {\bibfnamefont
				{D.}~\bibnamefont {Hazra}}, \bibinfo {author} {\bibfnamefont {S.~J.}\
				\bibnamefont {Hawkes}}, \bibinfo {author} {\bibfnamefont {C.~J.}\
				\bibnamefont {Hooker}}, \bibinfo {author} {\bibfnamefont {K.}~\bibnamefont
				{Krushelnick}}, \bibinfo {author} {\bibfnamefont {S.~P.~D.}\ \bibnamefont
				{Mangles}}, \bibinfo {author} {\bibfnamefont {V.~A.}\ \bibnamefont
				{Marshall}}, \bibinfo {author} {\bibfnamefont {C.~D.}\ \bibnamefont
				{Murphy}}, \bibinfo {author} {\bibfnamefont {Z.}~\bibnamefont {Najmudin}},
			\bibinfo {author} {\bibfnamefont {J.~A.}\ \bibnamefont {Nees}}, \bibinfo
			{author} {\bibfnamefont {J.}~\bibnamefont {Osterhoff}}, \bibinfo {author}
			{\bibfnamefont {B.}~\bibnamefont {Parry}}, \bibinfo {author} {\bibfnamefont
				{P.}~\bibnamefont {Pourmoussavi}}, \bibinfo {author} {\bibfnamefont {S.~V.}\
				\bibnamefont {Rahul}}, \bibinfo {author} {\bibfnamefont {P.~P.}\ \bibnamefont
				{Rajeev}}, \bibinfo {author} {\bibfnamefont {S.}~\bibnamefont {Rozario}},
			\bibinfo {author} {\bibfnamefont {J.~D.~E.}\ \bibnamefont {Scott}}, \bibinfo
			{author} {\bibfnamefont {R.~A.}\ \bibnamefont {Smith}}, \bibinfo {author}
			{\bibfnamefont {E.}~\bibnamefont {Springate}}, \bibinfo {author}
			{\bibfnamefont {Y.}~\bibnamefont {Tang}}, \bibinfo {author} {\bibfnamefont
				{S.}~\bibnamefont {Tata}}, \bibinfo {author} {\bibfnamefont {A.~G.~R.}\
				\bibnamefont {Thomas}}, \bibinfo {author} {\bibfnamefont {C.}~\bibnamefont
				{Thornton}}, \bibinfo {author} {\bibfnamefont {D.~R.}\ \bibnamefont {Symes}},
			\ and\ \bibinfo {author} {\bibfnamefont {M.~J.~V.}\ \bibnamefont
				{Streeter}},\ }\bibfield  {title} {\enquote {\bibinfo {title} {Laser
					wakefield acceleration with active feedback at 5 hz},}\ }\href {\doibase
			10.1103/PhysRevAccelBeams.22.041303} {\bibfield  {journal} {\bibinfo
				{journal} {Phys. Rev. Accel. Beams}\ }\textbf {\bibinfo {volume} {22}},\
			\bibinfo {pages} {041303} (\bibinfo {year} {2019})}\BibitemShut {NoStop}%
		\bibitem [{\citenamefont {He}\ \emph {et~al.}(2015)\citenamefont {He},
			\citenamefont {Hou}, \citenamefont {Lebailly}, \citenamefont {Nees},
			\citenamefont {Krushelnick},\ and\ \citenamefont {Thomas}}]{He2015}%
		\BibitemOpen
		\bibfield  {author} {\bibinfo {author} {\bibfnamefont {Z.-H.}\ \bibnamefont
				{He}}, \bibinfo {author} {\bibfnamefont {B.}~\bibnamefont {Hou}}, \bibinfo
			{author} {\bibfnamefont {V.}~\bibnamefont {Lebailly}}, \bibinfo {author}
			{\bibfnamefont {J.~A.}\ \bibnamefont {Nees}}, \bibinfo {author}
			{\bibfnamefont {K.}~\bibnamefont {Krushelnick}}, \ and\ \bibinfo {author}
			{\bibfnamefont {A.~G.~R.}\ \bibnamefont {Thomas}},\ }\bibfield  {title}
		{\enquote {\bibinfo {title} {Coherent control of plasma dynamics},}\
		}\href@noop {} {\bibfield  {journal} {\bibinfo  {journal} {Nat.
					Commun.}\ }\textbf {\bibinfo {volume} {6}}, \bibinfo {pages} {7156} (\bibinfo {year}
			{2015})}\BibitemShut {NoStop}%
		\bibitem [{\citenamefont {Rovige}\ \emph {et~al.}(2020)\citenamefont {Rovige},
			\citenamefont {Huijts}, \citenamefont {Andriyash}, \citenamefont {Vernier},
			\citenamefont {Tomkus}, \citenamefont {Girdauskas}, \citenamefont
			{Raciukaitis}, \citenamefont {Dudutis}, \citenamefont {Stankevic},
			\citenamefont {Gecys}, \citenamefont {Ouille}, \citenamefont {Cheng},
			\citenamefont {Lopez-Martens},\ and\ \citenamefont {Faure}}]{Rovige2020}%
		\BibitemOpen
		\bibfield  {author} {\bibinfo {author} {\bibfnamefont {L.}~\bibnamefont
				{Rovige}}, \bibinfo {author} {\bibfnamefont {J.}~\bibnamefont {Huijts}},
			\bibinfo {author} {\bibfnamefont {I.}~\bibnamefont {Andriyash}}, \bibinfo
			{author} {\bibfnamefont {A.}~\bibnamefont {Vernier}}, \bibinfo {author}
			{\bibfnamefont {V.}~\bibnamefont {Tomkus}}, \bibinfo {author} {\bibfnamefont
				{V.}~\bibnamefont {Girdauskas}}, \bibinfo {author} {\bibfnamefont
				{G.}~\bibnamefont {Raciukaitis}}, \bibinfo {author} {\bibfnamefont
				{J.}~\bibnamefont {Dudutis}}, \bibinfo {author} {\bibfnamefont
				{V.}~\bibnamefont {Stankevic}}, \bibinfo {author} {\bibfnamefont
				{P.}~\bibnamefont {Gecys}}, \bibinfo {author} {\bibfnamefont
				{M.}~\bibnamefont {Ouille}}, \bibinfo {author} {\bibfnamefont
				{Z.}~\bibnamefont {Cheng}}, \bibinfo {author} {\bibfnamefont
				{R.}~\bibnamefont {Lopez-Martens}}, \ and\ \bibinfo {author} {\bibfnamefont
				{J.}~\bibnamefont {Faure}},\ }\bibfield  {title} {\enquote {\bibinfo {title}
				{Demonstration of stable long-term operation of a kilohertz laser-plasma
					accelerator},}\ }\href {\doibase 10.1103/PhysRevAccelBeams.23.093401}
		{\bibfield  {journal} {\bibinfo  {journal} {Phys. Rev. Accel. Beams}\
			}\textbf {\bibinfo {volume} {23}},\ \bibinfo {pages} {093401} (\bibinfo
			{year} {2020})}\BibitemShut {NoStop}%
		\bibitem [{\citenamefont {Gustas}\ \emph {et~al.}(2018)\citenamefont {Gustas},
			\citenamefont {Gu\'enot}, \citenamefont {Vernier}, \citenamefont {Dutt},
			\citenamefont {B\"ohle}, \citenamefont {Lopez-Martens}, \citenamefont
			{Lifschitz},\ and\ \citenamefont {Faure}}]{Gustas2018}%
		\BibitemOpen
		\bibfield  {author} {\bibinfo {author} {\bibfnamefont {D.}~\bibnamefont
				{Gustas}}, \bibinfo {author} {\bibfnamefont {D.}~\bibnamefont {Gu\'enot}},
			\bibinfo {author} {\bibfnamefont {A.}~\bibnamefont {Vernier}}, \bibinfo
			{author} {\bibfnamefont {S.}~\bibnamefont {Dutt}}, \bibinfo {author}
			{\bibfnamefont {F.}~\bibnamefont {B\"ohle}}, \bibinfo {author} {\bibfnamefont
				{R.}~\bibnamefont {Lopez-Martens}}, \bibinfo {author} {\bibfnamefont
				{A.}~\bibnamefont {Lifschitz}}, \ and\ \bibinfo {author} {\bibfnamefont
				{J.}~\bibnamefont {Faure}},\ }\bibfield  {title} {\enquote {\bibinfo {title}
				{High-charge relativistic electron bunches from a khz laser-plasma
					accelerator},}\ }\href {\doibase 10.1103/PhysRevAccelBeams.21.013401}
		{\bibfield  {journal} {\bibinfo  {journal} {Phys. Rev. Accel. Beams}\
			}\textbf {\bibinfo {volume} {21}},\ \bibinfo {pages} {013401} (\bibinfo
			{year} {2018})}\BibitemShut {NoStop}%
		\bibitem [{\citenamefont {Sprangle}, \citenamefont {Tang},\ and\ \citenamefont
			{Esarey}(1987)}]{Sprangle1987}%
		\BibitemOpen
		\bibfield  {author} {\bibinfo {author} {\bibfnamefont {P.}~\bibnamefont
				{Sprangle}}, \bibinfo {author} {\bibfnamefont {C.-M.}\ \bibnamefont {Tang}},
			\ and\ \bibinfo {author} {\bibfnamefont {E.}~\bibnamefont {Esarey}},\
		}\bibfield  {title} {\enquote {\bibinfo {title} {Relativistic self-focusing
					of short-pulse radiation beams in plasmas},}\ }\href {\doibase
			10.1109/TPS.1987.4316677} {\bibfield  {journal} {\bibinfo  {journal} {IEEE
					Transactions on Plasma Science}\ }\textbf {\bibinfo {volume} {15}},\ \bibinfo
			{pages} {145--153} (\bibinfo {year} {1987})}\BibitemShut {NoStop}%
		\bibitem [{\citenamefont {Kokurewicz}\ \emph {et~al.}(2019)\citenamefont
			{Kokurewicz}, \citenamefont {Brunetti}, \citenamefont {Welsh}, \citenamefont
			{Wiggins}, \citenamefont {Boyd}, \citenamefont {Sorensen}, \citenamefont
			{Chalmers}, \citenamefont {Schettino}, \citenamefont {Subiel}, \citenamefont
			{DesRosiers},\ and\ \citenamefont {Jaroszynski}}]{Kokurewicz2019}%
		\BibitemOpen
		\bibfield  {author} {\bibinfo {author} {\bibfnamefont {K.}~\bibnamefont
				{Kokurewicz}}, \bibinfo {author} {\bibfnamefont {E.}~\bibnamefont
				{Brunetti}}, \bibinfo {author} {\bibfnamefont {G.~H.}\ \bibnamefont {Welsh}},
			\bibinfo {author} {\bibfnamefont {S.~M.}\ \bibnamefont {Wiggins}}, \bibinfo
			{author} {\bibfnamefont {M.}~\bibnamefont {Boyd}}, \bibinfo {author}
			{\bibfnamefont {A.}~\bibnamefont {Sorensen}}, \bibinfo {author}
			{\bibfnamefont {A.~J.}\ \bibnamefont {Chalmers}}, \bibinfo {author}
			{\bibfnamefont {G.}~\bibnamefont {Schettino}}, \bibinfo {author}
			{\bibfnamefont {A.}~\bibnamefont {Subiel}}, \bibinfo {author} {\bibfnamefont
				{C.}~\bibnamefont {DesRosiers}}, \ and\ \bibinfo {author} {\bibfnamefont
				{D.~A.}\ \bibnamefont {Jaroszynski}},\ }\bibfield  {title} {\enquote
			{\bibinfo {title} {Focused very high-energy electron beams as a novel
					radiotherapy modality for producing high-dose volumetric elements},}\
		}\href@noop {} {\bibfield  {journal} {\bibinfo  {journal} {Scientific
					Reports}\ }\textbf {\bibinfo {volume} {9}},\ \bibinfo {pages} {10837}
			(\bibinfo {year} {2019})}\BibitemShut {NoStop}%
		\bibitem [{\citenamefont {Palma}\ \emph {et~al.}(2015)\citenamefont {Palma},
			\citenamefont {Bazalova-Carter}, \citenamefont {Hardemark}, \citenamefont
			{Hynning}, \citenamefont {Qu}, \citenamefont {Loo},\ and\ \citenamefont
			{Maxim}}]{Palma2015}%
		\BibitemOpen
		\bibfield  {author} {\bibinfo {author} {\bibfnamefont {B.}~\bibnamefont
				{Palma}}, \bibinfo {author} {\bibfnamefont {M.}~\bibnamefont
				{Bazalova-Carter}}, \bibinfo {author} {\bibfnamefont {B.}~\bibnamefont
				{Hardemark}}, \bibinfo {author} {\bibfnamefont {E.}~\bibnamefont {Hynning}},
			\bibinfo {author} {\bibfnamefont {B.}~\bibnamefont {Qu}}, \bibinfo {author}
			{\bibfnamefont {B.}~\bibnamefont {Loo}}, \ and\ \bibinfo {author}
			{\bibfnamefont {P.}~\bibnamefont {Maxim}},\ }\bibfield  {title} {\enquote
			{\bibinfo {title} {Mo-fg-303-06: Evaluation of the performance of very
					high-energy electron (vhee) beams in radiotherapy: Five clinical cases},}\
		}\href {\doibase 10.1118/1.4925419} {\bibfield  {journal} {\bibinfo
				{journal} {Medical Physics}\ }\textbf {\bibinfo {volume} {42}},\ \bibinfo
			{pages} {3568--3568} (\bibinfo {year} {2015})},\ \Eprint
		{http://arxiv.org/abs/https://aapm.onlinelibrary.wiley.com/doi/pdf/10.1118/1.4925419}
		{https://aapm.onlinelibrary.wiley.com/doi/pdf/10.1118/1.4925419} \BibitemShut
		{NoStop}%
		\bibitem [{\citenamefont {Papiez}, \citenamefont {DesRosiers},\ and\
			\citenamefont {Moskvin}(2002)}]{Papiez2002}%
		\BibitemOpen
		\bibfield  {author} {\bibinfo {author} {\bibfnamefont {L.}~\bibnamefont
				{Papiez}}, \bibinfo {author} {\bibfnamefont {C.}~\bibnamefont {DesRosiers}},
			\ and\ \bibinfo {author} {\bibfnamefont {V.}~\bibnamefont {Moskvin}},\
		}\bibfield  {title} {\enquote {\bibinfo {title} {Very high energy electrons
					(50 – 250 mev) and radiation therapy},}\ }\href {\doibase
			10.1177/153303460200100202} {\bibfield  {journal} {\bibinfo  {journal}
				{Technology in Cancer Research \& Treatment}\ }\textbf {\bibinfo {volume}
				{1}},\ \bibinfo {pages} {105--110} (\bibinfo {year} {2002})},\ \bibinfo
		{note} {pMID: 12622516}\BibitemShut {NoStop}%
		\bibitem [{\citenamefont {DesRosiers}\ \emph {et~al.}(2008)\citenamefont
			{DesRosiers}, \citenamefont {Moskvin}, \citenamefont {Cao}, \citenamefont
			{Joshi},\ and\ \citenamefont {Langer}}]{DesRosiers2008}%
		\BibitemOpen
		\bibfield  {author} {\bibinfo {author} {\bibfnamefont {C.}~\bibnamefont
				{DesRosiers}}, \bibinfo {author} {\bibfnamefont {V.}~\bibnamefont {Moskvin}},
			\bibinfo {author} {\bibfnamefont {M.}~\bibnamefont {Cao}}, \bibinfo {author}
			{\bibfnamefont {C.~J.}\ \bibnamefont {Joshi}}, \ and\ \bibinfo {author}
			{\bibfnamefont {M.}~\bibnamefont {Langer}},\ }\bibfield  {title} {\enquote
			{\bibinfo {title} {{Laser-plasma generated very high energy electrons in
						radiation therapy of the prostate}},}\ }in\ \href {\doibase
			10.1117/12.761663} {\emph {\bibinfo {booktitle} {Commercial and Biomedical
					Applications of Ultrafast Lasers VIII}}},\ Vol.\ \bibinfo {volume} {6881},\
		\bibinfo {editor} {edited by\ \bibinfo {editor} {\bibfnamefont
				{J.}~\bibnamefont {Neev}}, \bibinfo {editor} {\bibfnamefont {S.}~\bibnamefont
				{Nolte}}, \bibinfo {editor} {\bibfnamefont {A.}~\bibnamefont {Heisterkamp}},
			\ and\ \bibinfo {editor} {\bibfnamefont {C.~B.}\ \bibnamefont {Schaffer}}},\
		\bibinfo {organization} {International Society for Optics and Photonics}\
		(\bibinfo  {publisher} {SPIE},\ \bibinfo {year} {2008})\ pp.\ \bibinfo
		{pages} {49 -- 62}\BibitemShut {NoStop}%
		\bibitem [{\citenamefont {Subiel}\ \emph {et~al.}(2014)\citenamefont {Subiel},
			\citenamefont {Moskvin}, \citenamefont {Welsh}, \citenamefont {Cipiccia},
			\citenamefont {Reboredo}, \citenamefont {Evans}, \citenamefont {Partridge},
			\citenamefont {DesRosiers}, \citenamefont {Anania}, \citenamefont {Cianchi},
			\citenamefont {Mostacci}, \citenamefont {Chiadroni}, \citenamefont
			{Di~Giovenale}, \citenamefont {Villa}, \citenamefont {Pompili}, \citenamefont
			{Ferrario}, \citenamefont {Belleveglia}, \citenamefont {Di~Pirro},
			\citenamefont {Gatti}, \citenamefont {Vaccarezza}, \citenamefont {Seitz},
			\citenamefont {Isaac}, \citenamefont {Brunetti}, \citenamefont {Wiggins},
			\citenamefont {Ersfeld}, \citenamefont {Islam}, \citenamefont {Mendonca},
			\citenamefont {Sorensen}, \citenamefont {Boyd},\ and\ \citenamefont
			{Jaroszynski}}]{Subiel2014}%
		\BibitemOpen
		\bibfield  {author} {\bibinfo {author} {\bibfnamefont {A.}~\bibnamefont
				{Subiel}}, \bibinfo {author} {\bibfnamefont {V.}~\bibnamefont {Moskvin}},
			\bibinfo {author} {\bibfnamefont {G.~H.}\ \bibnamefont {Welsh}}, \bibinfo
			{author} {\bibfnamefont {S.}~\bibnamefont {Cipiccia}}, \bibinfo {author}
			{\bibfnamefont {D.}~\bibnamefont {Reboredo}}, \bibinfo {author}
			{\bibfnamefont {P.}~\bibnamefont {Evans}}, \bibinfo {author} {\bibfnamefont
				{M.}~\bibnamefont {Partridge}}, \bibinfo {author} {\bibfnamefont
				{C.}~\bibnamefont {DesRosiers}}, \bibinfo {author} {\bibfnamefont {M.~P.}\
				\bibnamefont {Anania}}, \bibinfo {author} {\bibfnamefont {A.}~\bibnamefont
				{Cianchi}}, \bibinfo {author} {\bibfnamefont {A.}~\bibnamefont {Mostacci}},
			\bibinfo {author} {\bibfnamefont {E.}~\bibnamefont {Chiadroni}}, \bibinfo
			{author} {\bibfnamefont {D.}~\bibnamefont {Di~Giovenale}}, \bibinfo {author}
			{\bibfnamefont {F.}~\bibnamefont {Villa}}, \bibinfo {author} {\bibfnamefont
				{R.}~\bibnamefont {Pompili}}, \bibinfo {author} {\bibfnamefont
				{M.}~\bibnamefont {Ferrario}}, \bibinfo {author} {\bibfnamefont
				{M.}~\bibnamefont {Belleveglia}}, \bibinfo {author} {\bibfnamefont
				{G.}~\bibnamefont {Di~Pirro}}, \bibinfo {author} {\bibfnamefont
				{G.}~\bibnamefont {Gatti}}, \bibinfo {author} {\bibfnamefont
				{C.}~\bibnamefont {Vaccarezza}}, \bibinfo {author} {\bibfnamefont
				{B.}~\bibnamefont {Seitz}}, \bibinfo {author} {\bibfnamefont {R.~C.}\
				\bibnamefont {Isaac}}, \bibinfo {author} {\bibfnamefont {E.}~\bibnamefont
				{Brunetti}}, \bibinfo {author} {\bibfnamefont {S.~M.}\ \bibnamefont
				{Wiggins}}, \bibinfo {author} {\bibfnamefont {B.}~\bibnamefont {Ersfeld}},
			\bibinfo {author} {\bibfnamefont {M.~R.}\ \bibnamefont {Islam}}, \bibinfo
			{author} {\bibfnamefont {M.~S.}\ \bibnamefont {Mendonca}}, \bibinfo {author}
			{\bibfnamefont {A.}~\bibnamefont {Sorensen}}, \bibinfo {author}
			{\bibfnamefont {M.}~\bibnamefont {Boyd}}, \ and\ \bibinfo {author}
			{\bibfnamefont {D.~A.}\ \bibnamefont {Jaroszynski}},\ }\bibfield  {title}
		{\enquote {\bibinfo {title} {Dosimetry of very high energy electrons (vhee)
					for radiotherapy applications: using radiochromic film measurements and monte
					carlo simulations},}\ }\href@noop {} {\bibfield  {journal} {\bibinfo
				{journal} {Physics in Medicine and Biology}\ }\textbf {\bibinfo {volume}
				{59}},\ \bibinfo {pages} {5811} (\bibinfo {year} {2014})}\BibitemShut
		{NoStop}%
		\bibitem [{\citenamefont {Favaudon}\ \emph {et~al.}(2014)\citenamefont
			{Favaudon}, \citenamefont {Caplier}, \citenamefont {Monceau}, \citenamefont
			{Pouzoulet}, \citenamefont {Sayarath}, \citenamefont {Fouillade},
			\citenamefont {Poupon}, \citenamefont {Brito}, \citenamefont {Hup{\'e}},
			\citenamefont {Bourhis}, \citenamefont {Hall}, \citenamefont {Fontaine},\
			and\ \citenamefont {Vozenin}}]{Favaudon245ra93}%
		\BibitemOpen
		\bibfield  {author} {\bibinfo {author} {\bibfnamefont {V.}~\bibnamefont
				{Favaudon}}, \bibinfo {author} {\bibfnamefont {L.}~\bibnamefont {Caplier}},
			\bibinfo {author} {\bibfnamefont {V.}~\bibnamefont {Monceau}}, \bibinfo
			{author} {\bibfnamefont {F.}~\bibnamefont {Pouzoulet}}, \bibinfo {author}
			{\bibfnamefont {M.}~\bibnamefont {Sayarath}}, \bibinfo {author}
			{\bibfnamefont {C.}~\bibnamefont {Fouillade}}, \bibinfo {author}
			{\bibfnamefont {M.-F.}\ \bibnamefont {Poupon}}, \bibinfo {author}
			{\bibfnamefont {I.}~\bibnamefont {Brito}}, \bibinfo {author} {\bibfnamefont
				{P.}~\bibnamefont {Hup{\'e}}}, \bibinfo {author} {\bibfnamefont
				{J.}~\bibnamefont {Bourhis}}, \bibinfo {author} {\bibfnamefont
				{J.}~\bibnamefont {Hall}}, \bibinfo {author} {\bibfnamefont {J.-J.}\
				\bibnamefont {Fontaine}}, \ and\ \bibinfo {author} {\bibfnamefont {M.-C.}\
				\bibnamefont {Vozenin}},\ }\bibfield  {title} {\enquote {\bibinfo {title}
				{Ultrahigh dose-rate flash irradiation increases the differential response
					between normal and tumor tissue in mice},}\ }\href {\doibase
			10.1126/scitranslmed.3008973} {\bibfield  {journal} {\bibinfo  {journal}
				{Science Translational Medicine}\ }\textbf {\bibinfo {volume} {6}},\ \bibinfo
			{pages} {245ra93} (\bibinfo {year} {2014})}\ \BibitemShut
		{NoStop}%
		\bibitem [{\citenamefont {Durante}, \citenamefont {Br{\"a}uer-Krisch},\ and\
			\citenamefont {Hill}(2018)}]{Durante2018}%
		\BibitemOpen
		\bibfield  {author} {\bibinfo {author} {\bibfnamefont {M.}~\bibnamefont
				{Durante}}, \bibinfo {author} {\bibfnamefont {E.}~\bibnamefont
				{Br{\"a}uer-Krisch}}, \ and\ \bibinfo {author} {\bibfnamefont
				{M.}~\bibnamefont {Hill}},\ }\bibfield  {title} {\enquote {\bibinfo {title}
				{Faster and safer? flash ultra-high dose rate in radiotherapy},}\ }\href
		{\doibase 10.1259/bjr.20170628} {\bibfield  {journal} {\bibinfo  {journal}
				{Br J Radiol.}\ }\textbf {\bibinfo {volume} {91}},\ \bibinfo {pages}
			{20170628} (\bibinfo {year} {2018})}\BibitemShut {NoStop}%
		\bibitem [{\citenamefont {Staufer}\ \emph {et~al.}(2019)\citenamefont
			{Staufer}, \citenamefont {Bohlen}, \citenamefont {Poder}, \citenamefont
			{Brümmer}, \citenamefont {Blumendorf}, \citenamefont {Schmutzler},
			\citenamefont {Meisel}, \citenamefont {Osterhoff},\ and\ \citenamefont
			{Grüner}}]{Staufer2019}%
		\BibitemOpen
		\bibfield  {author} {\bibinfo {author} {\bibfnamefont {T.}~\bibnamefont
				{Staufer}}, \bibinfo {author} {\bibfnamefont {S.}~\bibnamefont {Bohlen}},
			\bibinfo {author} {\bibfnamefont {K.}~\bibnamefont {Poder}}, \bibinfo
			{author} {\bibfnamefont {T.}~\bibnamefont {Brümmer}}, \bibinfo {author}
			{\bibfnamefont {F.}~\bibnamefont {Blumendorf}}, \bibinfo {author}
			{\bibfnamefont {O.}~\bibnamefont {Schmutzler}}, \bibinfo {author}
			{\bibfnamefont {M.}~\bibnamefont {Meisel}}, \bibinfo {author} {\bibfnamefont
				{J.}~\bibnamefont {Osterhoff}}, \ and\ \bibinfo {author} {\bibfnamefont
				{F.}~\bibnamefont {Grüner}},\ }\bibfield  {title} {\enquote {\bibinfo
				{title} {{Development of a laser-wakefield Thomson x-ray source for x-ray
						fluorescence imaging}},}\ }in\ \href {\doibase 10.1117/12.2520685} {\emph
			{\bibinfo {booktitle} {Laser Acceleration of Electrons, Protons, and Ions
					V}}},\ Vol.\ \bibinfo {volume} {11037},\ \bibinfo {editor} {edited by\
			\bibinfo {editor} {\bibfnamefont {E.}~\bibnamefont {Esarey}}, \bibinfo
			{editor} {\bibfnamefont {C.~B.}\ \bibnamefont {Schroeder}}, \ and\ \bibinfo
			{editor} {\bibfnamefont {J.}~\bibnamefont {Schreiber}}},\ \bibinfo
		{organization} {International Society for Optics and Photonics}\ (\bibinfo
		{publisher} {SPIE},\ \bibinfo {year} {2019})\ pp.\ \bibinfo {pages} {38 --
			47}\BibitemShut {NoStop}%
		\bibitem [{\citenamefont {Cheong}\ \emph {et~al.}(2010)\citenamefont {Cheong},
			\citenamefont {Jones}, \citenamefont {Siddiqi}, \citenamefont {Liu},
			\citenamefont {Manohar},\ and\ \citenamefont {Cho}}]{Cheong2010}%
		\BibitemOpen
		\bibfield  {author} {\bibinfo {author} {\bibfnamefont {S.-K.}\ \bibnamefont
				{Cheong}}, \bibinfo {author} {\bibfnamefont {B.~L.}\ \bibnamefont {Jones}},
			\bibinfo {author} {\bibfnamefont {A.~K.}\ \bibnamefont {Siddiqi}}, \bibinfo
			{author} {\bibfnamefont {F.}~\bibnamefont {Liu}}, \bibinfo {author}
			{\bibfnamefont {N.}~\bibnamefont {Manohar}}, \ and\ \bibinfo {author}
			{\bibfnamefont {S.~H.}\ \bibnamefont {Cho}},\ }\bibfield  {title} {\enquote
			{\bibinfo {title} {X-ray fluorescence computed tomography (xfct) imaging of
					gold nanoparticle-loaded objects using 110 kvp x-rays},}\ }\href@noop {}
		{\bibfield  {journal} {\bibinfo  {journal} {Physics in Medicine and Biology}\
			}\textbf {\bibinfo {volume} {55}},\ \bibinfo {pages} {647} (\bibinfo {year}
			{2010})}\BibitemShut {NoStop}%
		\bibitem [{\citenamefont {Manohar}\ \emph {et~al.}(2016)\citenamefont
			{Manohar}, \citenamefont {Reynoso}, \citenamefont {Diagaradjane},
			\citenamefont {Krishnan},\ and\ \citenamefont {Cho}}]{Manohar2016}%
		\BibitemOpen
		\bibfield  {author} {\bibinfo {author} {\bibfnamefont {N.}~\bibnamefont
				{Manohar}}, \bibinfo {author} {\bibfnamefont {F.~J.}\ \bibnamefont
				{Reynoso}}, \bibinfo {author} {\bibfnamefont {P.}~\bibnamefont
				{Diagaradjane}}, \bibinfo {author} {\bibfnamefont {S.}~\bibnamefont
				{Krishnan}}, \ and\ \bibinfo {author} {\bibfnamefont {S.~H.}\ \bibnamefont
				{Cho}},\ }\bibfield  {title} {\enquote {\bibinfo {title} {Quantitative
					imaging of gold nanoparticle distribution in a tumor-bearing mouse using
					benchtop x-ray fluorescence computed tomography},}\ }\href@noop {} {\bibfield
			{journal} {\bibinfo  {journal} {Scientific Reports}\ }\textbf {\bibinfo
				{volume} {6}},\ \bibinfo {pages} {22079} (\bibinfo {year}
			{2016})}\BibitemShut {NoStop}%
		\bibitem [{\citenamefont {Grüner}\ \emph {et~al.}(2018)\citenamefont
			{Grüner}, \citenamefont {Blumendorf}, \citenamefont {Schmutzler},
			\citenamefont {Staufer}, \citenamefont {Bradbury}, \citenamefont {Wiesner},
			\citenamefont {Rosentreter}, \citenamefont {Loers}, \citenamefont {Lutz},
			\citenamefont {Richter}, \citenamefont {Fischer}, \citenamefont {Schulz},
			\citenamefont {Steiner}, \citenamefont {Warmer}, \citenamefont {Burkhardt},
			\citenamefont {Meents}, \citenamefont {Kupinski},\ and\ \citenamefont
			{Hoeschen}}]{Gruner2018}%
		\BibitemOpen
		\bibfield  {author} {\bibinfo {author} {\bibfnamefont {F.}~\bibnamefont
				{Grüner}}, \bibinfo {author} {\bibfnamefont {F.}~\bibnamefont {Blumendorf}},
			\bibinfo {author} {\bibfnamefont {O.}~\bibnamefont {Schmutzler}}, \bibinfo
			{author} {\bibfnamefont {T.}~\bibnamefont {Staufer}}, \bibinfo {author}
			{\bibfnamefont {M.}~\bibnamefont {Bradbury}}, \bibinfo {author}
			{\bibfnamefont {U.}~\bibnamefont {Wiesner}}, \bibinfo {author} {\bibfnamefont
				{T.}~\bibnamefont {Rosentreter}}, \bibinfo {author} {\bibfnamefont
				{G.}~\bibnamefont {Loers}}, \bibinfo {author} {\bibfnamefont
				{D.}~\bibnamefont {Lutz}}, \bibinfo {author} {\bibfnamefont {B.}~\bibnamefont
				{Richter}}, \bibinfo {author} {\bibfnamefont {M.}~\bibnamefont {Fischer}},
			\bibinfo {author} {\bibfnamefont {F.}~\bibnamefont {Schulz}}, \bibinfo
			{author} {\bibfnamefont {S.}~\bibnamefont {Steiner}}, \bibinfo {author}
			{\bibfnamefont {M.}~\bibnamefont {Warmer}}, \bibinfo {author} {\bibfnamefont
				{A.}~\bibnamefont {Burkhardt}}, \bibinfo {author} {\bibfnamefont
				{A.}~\bibnamefont {Meents}}, \bibinfo {author} {\bibfnamefont
				{M.}~\bibnamefont {Kupinski}}, \ and\ \bibinfo {author} {\bibfnamefont
				{C.}~\bibnamefont {Hoeschen}},\ }\bibfield  {title} {\enquote {\bibinfo
				{title} {Localising functionalised gold-nanoparticles in murine spinal cords
					by x-ray fluorescence imaging and background-reduction through spatial
					filtering for human-sized objects},}\ }\href@noop {} {\bibfield  {journal}
			{\bibinfo  {journal} {Scientific Reports}\ }\textbf {\bibinfo {volume} {8}},\
			\bibinfo {pages} {16561} (\bibinfo {year} {2018})}\BibitemShut {NoStop}%
		\bibitem [{\citenamefont {Schmutzler}\ \emph {et~al.}(2021)\citenamefont
			{Schmutzler}, \citenamefont {Graf}, \citenamefont {Behm}, \citenamefont
			{Mansour}, \citenamefont {Blumendorf}, \citenamefont {Staufer}, \citenamefont
			{Körnig}, \citenamefont {Salah}, \citenamefont {Kang}, \citenamefont
			{Peters}, \citenamefont {Liu}, \citenamefont {Feliu}, \citenamefont {Parak},
			\citenamefont {Burkhardt}, \citenamefont {Gargioni}, \citenamefont {Gennis},
			\citenamefont {Chandralingam}, \citenamefont {Höeg}, \citenamefont {Maison},
			\citenamefont {Rothkamm}, \citenamefont {Schulz},\ and\ \citenamefont
			{Grüner}}]{Schmutzler2021}%
		\BibitemOpen
		\bibfield  {author} {\bibinfo {author} {\bibfnamefont {O.}~\bibnamefont
				{Schmutzler}}, \bibinfo {author} {\bibfnamefont {S.}~\bibnamefont {Graf}},
			\bibinfo {author} {\bibfnamefont {N.}~\bibnamefont {Behm}}, \bibinfo {author}
			{\bibfnamefont {W.~Y.}\ \bibnamefont {Mansour}}, \bibinfo {author}
			{\bibfnamefont {F.}~\bibnamefont {Blumendorf}}, \bibinfo {author}
			{\bibfnamefont {T.}~\bibnamefont {Staufer}}, \bibinfo {author} {\bibfnamefont
				{C.}~\bibnamefont {Körnig}}, \bibinfo {author} {\bibfnamefont
				{D.}~\bibnamefont {Salah}}, \bibinfo {author} {\bibfnamefont
				{Y.}~\bibnamefont {Kang}}, \bibinfo {author} {\bibfnamefont {J.~N.}\
				\bibnamefont {Peters}}, \bibinfo {author} {\bibfnamefont {Y.}~\bibnamefont
				{Liu}}, \bibinfo {author} {\bibfnamefont {N.}~\bibnamefont {Feliu}}, \bibinfo
			{author} {\bibfnamefont {W.~J.}\ \bibnamefont {Parak}}, \bibinfo {author}
			{\bibfnamefont {A.}~\bibnamefont {Burkhardt}}, \bibinfo {author}
			{\bibfnamefont {E.}~\bibnamefont {Gargioni}}, \bibinfo {author}
			{\bibfnamefont {S.}~\bibnamefont {Gennis}}, \bibinfo {author} {\bibfnamefont
				{S.}~\bibnamefont {Chandralingam}}, \bibinfo {author} {\bibfnamefont
				{F.}~\bibnamefont {Höeg}}, \bibinfo {author} {\bibfnamefont
				{W.}~\bibnamefont {Maison}}, \bibinfo {author} {\bibfnamefont
				{K.}~\bibnamefont {Rothkamm}}, \bibinfo {author} {\bibfnamefont
				{F.}~\bibnamefont {Schulz}}, \ and\ \bibinfo {author} {\bibfnamefont
				{F.}~\bibnamefont {Grüner}},\ }\bibfield  {title} {\enquote {\bibinfo
				{title} {X-ray fluorescence uptake measurement of functionalized gold
					nanoparticles in tumor cell microsamples},}\ }\href {\doibase
			10.3390/ijms22073691} {\bibfield  {journal} {\bibinfo  {journal}
				{International Journal of Molecular Sciences}\ }\textbf {\bibinfo {volume}
				{22}} (\bibinfo {year} {2021}),\ 10.3390/ijms22073691}\BibitemShut {NoStop}%
		\bibitem [{\citenamefont {Sanchez-Cano}\ \emph {et~al.}(2021)\citenamefont
			{Sanchez-Cano}, \citenamefont {Alvarez-Puebla}, \citenamefont {Abendroth},
			\citenamefont {Beck}, \citenamefont {Blick}, \citenamefont {Cao},
			\citenamefont {Caruso}, \citenamefont {Chakraborty}, \citenamefont {Chapman},
			\citenamefont {Chen}, \citenamefont {Cohen}, \citenamefont {Conceição},
			\citenamefont {Cormode}, \citenamefont {Cui}, \citenamefont {Dawson},
			\citenamefont {Falkenberg}, \citenamefont {Fan}, \citenamefont {Feliu},
			\citenamefont {Gao}, \citenamefont {Gargioni}, \citenamefont {Glüer},
			\citenamefont {Grüner}, \citenamefont {Hassan}, \citenamefont {Hu},
			\citenamefont {Huang}, \citenamefont {Huber}, \citenamefont {Huse},
			\citenamefont {Kang}, \citenamefont {Khademhosseini}, \citenamefont {Keller},
			\citenamefont {Körnig}, \citenamefont {Kotov}, \citenamefont {Koziej},
			\citenamefont {Liang}, \citenamefont {Liu}, \citenamefont {Liu},
			\citenamefont {Liu}, \citenamefont {Liu}, \citenamefont {Liz-Marzán},
			\citenamefont {Ma}, \citenamefont {Machicote}, \citenamefont {Maison},
			\citenamefont {Mancuso}, \citenamefont {Megahed}, \citenamefont {Nickel},
			\citenamefont {Otto}, \citenamefont {Palencia}, \citenamefont {Pascarelli},
			\citenamefont {Pearson}, \citenamefont {Peñate-Medina}, \citenamefont {Qi},
			\citenamefont {Rädler}, \citenamefont {Richardson}, \citenamefont
			{Rosenhahn}, \citenamefont {Rothkamm}, \citenamefont {Rübhausen},
			\citenamefont {Sanyal}, \citenamefont {Schaak}, \citenamefont {Schlemmer},
			\citenamefont {Schmidt}, \citenamefont {Schmutzler}, \citenamefont
			{Schotten}, \citenamefont {Schulz}, \citenamefont {Sood}, \citenamefont
			{Spiers}, \citenamefont {Staufer}, \citenamefont {Stemer}, \citenamefont
			{Stierle}, \citenamefont {Sun}, \citenamefont {Tsakanova}, \citenamefont
			{Weiss}, \citenamefont {Weller}, \citenamefont {Westermeier}, \citenamefont
			{Xu}, \citenamefont {Yan}, \citenamefont {Zeng}, \citenamefont {Zhao},
			\citenamefont {Zhao}, \citenamefont {Zhu}, \citenamefont {Zhu},\ and\
			\citenamefont {Parak}}]{Sanchez-Cano2021}%
		\BibitemOpen
		\bibfield  {author} {\bibinfo {author} {\bibfnamefont {C.}~\bibnamefont
				{Sanchez-Cano}}, \bibinfo {author} {\bibfnamefont {R.~A.}\ \bibnamefont
				{Alvarez-Puebla}}, \bibinfo {author} {\bibfnamefont {J.~M.}\ \bibnamefont
				{Abendroth}}, \bibinfo {author} {\bibfnamefont {T.}~\bibnamefont {Beck}},
			\bibinfo {author} {\bibfnamefont {R.}~\bibnamefont {Blick}}, \bibinfo
			{author} {\bibfnamefont {Y.}~\bibnamefont {Cao}}, \bibinfo {author}
			{\bibfnamefont {F.}~\bibnamefont {Caruso}}, \bibinfo {author} {\bibfnamefont
				{I.}~\bibnamefont {Chakraborty}}, \bibinfo {author} {\bibfnamefont {H.~N.}\
				\bibnamefont {Chapman}}, \bibinfo {author} {\bibfnamefont {C.}~\bibnamefont
				{Chen}}, \bibinfo {author} {\bibfnamefont {B.~E.}\ \bibnamefont {Cohen}},
			\bibinfo {author} {\bibfnamefont {A.~L.~C.}\ \bibnamefont {Conceição}},
			\bibinfo {author} {\bibfnamefont {D.~P.}\ \bibnamefont {Cormode}}, \bibinfo
			{author} {\bibfnamefont {D.}~\bibnamefont {Cui}}, \bibinfo {author}
			{\bibfnamefont {K.~A.}\ \bibnamefont {Dawson}}, \bibinfo {author}
			{\bibfnamefont {G.}~\bibnamefont {Falkenberg}}, \bibinfo {author}
			{\bibfnamefont {C.}~\bibnamefont {Fan}}, \bibinfo {author} {\bibfnamefont
				{N.}~\bibnamefont {Feliu}}, \bibinfo {author} {\bibfnamefont
				{M.}~\bibnamefont {Gao}}, \bibinfo {author} {\bibfnamefont {E.}~\bibnamefont
				{Gargioni}}, \bibinfo {author} {\bibfnamefont {C.-C.}\ \bibnamefont
				{Glüer}}, \bibinfo {author} {\bibfnamefont {F.}~\bibnamefont {Grüner}},
			\bibinfo {author} {\bibfnamefont {M.}~\bibnamefont {Hassan}}, \bibinfo
			{author} {\bibfnamefont {Y.}~\bibnamefont {Hu}}, \bibinfo {author}
			{\bibfnamefont {Y.}~\bibnamefont {Huang}}, \bibinfo {author} {\bibfnamefont
				{S.}~\bibnamefont {Huber}}, \bibinfo {author} {\bibfnamefont
				{N.}~\bibnamefont {Huse}}, \bibinfo {author} {\bibfnamefont {Y.}~\bibnamefont
				{Kang}}, \bibinfo {author} {\bibfnamefont {A.}~\bibnamefont
				{Khademhosseini}}, \bibinfo {author} {\bibfnamefont {T.~F.}\ \bibnamefont
				{Keller}}, \bibinfo {author} {\bibfnamefont {C.}~\bibnamefont {Körnig}},
			\bibinfo {author} {\bibfnamefont {N.~A.}\ \bibnamefont {Kotov}}, \bibinfo
			{author} {\bibfnamefont {D.}~\bibnamefont {Koziej}}, \bibinfo {author}
			{\bibfnamefont {X.-J.}\ \bibnamefont {Liang}}, \bibinfo {author}
			{\bibfnamefont {B.}~\bibnamefont {Liu}}, \bibinfo {author} {\bibfnamefont
				{S.}~\bibnamefont {Liu}}, \bibinfo {author} {\bibfnamefont {Y.}~\bibnamefont
				{Liu}}, \bibinfo {author} {\bibfnamefont {Z.}~\bibnamefont {Liu}}, \bibinfo
			{author} {\bibfnamefont {L.~M.}\ \bibnamefont {Liz-Marzán}}, \bibinfo
			{author} {\bibfnamefont {X.}~\bibnamefont {Ma}}, \bibinfo {author}
			{\bibfnamefont {A.}~\bibnamefont {Machicote}}, \bibinfo {author}
			{\bibfnamefont {W.}~\bibnamefont {Maison}}, \bibinfo {author} {\bibfnamefont
				{A.~P.}\ \bibnamefont {Mancuso}}, \bibinfo {author} {\bibfnamefont
				{S.}~\bibnamefont {Megahed}}, \bibinfo {author} {\bibfnamefont
				{B.}~\bibnamefont {Nickel}}, \bibinfo {author} {\bibfnamefont
				{F.}~\bibnamefont {Otto}}, \bibinfo {author} {\bibfnamefont {C.}~\bibnamefont
				{Palencia}}, \bibinfo {author} {\bibfnamefont {S.}~\bibnamefont
				{Pascarelli}}, \bibinfo {author} {\bibfnamefont {A.}~\bibnamefont {Pearson}},
			\bibinfo {author} {\bibfnamefont {O.}~\bibnamefont {Peñate-Medina}},
			\bibinfo {author} {\bibfnamefont {B.}~\bibnamefont {Qi}}, \bibinfo {author}
			{\bibfnamefont {J.}~\bibnamefont {Rädler}}, \bibinfo {author} {\bibfnamefont
				{J.~J.}\ \bibnamefont {Richardson}}, \bibinfo {author} {\bibfnamefont
				{A.}~\bibnamefont {Rosenhahn}}, \bibinfo {author} {\bibfnamefont
				{K.}~\bibnamefont {Rothkamm}}, \bibinfo {author} {\bibfnamefont
				{M.}~\bibnamefont {Rübhausen}}, \bibinfo {author} {\bibfnamefont {M.~K.}\
				\bibnamefont {Sanyal}}, \bibinfo {author} {\bibfnamefont {R.~E.}\
				\bibnamefont {Schaak}}, \bibinfo {author} {\bibfnamefont {H.-P.}\
				\bibnamefont {Schlemmer}}, \bibinfo {author} {\bibfnamefont {M.}~\bibnamefont
				{Schmidt}}, \bibinfo {author} {\bibfnamefont {O.}~\bibnamefont {Schmutzler}},
			\bibinfo {author} {\bibfnamefont {T.}~\bibnamefont {Schotten}}, \bibinfo
			{author} {\bibfnamefont {F.}~\bibnamefont {Schulz}}, \bibinfo {author}
			{\bibfnamefont {A.~K.}\ \bibnamefont {Sood}}, \bibinfo {author}
			{\bibfnamefont {K.~M.}\ \bibnamefont {Spiers}}, \bibinfo {author}
			{\bibfnamefont {T.}~\bibnamefont {Staufer}}, \bibinfo {author} {\bibfnamefont
				{D.~M.}\ \bibnamefont {Stemer}}, \bibinfo {author} {\bibfnamefont
				{A.}~\bibnamefont {Stierle}}, \bibinfo {author} {\bibfnamefont
				{X.}~\bibnamefont {Sun}}, \bibinfo {author} {\bibfnamefont {G.}~\bibnamefont
				{Tsakanova}}, \bibinfo {author} {\bibfnamefont {P.~S.}\ \bibnamefont
				{Weiss}}, \bibinfo {author} {\bibfnamefont {H.}~\bibnamefont {Weller}},
			\bibinfo {author} {\bibfnamefont {F.}~\bibnamefont {Westermeier}}, \bibinfo
			{author} {\bibfnamefont {M.}~\bibnamefont {Xu}}, \bibinfo {author}
			{\bibfnamefont {H.}~\bibnamefont {Yan}}, \bibinfo {author} {\bibfnamefont
				{Y.}~\bibnamefont {Zeng}}, \bibinfo {author} {\bibfnamefont {Y.}~\bibnamefont
				{Zhao}}, \bibinfo {author} {\bibfnamefont {Y.}~\bibnamefont {Zhao}}, \bibinfo
			{author} {\bibfnamefont {D.}~\bibnamefont {Zhu}}, \bibinfo {author}
			{\bibfnamefont {Y.}~\bibnamefont {Zhu}}, \ and\ \bibinfo {author}
			{\bibfnamefont {W.~J.}\ \bibnamefont {Parak}},\ }\bibfield  {title} {\enquote
			{\bibinfo {title} {X-ray-based techniques to study the nano–bio
					interface},}\ }\href {\doibase 10.1021/acsnano.0c09563} {\bibfield  {journal}
			{\bibinfo  {journal} {ACS Nano}\ }\textbf {\bibinfo {volume} {15}},\ \bibinfo
			{pages} {3754--3807} (\bibinfo {year} {2021})},\ \bibinfo {note} {pMID:
			33650433},\ \Eprint
		{http://arxiv.org/abs/https://doi.org/10.1021/acsnano.0c09563}
		{https://doi.org/10.1021/acsnano.0c09563} \BibitemShut {NoStop}%
		\bibitem [{\citenamefont {Kahl}\ \emph {et~al.}(2021)\citenamefont {Kahl},
			\citenamefont {Staufer}, \citenamefont {Körnig}, \citenamefont {Schmutzler},
			\citenamefont {Rothkamm},\ and\ \citenamefont {Grüner}}]{Kahl2021}%
		\BibitemOpen
		\bibfield  {author} {\bibinfo {author} {\bibfnamefont {H.}~\bibnamefont
				{Kahl}}, \bibinfo {author} {\bibfnamefont {T.}~\bibnamefont {Staufer}},
			\bibinfo {author} {\bibfnamefont {C.}~\bibnamefont {Körnig}}, \bibinfo
			{author} {\bibfnamefont {O.}~\bibnamefont {Schmutzler}}, \bibinfo {author}
			{\bibfnamefont {K.}~\bibnamefont {Rothkamm}}, \ and\ \bibinfo {author}
			{\bibfnamefont {F.}~\bibnamefont {Grüner}},\ }\bibfield  {title} {\enquote
			{\bibinfo {title} {Feasibility of monitoring tumor response by tracking
					nanoparticle-labelled t cells using x-ray fluorescence imaging—a numerical
					study},}\ }\href {\doibase 10.3390/ijms22168736} {\bibfield  {journal}
			{\bibinfo  {journal} {International Journal of Molecular Sciences}\ }\textbf
			{\bibinfo {volume} {22}} (\bibinfo {year} {2021}),\
			10.3390/ijms22168736}\BibitemShut {NoStop}%
		\bibitem [{\citenamefont {Cole}\ \emph
			{et~al.}(2018{\natexlab{b}})\citenamefont {Cole}, \citenamefont {Behm},
			\citenamefont {Gerstmayr}, \citenamefont {Blackburn}, \citenamefont {Wood},
			\citenamefont {Baird}, \citenamefont {Duff}, \citenamefont {Harvey},
			\citenamefont {Ilderton}, \citenamefont {Joglekar}, \citenamefont
			{Krushelnick}, \citenamefont {Kuschel}, \citenamefont {Marklund},
			\citenamefont {McKenna}, \citenamefont {Murphy}, \citenamefont {Poder},
			\citenamefont {Ridgers}, \citenamefont {Samarin}, \citenamefont {Sarri},
			\citenamefont {Symes}, \citenamefont {Thomas}, \citenamefont {Warwick},
			\citenamefont {Zepf}, \citenamefont {Najmudin},\ and\ \citenamefont
			{Mangles}}]{Cole2018}%
		\BibitemOpen
		\bibfield  {author} {\bibinfo {author} {\bibfnamefont {J.~M.}\ \bibnamefont
				{Cole}}, \bibinfo {author} {\bibfnamefont {K.~T.}\ \bibnamefont {Behm}},
			\bibinfo {author} {\bibfnamefont {E.}~\bibnamefont {Gerstmayr}}, \bibinfo
			{author} {\bibfnamefont {T.~G.}\ \bibnamefont {Blackburn}}, \bibinfo {author}
			{\bibfnamefont {J.~C.}\ \bibnamefont {Wood}}, \bibinfo {author}
			{\bibfnamefont {C.~D.}\ \bibnamefont {Baird}}, \bibinfo {author}
			{\bibfnamefont {M.~J.}\ \bibnamefont {Duff}}, \bibinfo {author}
			{\bibfnamefont {C.}~\bibnamefont {Harvey}}, \bibinfo {author} {\bibfnamefont
				{A.}~\bibnamefont {Ilderton}}, \bibinfo {author} {\bibfnamefont {A.~S.}\
				\bibnamefont {Joglekar}}, \bibinfo {author} {\bibfnamefont {K.}~\bibnamefont
				{Krushelnick}}, \bibinfo {author} {\bibfnamefont {S.}~\bibnamefont
				{Kuschel}}, \bibinfo {author} {\bibfnamefont {M.}~\bibnamefont {Marklund}},
			\bibinfo {author} {\bibfnamefont {P.}~\bibnamefont {McKenna}}, \bibinfo
			{author} {\bibfnamefont {C.~D.}\ \bibnamefont {Murphy}}, \bibinfo {author}
			{\bibfnamefont {K.}~\bibnamefont {Poder}}, \bibinfo {author} {\bibfnamefont
				{C.~P.}\ \bibnamefont {Ridgers}}, \bibinfo {author} {\bibfnamefont {G.~M.}\
				\bibnamefont {Samarin}}, \bibinfo {author} {\bibfnamefont {G.}~\bibnamefont
				{Sarri}}, \bibinfo {author} {\bibfnamefont {D.~R.}\ \bibnamefont {Symes}},
			\bibinfo {author} {\bibfnamefont {A.~G.~R.}\ \bibnamefont {Thomas}}, \bibinfo
			{author} {\bibfnamefont {J.}~\bibnamefont {Warwick}}, \bibinfo {author}
			{\bibfnamefont {M.}~\bibnamefont {Zepf}}, \bibinfo {author} {\bibfnamefont
				{Z.}~\bibnamefont {Najmudin}}, \ and\ \bibinfo {author} {\bibfnamefont
				{S.~P.~D.}\ \bibnamefont {Mangles}},\ }\bibfield  {title} {\enquote {\bibinfo
				{title} {Experimental evidence of radiation reaction in the collision of a
					high-intensity laser pulse with a laser-wakefield accelerated electron
					beam},}\ }\href {\doibase 10.1103/PhysRevX.8.011020} {\bibfield  {journal}
			{\bibinfo  {journal} {Phys. Rev. X}\ }\textbf {\bibinfo {volume} {8}},\
			\bibinfo {pages} {011020} (\bibinfo {year} {2018}{\natexlab{b}})}\BibitemShut
		{NoStop}%
		\bibitem [{\citenamefont {Poder}\ \emph {et~al.}(2018)\citenamefont {Poder},
			\citenamefont {Tamburini}, \citenamefont {Sarri}, \citenamefont {Di~Piazza},
			\citenamefont {Kuschel}, \citenamefont {Baird}, \citenamefont {Behm},
			\citenamefont {Bohlen}, \citenamefont {Cole}, \citenamefont {Corvan},
			\citenamefont {Duff}, \citenamefont {Gerstmayr}, \citenamefont {Keitel},
			\citenamefont {Krushelnick}, \citenamefont {Mangles}, \citenamefont
			{McKenna}, \citenamefont {Murphy}, \citenamefont {Najmudin}, \citenamefont
			{Ridgers}, \citenamefont {Samarin}, \citenamefont {Symes}, \citenamefont
			{Thomas}, \citenamefont {Warwick},\ and\ \citenamefont {Zepf}}]{Poder2018}%
		\BibitemOpen
		\bibfield  {author} {\bibinfo {author} {\bibfnamefont {K.}~\bibnamefont
				{Poder}}, \bibinfo {author} {\bibfnamefont {M.}~\bibnamefont {Tamburini}},
			\bibinfo {author} {\bibfnamefont {G.}~\bibnamefont {Sarri}}, \bibinfo
			{author} {\bibfnamefont {A.}~\bibnamefont {Di~Piazza}}, \bibinfo {author}
			{\bibfnamefont {S.}~\bibnamefont {Kuschel}}, \bibinfo {author} {\bibfnamefont
				{C.~D.}\ \bibnamefont {Baird}}, \bibinfo {author} {\bibfnamefont
				{K.}~\bibnamefont {Behm}}, \bibinfo {author} {\bibfnamefont {S.}~\bibnamefont
				{Bohlen}}, \bibinfo {author} {\bibfnamefont {J.~M.}\ \bibnamefont {Cole}},
			\bibinfo {author} {\bibfnamefont {D.~J.}\ \bibnamefont {Corvan}}, \bibinfo
			{author} {\bibfnamefont {M.}~\bibnamefont {Duff}}, \bibinfo {author}
			{\bibfnamefont {E.}~\bibnamefont {Gerstmayr}}, \bibinfo {author}
			{\bibfnamefont {C.~H.}\ \bibnamefont {Keitel}}, \bibinfo {author}
			{\bibfnamefont {K.}~\bibnamefont {Krushelnick}}, \bibinfo {author}
			{\bibfnamefont {S.~P.~D.}\ \bibnamefont {Mangles}}, \bibinfo {author}
			{\bibfnamefont {P.}~\bibnamefont {McKenna}}, \bibinfo {author} {\bibfnamefont
				{C.~D.}\ \bibnamefont {Murphy}}, \bibinfo {author} {\bibfnamefont
				{Z.}~\bibnamefont {Najmudin}}, \bibinfo {author} {\bibfnamefont {C.~P.}\
				\bibnamefont {Ridgers}}, \bibinfo {author} {\bibfnamefont {G.~M.}\
				\bibnamefont {Samarin}}, \bibinfo {author} {\bibfnamefont {D.~R.}\
				\bibnamefont {Symes}}, \bibinfo {author} {\bibfnamefont {A.~G.~R.}\
				\bibnamefont {Thomas}}, \bibinfo {author} {\bibfnamefont {J.}~\bibnamefont
				{Warwick}}, \ and\ \bibinfo {author} {\bibfnamefont {M.}~\bibnamefont
				{Zepf}},\ }\bibfield  {title} {\enquote {\bibinfo {title} {Experimental
					signatures of the quantum nature of radiation reaction in the field of an
					ultraintense laser},}\ }\href {\doibase 10.1103/PhysRevX.8.031004} {\bibfield
			{journal} {\bibinfo  {journal} {Phys. Rev. X}\ }\textbf {\bibinfo {volume}
				{8}},\ \bibinfo {pages} {031004} (\bibinfo {year} {2018})}\BibitemShut
		{NoStop}%
		\bibitem [{\citenamefont {Lipka}\ \emph {et~al.}(2011)\citenamefont {Lipka},
			\citenamefont {Kleen}, \citenamefont {Lund-Nielsen}, \citenamefont {Nlle},
			\citenamefont {Vilcins},\ and\ \citenamefont {Vogel}}]{Lipka2011}%
		\BibitemOpen
		\bibfield  {author} {\bibinfo {author} {\bibfnamefont {D.}~\bibnamefont
				{Lipka}}, \bibinfo {author} {\bibfnamefont {W.}~\bibnamefont {Kleen}},
			\bibinfo {author} {\bibfnamefont {J.}~\bibnamefont {Lund-Nielsen}}, \bibinfo
			{author} {\bibfnamefont {D.}~\bibnamefont {Nlle}}, \bibinfo {author}
			{\bibfnamefont {S.}~\bibnamefont {Vilcins}}, \ and\ \bibinfo {author}
			{\bibfnamefont {V.}~\bibnamefont {Vogel}},\ }\bibfield  {title} {\enquote
			{\bibinfo {title} {Dark current monitor for the european xfel},}\ }\href@noop
		{} {\bibfield  {journal} {\bibinfo  {journal} {Proc. IPAC 2011 (Hamburg)}\ ,\
				\bibinfo {pages} {572--574}} (\bibinfo {year} {2011})}\BibitemShut {NoStop}%
		\bibitem [{\citenamefont {Lipka}, \citenamefont {Lund-Nielsen},\ and\
			\citenamefont {Seebach}(2013)}]{Lipka2013}%
		\BibitemOpen
		\bibfield  {author} {\bibinfo {author} {\bibfnamefont {D.}~\bibnamefont
				{Lipka}}, \bibinfo {author} {\bibfnamefont {J.}~\bibnamefont {Lund-Nielsen}},
			\ and\ \bibinfo {author} {\bibfnamefont {M.}~\bibnamefont {Seebach}},\
		}\bibfield  {title} {\enquote {\bibinfo {title} {Resonator for charge
					measurement at regea},}\ }\href@noop {} {\bibfield  {journal} {\bibinfo
				{journal} {Proc. IBIC 2013 (Oxford)}\ ,\ \bibinfo {pages} {872--875}}
			(\bibinfo {year} {2013})}\BibitemShut {NoStop}%
		\bibitem [{\citenamefont {Kurz}\ \emph {et~al.}(2018)\citenamefont {Kurz},
			\citenamefont {Couperus}, \citenamefont {Krämer}, \citenamefont {Ding},
			\citenamefont {Kuschel}, \citenamefont {Köhler}, \citenamefont {Zarini},
			\citenamefont {Hollatz}, \citenamefont {Schinkel}, \citenamefont {D’Arcy},
			\citenamefont {Schwinkendorf}, \citenamefont {Osterhoff}, \citenamefont
			{Irman}, \citenamefont {Schramm},\ and\ \citenamefont {Karsch}}]{Kurz2018}%
		\BibitemOpen
		\bibfield  {author} {\bibinfo {author} {\bibfnamefont {T.}~\bibnamefont
				{Kurz}}, \bibinfo {author} {\bibfnamefont {J.~P.}\ \bibnamefont {Couperus}},
			\bibinfo {author} {\bibfnamefont {J.~M.}\ \bibnamefont {Krämer}}, \bibinfo
			{author} {\bibfnamefont {H.}~\bibnamefont {Ding}}, \bibinfo {author}
			{\bibfnamefont {S.}~\bibnamefont {Kuschel}}, \bibinfo {author} {\bibfnamefont
				{A.}~\bibnamefont {Köhler}}, \bibinfo {author} {\bibfnamefont
				{O.}~\bibnamefont {Zarini}}, \bibinfo {author} {\bibfnamefont
				{D.}~\bibnamefont {Hollatz}}, \bibinfo {author} {\bibfnamefont
				{D.}~\bibnamefont {Schinkel}}, \bibinfo {author} {\bibfnamefont
				{R.}~\bibnamefont {D’Arcy}}, \bibinfo {author} {\bibfnamefont {J.-P.}\
				\bibnamefont {Schwinkendorf}}, \bibinfo {author} {\bibfnamefont
				{J.}~\bibnamefont {Osterhoff}}, \bibinfo {author} {\bibfnamefont
				{A.}~\bibnamefont {Irman}}, \bibinfo {author} {\bibfnamefont
				{U.}~\bibnamefont {Schramm}}, \ and\ \bibinfo {author} {\bibfnamefont
				{S.}~\bibnamefont {Karsch}},\ }\bibfield  {title} {\enquote {\bibinfo {title}
				{Calibration and cross-laboratory implementation of scintillating screens for
					electron bunch charge determination},}\ }\href {\doibase 10.1063/1.5041755}
		{\bibfield  {journal} {\bibinfo  {journal} {Review of Scientific
					Instruments}\ }\textbf {\bibinfo {volume} {89}},\ \bibinfo {pages} {093303}
			(\bibinfo {year} {2018})},\ \Eprint
		{http://arxiv.org/abs/https://doi.org/10.1063/1.5041755}
		{https://doi.org/10.1063/1.5041755} \BibitemShut {NoStop}%
		\bibitem [{\citenamefont {Schwinkendorf}\ \emph {et~al.}(2019)\citenamefont
			{Schwinkendorf}, \citenamefont {Bohlen}, \citenamefont {Cabada{\u{g}}},
			\citenamefont {Ding}, \citenamefont {Irman}, \citenamefont {Karsch},
			\citenamefont {Köhler}, \citenamefont {Krämer}, \citenamefont {Kurz},
			\citenamefont {Kuschel}, \citenamefont {Osterhoff}, \citenamefont {Schaper},
			\citenamefont {Schinkel}, \citenamefont {Schramm}, \citenamefont {Zarini},\
			and\ \citenamefont {D'Arcy}}]{Schwinkendorf2019}%
		\BibitemOpen
		\bibfield  {author} {\bibinfo {author} {\bibfnamefont {J.-P.}\ \bibnamefont
				{Schwinkendorf}}, \bibinfo {author} {\bibfnamefont {S.}~\bibnamefont
				{Bohlen}}, \bibinfo {author} {\bibfnamefont {J.~C.}\ \bibnamefont
				{Cabada{\u{g}}}}, \bibinfo {author} {\bibfnamefont {H.}~\bibnamefont {Ding}},
			\bibinfo {author} {\bibfnamefont {A.}~\bibnamefont {Irman}}, \bibinfo
			{author} {\bibfnamefont {S.}~\bibnamefont {Karsch}}, \bibinfo {author}
			{\bibfnamefont {A.}~\bibnamefont {Köhler}}, \bibinfo {author} {\bibfnamefont
				{J.}~\bibnamefont {Krämer}}, \bibinfo {author} {\bibfnamefont
				{T.}~\bibnamefont {Kurz}}, \bibinfo {author} {\bibfnamefont {S.}~\bibnamefont
				{Kuschel}}, \bibinfo {author} {\bibfnamefont {J.}~\bibnamefont {Osterhoff}},
			\bibinfo {author} {\bibfnamefont {L.}~\bibnamefont {Schaper}}, \bibinfo
			{author} {\bibfnamefont {D.}~\bibnamefont {Schinkel}}, \bibinfo {author}
			{\bibfnamefont {U.}~\bibnamefont {Schramm}}, \bibinfo {author} {\bibfnamefont
				{O.}~\bibnamefont {Zarini}}, \ and\ \bibinfo {author} {\bibfnamefont
				{R.}~\bibnamefont {D'Arcy}},\ }\bibfield  {title} {\enquote {\bibinfo {title}
				{Charge calibration of {DRZ} scintillation phosphor screens},}\ }\href
		{\doibase 10.1088/1748-0221/14/09/p09025} {\bibfield  {journal} {\bibinfo
				{journal} {Journal of Instrumentation}\ }\textbf {\bibinfo {volume} {14}},\
			\bibinfo {pages} {P09025--P09025} (\bibinfo {year} {2019})}\BibitemShut
		{NoStop}%
		\bibitem [{\citenamefont {Fourmaux}\ \emph {et~al.}(2009)\citenamefont
			{Fourmaux}, \citenamefont {Serbanescu}, \citenamefont {Lecherbourg},
			\citenamefont {Payeur}, \citenamefont {Martin},\ and\ \citenamefont
			{Kieffer}}]{Fourmaux2009}%
		\BibitemOpen
		\bibfield  {author} {\bibinfo {author} {\bibfnamefont {S.}~\bibnamefont
				{Fourmaux}}, \bibinfo {author} {\bibfnamefont {C.}~\bibnamefont
				{Serbanescu}}, \bibinfo {author} {\bibfnamefont {L.}~\bibnamefont
				{Lecherbourg}}, \bibinfo {author} {\bibfnamefont {S.}~\bibnamefont {Payeur}},
			\bibinfo {author} {\bibfnamefont {F.}~\bibnamefont {Martin}}, \ and\ \bibinfo
			{author} {\bibfnamefont {J.~C.}\ \bibnamefont {Kieffer}},\ }\bibfield
		{title} {\enquote {\bibinfo {title} {Investigation of the thermally induced
					laser beam distortion associated with vacuum compressor gratings in high
					energy and high average power femtosecond laser systems},}\ }\href {\doibase
			10.1364/OE.17.000178} {\bibfield  {journal} {\bibinfo  {journal} {Opt.
					Express}\ }\textbf {\bibinfo {volume} {17}},\ \bibinfo {pages} {178--184}
			(\bibinfo {year} {2009})}\BibitemShut {NoStop}%
		\bibitem [{\citenamefont {Leroux}\ \emph {et~al.}(2018)\citenamefont {Leroux},
			\citenamefont {Jolly}, \citenamefont {Schnepp}, \citenamefont {Eichner},
			\citenamefont {Jalas}, \citenamefont {Kirchen}, \citenamefont {Messner},
			\citenamefont {Werle}, \citenamefont {Winkler},\ and\ \citenamefont
			{Maier}}]{Leroux2018}%
		\BibitemOpen
		\bibfield  {author} {\bibinfo {author} {\bibfnamefont {V.}~\bibnamefont
				{Leroux}}, \bibinfo {author} {\bibfnamefont {S.~W.}\ \bibnamefont {Jolly}},
			\bibinfo {author} {\bibfnamefont {M.}~\bibnamefont {Schnepp}}, \bibinfo
			{author} {\bibfnamefont {T.}~\bibnamefont {Eichner}}, \bibinfo {author}
			{\bibfnamefont {S.}~\bibnamefont {Jalas}}, \bibinfo {author} {\bibfnamefont
				{M.}~\bibnamefont {Kirchen}}, \bibinfo {author} {\bibfnamefont
				{P.}~\bibnamefont {Messner}}, \bibinfo {author} {\bibfnamefont
				{C.}~\bibnamefont {Werle}}, \bibinfo {author} {\bibfnamefont
				{P.}~\bibnamefont {Winkler}}, \ and\ \bibinfo {author} {\bibfnamefont
				{A.~R.}\ \bibnamefont {Maier}},\ }\bibfield  {title} {\enquote {\bibinfo
				{title} {Wavefront degradation of a 200 tw laser from heat-induced
					deformation of in-vacuum compressor gratings},}\ }\href {\doibase
			10.1364/OE.26.013061} {\bibfield  {journal} {\bibinfo  {journal} {Opt.
					Express}\ }\textbf {\bibinfo {volume} {26}},\ \bibinfo {pages} {13061--13071}
			(\bibinfo {year} {2018})}\BibitemShut {NoStop}%
		\bibitem [{\citenamefont {Dresselhaus}(2020)}]{Dresselhaus2020}%
		\BibitemOpen
		\bibfield  {author} {\bibinfo {author} {\bibfnamefont {J.~L.}\ \bibnamefont
				{Dresselhaus}},\ }\emph {\bibinfo {title} {{Paving the way of high repetition
					rate laser wakefield acceleration by solving heat induced grating deformation
					issues}}},\ \href {\doibase 10.3204/PUBDB-2020-05063} {Master's thesis},\
		\bibinfo  {school} {University of Hamburg} (\bibinfo {year}
		{2020})\BibitemShut {NoStop}%
		\bibitem [{\citenamefont {Chen}\ \emph {et~al.}(2006)\citenamefont {Chen},
			\citenamefont {Sheng}, \citenamefont {Ma},\ and\ \citenamefont
			{Zhang}}]{Chen2006}%
		\BibitemOpen
		\bibfield  {author} {\bibinfo {author} {\bibfnamefont {M.}~\bibnamefont
				{Chen}}, \bibinfo {author} {\bibfnamefont {Z.-M.}\ \bibnamefont {Sheng}},
			\bibinfo {author} {\bibfnamefont {Y.-Y.}\ \bibnamefont {Ma}}, \ and\ \bibinfo
			{author} {\bibfnamefont {J.}~\bibnamefont {Zhang}},\ }\bibfield  {title}
		{\enquote {\bibinfo {title} {Electron injection and trapping in a laser
					wakefield by field ionization to high-charge states of gases},}\ }\href
		{\doibase 10.1063/1.2179194} {\bibfield  {journal} {\bibinfo  {journal}
				{Journal of Applied Physics}\ }\textbf {\bibinfo {volume} {99}},\ \bibinfo
			{pages} {056109} (\bibinfo {year} {2006})},\ \Eprint
		{http://arxiv.org/abs/https://doi.org/10.1063/1.2179194}
		{https://doi.org/10.1063/1.2179194} \BibitemShut {NoStop}%
		\bibitem [{\citenamefont {Oz}\ \emph {et~al.}(2007)\citenamefont {Oz},
			\citenamefont {Deng}, \citenamefont {Katsouleas}, \citenamefont {Muggli},
			\citenamefont {Barnes}, \citenamefont {Blumenfeld}, \citenamefont {Decker},
			\citenamefont {Emma}, \citenamefont {Hogan}, \citenamefont {Ischebeck},
			\citenamefont {Iverson}, \citenamefont {Kirby}, \citenamefont {Krejcik},
			\citenamefont {O'Connell}, \citenamefont {Siemann}, \citenamefont {Walz},
			\citenamefont {Auerbach}, \citenamefont {Clayton}, \citenamefont {Huang},
			\citenamefont {Johnson}, \citenamefont {Joshi}, \citenamefont {Lu},
			\citenamefont {Marsh}, \citenamefont {Mori},\ and\ \citenamefont
			{Zhou}}]{Oz2007}%
		\BibitemOpen
		\bibfield  {author} {\bibinfo {author} {\bibfnamefont {E.}~\bibnamefont
				{Oz}}, \bibinfo {author} {\bibfnamefont {S.}~\bibnamefont {Deng}}, \bibinfo
			{author} {\bibfnamefont {T.}~\bibnamefont {Katsouleas}}, \bibinfo {author}
			{\bibfnamefont {P.}~\bibnamefont {Muggli}}, \bibinfo {author} {\bibfnamefont
				{C.~D.}\ \bibnamefont {Barnes}}, \bibinfo {author} {\bibfnamefont
				{I.}~\bibnamefont {Blumenfeld}}, \bibinfo {author} {\bibfnamefont {F.~J.}\
				\bibnamefont {Decker}}, \bibinfo {author} {\bibfnamefont {P.}~\bibnamefont
				{Emma}}, \bibinfo {author} {\bibfnamefont {M.~J.}\ \bibnamefont {Hogan}},
			\bibinfo {author} {\bibfnamefont {R.}~\bibnamefont {Ischebeck}}, \bibinfo
			{author} {\bibfnamefont {R.~H.}\ \bibnamefont {Iverson}}, \bibinfo {author}
			{\bibfnamefont {N.}~\bibnamefont {Kirby}}, \bibinfo {author} {\bibfnamefont
				{P.}~\bibnamefont {Krejcik}}, \bibinfo {author} {\bibfnamefont
				{C.}~\bibnamefont {O'Connell}}, \bibinfo {author} {\bibfnamefont {R.~H.}\
				\bibnamefont {Siemann}}, \bibinfo {author} {\bibfnamefont {D.}~\bibnamefont
				{Walz}}, \bibinfo {author} {\bibfnamefont {D.}~\bibnamefont {Auerbach}},
			\bibinfo {author} {\bibfnamefont {C.~E.}\ \bibnamefont {Clayton}}, \bibinfo
			{author} {\bibfnamefont {C.}~\bibnamefont {Huang}}, \bibinfo {author}
			{\bibfnamefont {D.~K.}\ \bibnamefont {Johnson}}, \bibinfo {author}
			{\bibfnamefont {C.}~\bibnamefont {Joshi}}, \bibinfo {author} {\bibfnamefont
				{W.}~\bibnamefont {Lu}}, \bibinfo {author} {\bibfnamefont {K.~A.}\
				\bibnamefont {Marsh}}, \bibinfo {author} {\bibfnamefont {W.~B.}\ \bibnamefont
				{Mori}}, \ and\ \bibinfo {author} {\bibfnamefont {M.}~\bibnamefont {Zhou}},\
		}\bibfield  {title} {\enquote {\bibinfo {title} {Ionization-induced electron
					trapping in ultrarelativistic plasma wakes},}\ }\href {\doibase
			10.1103/PhysRevLett.98.084801} {\bibfield  {journal} {\bibinfo  {journal}
				{Phys. Rev. Lett.}\ }\textbf {\bibinfo {volume} {98}},\ \bibinfo {pages}
			{084801} (\bibinfo {year} {2007})}\BibitemShut {NoStop}%
		\bibitem [{\citenamefont {Pak}\ \emph {et~al.}(2010)\citenamefont {Pak},
			\citenamefont {Marsh}, \citenamefont {Martins}, \citenamefont {Lu},
			\citenamefont {Mori},\ and\ \citenamefont {Joshi}}]{Pak2010}%
		\BibitemOpen
		\bibfield  {author} {\bibinfo {author} {\bibfnamefont {A.}~\bibnamefont
				{Pak}}, \bibinfo {author} {\bibfnamefont {K.~A.}\ \bibnamefont {Marsh}},
			\bibinfo {author} {\bibfnamefont {S.~F.}\ \bibnamefont {Martins}}, \bibinfo
			{author} {\bibfnamefont {W.}~\bibnamefont {Lu}}, \bibinfo {author}
			{\bibfnamefont {W.~B.}\ \bibnamefont {Mori}}, \ and\ \bibinfo {author}
			{\bibfnamefont {C.}~\bibnamefont {Joshi}},\ }\bibfield  {title} {\enquote
			{\bibinfo {title} {Injection and trapping of tunnel-ionized electrons into
					laser-produced wakes},}\ }\href {\doibase 10.1103/PhysRevLett.104.025003}
		{\bibfield  {journal} {\bibinfo  {journal} {Phys. Rev. Lett.}\ }\textbf
			{\bibinfo {volume} {104}},\ \bibinfo {pages} {025003} (\bibinfo {year}
			{2010})}\BibitemShut {NoStop}%
		\bibitem [{\citenamefont {McGuffey}\ \emph {et~al.}(2010)\citenamefont
			{McGuffey}, \citenamefont {Thomas}, \citenamefont {Schumaker}, \citenamefont
			{Matsuoka}, \citenamefont {Chvykov}, \citenamefont {Dollar}, \citenamefont
			{Kalintchenko}, \citenamefont {Yanovsky}, \citenamefont {Maksimchuk},
			\citenamefont {Krushelnick}, \citenamefont {Bychenkov}, \citenamefont
			{Glazyrin},\ and\ \citenamefont {Karpeev}}]{McGuffey2010}%
		\BibitemOpen
		\bibfield  {author} {\bibinfo {author} {\bibfnamefont {C.}~\bibnamefont
				{McGuffey}}, \bibinfo {author} {\bibfnamefont {A.~G.~R.}\ \bibnamefont
				{Thomas}}, \bibinfo {author} {\bibfnamefont {W.}~\bibnamefont {Schumaker}},
			\bibinfo {author} {\bibfnamefont {T.}~\bibnamefont {Matsuoka}}, \bibinfo
			{author} {\bibfnamefont {V.}~\bibnamefont {Chvykov}}, \bibinfo {author}
			{\bibfnamefont {F.~J.}\ \bibnamefont {Dollar}}, \bibinfo {author}
			{\bibfnamefont {G.}~\bibnamefont {Kalintchenko}}, \bibinfo {author}
			{\bibfnamefont {V.}~\bibnamefont {Yanovsky}}, \bibinfo {author}
			{\bibfnamefont {A.}~\bibnamefont {Maksimchuk}}, \bibinfo {author}
			{\bibfnamefont {K.}~\bibnamefont {Krushelnick}}, \bibinfo {author}
			{\bibfnamefont {V.~Y.}\ \bibnamefont {Bychenkov}}, \bibinfo {author}
			{\bibfnamefont {I.~V.}\ \bibnamefont {Glazyrin}}, \ and\ \bibinfo {author}
			{\bibfnamefont {A.~V.}\ \bibnamefont {Karpeev}},\ }\bibfield  {title}
		{\enquote {\bibinfo {title} {Ionization induced trapping in a laser wakefield
					accelerator},}\ }\href {\doibase 10.1103/PhysRevLett.104.025004} {\bibfield
			{journal} {\bibinfo  {journal} {Phys. Rev. Lett.}\ }\textbf {\bibinfo
				{volume} {104}},\ \bibinfo {pages} {025004} (\bibinfo {year}
			{2010})}\BibitemShut {NoStop}%
		\bibitem [{\citenamefont {Clayton}\ \emph {et~al.}(2010)\citenamefont
			{Clayton}, \citenamefont {Ralph}, \citenamefont {Albert}, \citenamefont
			{Fonseca}, \citenamefont {Glenzer}, \citenamefont {Joshi}, \citenamefont
			{Lu}, \citenamefont {Marsh}, \citenamefont {Martins}, \citenamefont {Mori},
			\citenamefont {Pak}, \citenamefont {Tsung}, \citenamefont {Pollock},
			\citenamefont {Ross}, \citenamefont {Silva},\ and\ \citenamefont
			{Froula}}]{Clayton2010}%
		\BibitemOpen
		\bibfield  {author} {\bibinfo {author} {\bibfnamefont {C.~E.}\ \bibnamefont
				{Clayton}}, \bibinfo {author} {\bibfnamefont {J.~E.}\ \bibnamefont {Ralph}},
			\bibinfo {author} {\bibfnamefont {F.}~\bibnamefont {Albert}}, \bibinfo
			{author} {\bibfnamefont {R.~A.}\ \bibnamefont {Fonseca}}, \bibinfo {author}
			{\bibfnamefont {S.~H.}\ \bibnamefont {Glenzer}}, \bibinfo {author}
			{\bibfnamefont {C.}~\bibnamefont {Joshi}}, \bibinfo {author} {\bibfnamefont
				{W.}~\bibnamefont {Lu}}, \bibinfo {author} {\bibfnamefont {K.~A.}\
				\bibnamefont {Marsh}}, \bibinfo {author} {\bibfnamefont {S.~F.}\ \bibnamefont
				{Martins}}, \bibinfo {author} {\bibfnamefont {W.~B.}\ \bibnamefont {Mori}},
			\bibinfo {author} {\bibfnamefont {A.}~\bibnamefont {Pak}}, \bibinfo {author}
			{\bibfnamefont {F.~S.}\ \bibnamefont {Tsung}}, \bibinfo {author}
			{\bibfnamefont {B.~B.}\ \bibnamefont {Pollock}}, \bibinfo {author}
			{\bibfnamefont {J.~S.}\ \bibnamefont {Ross}}, \bibinfo {author}
			{\bibfnamefont {L.~O.}\ \bibnamefont {Silva}}, \ and\ \bibinfo {author}
			{\bibfnamefont {D.~H.}\ \bibnamefont {Froula}},\ }\bibfield  {title}
		{\enquote {\bibinfo {title} {Self-guided laser wakefield acceleration beyond
					1 gev using ionization-induced injection},}\ }\href {\doibase
			10.1103/PhysRevLett.105.105003} {\bibfield  {journal} {\bibinfo  {journal}
				{Phys. Rev. Lett.}\ }\textbf {\bibinfo {volume} {105}},\ \bibinfo {pages}
			{105003} (\bibinfo {year} {2010})}\BibitemShut {NoStop}%
		\bibitem [{\citenamefont {Lehe}\ \emph {et~al.}(2016)\citenamefont {Lehe},
			\citenamefont {Kirchen}, \citenamefont {Andriyash}, \citenamefont {Godfrey},\
			and\ \citenamefont {Vay}}]{LEHE2016}%
		\BibitemOpen
		\bibfield  {author} {\bibinfo {author} {\bibfnamefont {R.}~\bibnamefont
				{Lehe}}, \bibinfo {author} {\bibfnamefont {M.}~\bibnamefont {Kirchen}},
			\bibinfo {author} {\bibfnamefont {I.~A.}\ \bibnamefont {Andriyash}}, \bibinfo
			{author} {\bibfnamefont {B.~B.}\ \bibnamefont {Godfrey}}, \ and\ \bibinfo
			{author} {\bibfnamefont {J.-L.}\ \bibnamefont {Vay}},\ }\bibfield  {title}
		{\enquote {\bibinfo {title} {A spectral, quasi-cylindrical and
					dispersion-free particle-in-cell algorithm},}\ }\href {\doibase
			https://doi.org/10.1016/j.cpc.2016.02.007} {\bibfield  {journal} {\bibinfo
				{journal} {Computer Physics Communications}\ }\textbf {\bibinfo {volume}
				{203}},\ \bibinfo {pages} {66--82} (\bibinfo {year} {2016})}\BibitemShut
		{NoStop}%
		\bibitem [{\citenamefont {Lu}\ \emph {et~al.}(2007)\citenamefont {Lu},
			\citenamefont {Tzoufras}, \citenamefont {Joshi}, \citenamefont {Tsung},
			\citenamefont {Mori}, \citenamefont {Vieira}, \citenamefont {Fonseca},\ and\
			\citenamefont {Silva}}]{Lu2007}%
		\BibitemOpen
		\bibfield  {author} {\bibinfo {author} {\bibfnamefont {W.}~\bibnamefont
				{Lu}}, \bibinfo {author} {\bibfnamefont {M.}~\bibnamefont {Tzoufras}},
			\bibinfo {author} {\bibfnamefont {C.}~\bibnamefont {Joshi}}, \bibinfo
			{author} {\bibfnamefont {F.~S.}\ \bibnamefont {Tsung}}, \bibinfo {author}
			{\bibfnamefont {W.~B.}\ \bibnamefont {Mori}}, \bibinfo {author}
			{\bibfnamefont {J.}~\bibnamefont {Vieira}}, \bibinfo {author} {\bibfnamefont
				{R.~A.}\ \bibnamefont {Fonseca}}, \ and\ \bibinfo {author} {\bibfnamefont
				{L.~O.}\ \bibnamefont {Silva}},\ }\bibfield  {title} {\enquote {\bibinfo
				{title} {Generating multi-gev electron bunches using single stage laser
					wakefield acceleration in a 3d nonlinear regime},}\ }\href {\doibase
			10.1103/PhysRevSTAB.10.061301} {\bibfield  {journal} {\bibinfo  {journal}
				{Phys. Rev. ST Accel. Beams}\ }\textbf {\bibinfo {volume} {10}},\ \bibinfo
			{pages} {061301} (\bibinfo {year} {2007})}\BibitemShut {NoStop}%
		\bibitem [{\citenamefont {Hansson}\ \emph {et~al.}(2014)\citenamefont
			{Hansson}, \citenamefont {Senje}, \citenamefont {Persson}, \citenamefont
			{Lundh}, \citenamefont {Wahlstr\"om}, \citenamefont {Desforges},
			\citenamefont {Ju}, \citenamefont {Audet}, \citenamefont {Cros},
			\citenamefont {Dobosz~Dufr\'enoy},\ and\ \citenamefont
			{Monot}}]{Hansson2014}%
		\BibitemOpen
		\bibfield  {author} {\bibinfo {author} {\bibfnamefont {M.}~\bibnamefont
				{Hansson}}, \bibinfo {author} {\bibfnamefont {L.}~\bibnamefont {Senje}},
			\bibinfo {author} {\bibfnamefont {A.}~\bibnamefont {Persson}}, \bibinfo
			{author} {\bibfnamefont {O.}~\bibnamefont {Lundh}}, \bibinfo {author}
			{\bibfnamefont {C.-G.}\ \bibnamefont {Wahlstr\"om}}, \bibinfo {author}
			{\bibfnamefont {F.~G.}\ \bibnamefont {Desforges}}, \bibinfo {author}
			{\bibfnamefont {J.}~\bibnamefont {Ju}}, \bibinfo {author} {\bibfnamefont
				{T.~L.}\ \bibnamefont {Audet}}, \bibinfo {author} {\bibfnamefont
				{B.}~\bibnamefont {Cros}}, \bibinfo {author} {\bibfnamefont {S.}~\bibnamefont
				{Dobosz~Dufr\'enoy}}, \ and\ \bibinfo {author} {\bibfnamefont
				{P.}~\bibnamefont {Monot}},\ }\bibfield  {title} {\enquote {\bibinfo {title}
				{Enhanced stability of laser wakefield acceleration using dielectric
					capillary tubes},}\ }\href {\doibase 10.1103/PhysRevSTAB.17.031303}
		{\bibfield  {journal} {\bibinfo  {journal} {Phys. Rev. ST Accel. Beams}\
			}\textbf {\bibinfo {volume} {17}},\ \bibinfo {pages} {031303} (\bibinfo
			{year} {2014})}\BibitemShut {NoStop}%
		\bibitem [{\citenamefont {Karsch}\ \emph {et~al.}(2007)\citenamefont {Karsch},
			\citenamefont {Osterhoff}, \citenamefont {Popp}, \citenamefont
			{Rowlands-Rees}, \citenamefont {Major}, \citenamefont {Fuchs}, \citenamefont
			{Marx}, \citenamefont {Hörlein}, \citenamefont {Schmid}, \citenamefont
			{Veisz}, \citenamefont {Becker}, \citenamefont {Schramm}, \citenamefont
			{Hidding}, \citenamefont {Pretzler}, \citenamefont {Habs}, \citenamefont
			{Grüner}, \citenamefont {Krausz},\ and\ \citenamefont
			{Hooker}}]{Karsch2007}%
		\BibitemOpen
		\bibfield  {author} {\bibinfo {author} {\bibfnamefont {S.}~\bibnamefont
				{Karsch}}, \bibinfo {author} {\bibfnamefont {J.}~\bibnamefont {Osterhoff}},
			\bibinfo {author} {\bibfnamefont {A.}~\bibnamefont {Popp}}, \bibinfo {author}
			{\bibfnamefont {T.~P.}\ \bibnamefont {Rowlands-Rees}}, \bibinfo {author}
			{\bibfnamefont {Z.}~\bibnamefont {Major}}, \bibinfo {author} {\bibfnamefont
				{M.}~\bibnamefont {Fuchs}}, \bibinfo {author} {\bibfnamefont
				{B.}~\bibnamefont {Marx}}, \bibinfo {author} {\bibfnamefont {R.}~\bibnamefont
				{Hörlein}}, \bibinfo {author} {\bibfnamefont {K.}~\bibnamefont {Schmid}},
			\bibinfo {author} {\bibfnamefont {L.}~\bibnamefont {Veisz}}, \bibinfo
			{author} {\bibfnamefont {S.}~\bibnamefont {Becker}}, \bibinfo {author}
			{\bibfnamefont {U.}~\bibnamefont {Schramm}}, \bibinfo {author} {\bibfnamefont
				{B.}~\bibnamefont {Hidding}}, \bibinfo {author} {\bibfnamefont
				{G.}~\bibnamefont {Pretzler}}, \bibinfo {author} {\bibfnamefont
				{D.}~\bibnamefont {Habs}}, \bibinfo {author} {\bibfnamefont {F.}~\bibnamefont
				{Grüner}}, \bibinfo {author} {\bibfnamefont {F.}~\bibnamefont {Krausz}}, \
			and\ \bibinfo {author} {\bibfnamefont {S.~M.}\ \bibnamefont {Hooker}},\
		}\bibfield  {title} {\enquote {\bibinfo {title} {{GeV}-scale electron
					acceleration in a gas-filled capillary discharge waveguide},}\ }\href
		{\doibase 10.1088/1367-2630/9/11/415} {\bibinfo  {journal} {New J. Phys.}\
		}  {\ \textbf {\bibinfo {volume} {9}},\
			\bibinfo {pages} {415} (\bibinfo {year} {2007})}\BibitemShut {NoStop}%
	\end{thebibliography}
	
	%

\end{document}